\documentclass[aps, prc, reprint, superscriptaddress, twocolumn]{revtex4-1}
\usepackage{amsmath}
\usepackage{amsfonts}
\usepackage{amssymb}
\usepackage{lineno}
\usepackage{siunitx}

\usepackage{float}
\usepackage{graphicx}
\usepackage[caption=false]{subfig}

\usepackage[colorlinks]{hyperref}
\hypersetup{linkcolor=blue, citecolor=blue, filecolor=blue, urlcolor=blue}
\usepackage[all]{hypcap}

\usepackage[utf8]{inputenc}
\usepackage{lipsum}
\usepackage{multirow}

\newcommand{\sNN}{$\sqrt{s_{\mathrm{NN}}}$}
\newcommand{\nhitsdedx}{$\mathrm{N}_{\mathrm{hits}}^{dE/dx}$}
\newcommand{\nhitsfit}{$\mathrm{N_{hits}^{fit}}$}

\begin{document}
\title{Production of Protons and Light Nuclei in Au+Au Collisions at {\sNN} = \SI{3}{GeV} with the STAR Detector}
\affiliation{Abilene Christian University, Abilene, Texas   79699}
\affiliation{AGH University of Krakow, FPACS, Cracow 30-059, Poland}
\affiliation{Argonne National Laboratory, Argonne, Illinois 60439}
\affiliation{American University in Cairo, New Cairo 11835, Egypt}
\affiliation{Ball State University, Muncie, Indiana, 47306}
\affiliation{Brookhaven National Laboratory, Upton, New York 11973}
\affiliation{University of Calabria \& INFN-Cosenza, Rende 87036, Italy}
\affiliation{University of California, Berkeley, California 94720}
\affiliation{University of California, Davis, California 95616}
\affiliation{University of California, Los Angeles, California 90095}
\affiliation{University of California, Riverside, California 92521}
\affiliation{Central China Normal University, Wuhan, Hubei 430079 }
\affiliation{University of Illinois at Chicago, Chicago, Illinois 60607}
\affiliation{Chongqing University, Chongqing, 401331}
\affiliation{Creighton University, Omaha, Nebraska 68178}
\affiliation{Czech Technical University in Prague, FNSPE, Prague 115 19, Czech Republic}
\affiliation{Technische Universit\"at Darmstadt, Darmstadt 64289, Germany}
\affiliation{National Institute of Technology Durgapur, Durgapur - 713209, India}
\affiliation{ELTE E\"otv\"os Lor\'and University, Budapest, Hungary H-1117}
\affiliation{Frankfurt Institute for Advanced Studies FIAS, Frankfurt 60438, Germany}
\affiliation{Fudan University, Shanghai, 200433 }
\affiliation{Guangxi Normal University, Guilin, 541004}
\affiliation{University of Heidelberg, Heidelberg 69120, Germany }
\affiliation{University of Houston, Houston, Texas 77204}
\affiliation{Huzhou University, Huzhou, Zhejiang  313000}
\affiliation{Indian Institute of Science Education and Research (IISER), Berhampur 760010 , India}
\affiliation{Indian Institute of Science Education and Research (IISER) Tirupati, Tirupati 517507, India}
\affiliation{Indian Institute Technology, Patna, Bihar 801106, India}
\affiliation{Indiana University, Bloomington, Indiana 47408}
\affiliation{Institute of Modern Physics, Chinese Academy of Sciences, Lanzhou, Gansu 730000 }
\affiliation{University of Jammu, Jammu 180001, India}
\affiliation{Kent State University, Kent, Ohio 44242}
\affiliation{University of Kentucky, Lexington, Kentucky 40506-0055}
\affiliation{Lawrence Berkeley National Laboratory, Berkeley, California 94720}
\affiliation{Lehigh University, Bethlehem, Pennsylvania 18015}
\affiliation{Max-Planck-Institut f\"ur Physik, Munich 80805, Germany}
\affiliation{Michigan State University, East Lansing, Michigan 48824}
\affiliation{National Institute of Science Education and Research, HBNI, Jatni 752050, India}
\affiliation{National Cheng Kung University, Tainan 70101 }
\affiliation{Nuclear Physics Institute of the CAS, Rez 250 68, Czech Republic}
\affiliation{The Ohio State University, Columbus, Ohio 43210}
\affiliation{Institute of Nuclear Physics PAN, Cracow 31-342, Poland}
\affiliation{Panjab University, Chandigarh 160014, India}
\affiliation{Purdue University, West Lafayette, Indiana 47907}
\affiliation{Rice University, Houston, Texas 77251}
\affiliation{Rutgers University, Piscataway, New Jersey 08854}
\affiliation{University of Science and Technology of China, Hefei, Anhui 230026}
\affiliation{South China Normal University, Guangzhou, Guangdong 510631}
\affiliation{Sejong University, Seoul, 05006, South Korea}
\affiliation{Shandong University, Qingdao, Shandong 266237}
\affiliation{Shanghai Institute of Applied Physics, Chinese Academy of Sciences, Shanghai 201800}
\affiliation{Southern Connecticut State University, New Haven, Connecticut 06515}
\affiliation{State University of New York, Stony Brook, New York 11794}
\affiliation{Instituto de Alta Investigaci\'on, Universidad de Tarapac\'a, Arica 1000000, Chile}
\affiliation{Temple University, Philadelphia, Pennsylvania 19122}
\affiliation{Texas A\&M University, College Station, Texas 77843}
\affiliation{University of Texas, Austin, Texas 78712}
\affiliation{Tsinghua University, Beijing 100084}
\affiliation{University of Tsukuba, Tsukuba, Ibaraki 305-8571, Japan}
\affiliation{University of Chinese Academy of Sciences, Beijing, 101408}
\affiliation{United States Naval Academy, Annapolis, Maryland 21402}
\affiliation{Valparaiso University, Valparaiso, Indiana 46383}
\affiliation{Variable Energy Cyclotron Centre, Kolkata 700064, India}
\affiliation{Warsaw University of Technology, Warsaw 00-661, Poland}
\affiliation{Wayne State University, Detroit, Michigan 48201}
\affiliation{Wuhan University of Science and Technology, Wuhan, Hubei 430065}
\affiliation{Yale University, New Haven, Connecticut 06520}

\author{M.~I.~Abdulhamid}\affiliation{American University in Cairo, New Cairo 11835, Egypt}
\author{B.~E.~Aboona}\affiliation{Texas A\&M University, College Station, Texas 77843}
\author{J.~Adam}\affiliation{Czech Technical University in Prague, FNSPE, Prague 115 19, Czech Republic}
\author{L.~Adamczyk}\affiliation{AGH University of Krakow, FPACS, Cracow 30-059, Poland}
\author{J.~R.~Adams}\affiliation{The Ohio State University, Columbus, Ohio 43210}
\author{I.~Aggarwal}\affiliation{Panjab University, Chandigarh 160014, India}
\author{M.~M.~Aggarwal}\affiliation{Panjab University, Chandigarh 160014, India}
\author{Z.~Ahammed}\affiliation{Variable Energy Cyclotron Centre, Kolkata 700064, India}
\author{E.~C.~Aschenauer}\affiliation{Brookhaven National Laboratory, Upton, New York 11973}
\author{S.~Aslam}\affiliation{Indian Institute Technology, Patna, Bihar 801106, India}
% ...

\collaboration{STAR Collaboration}\noaffiliation

\begin{abstract}
We report the systematic measurement of protons and light nuclei production in Au+Au collisions at $\sqrt{s_{\mathrm{NN}}}$ = \SI{3}{GeV} by the STAR experiment at the Relativistic Heavy Ion Collider (RHIC). The transverse momentum ($p_{T}$) spectra of protons ($p$), deuterons ($d$), tritons ($t$), $^{3}\mathrm{He}$, and $^{4}\mathrm{He}$ have been measured from mid-rapidity to target rapidity for different collision centralities. We present the rapidity and centrality dependence of particle yields ($dN/dy$), average transverse momentum ($\langle p_{T}\rangle$),  yield ratios ($d/p$, $t/p$,$^{3}\mathrm{He}/p$, $^{4}\mathrm{He}/p$), as well as the coalescence parameters ($B_2$, $B_3$). The 4$\pi$ yields for various particles are determined by utilizing the measured rapidity distributions, $dN/dy$. Furthermore, we present the energy, centrality, and rapidity dependence of the compound yield ratios ($N_{p} \times N_{t} / N_{d}^{2}$) and compare them with various model calculations. 
The physics implications of these results on the production mechanism of light nuclei and the QCD phase structure are discussed.  

\end{abstract}

\maketitle

\section{Introduction}
Relativistic heavy-ion collisions provide a unique experimental tool to investigate the Quantum Chromodynamics (QCD) phase diagram and the properties of strongly interacting nuclear matter under extreme conditions~\cite{BRAHMS:2004adc,PHENIX:2004vcz,PHOBOS:2004zne,STAR:2005gfr,ALICE:2022wpn,Chen:2018tnh,Luo:2022mtp,Chen:2024zwk}. At vanishing baryon chemical potential ($\mu_{B} = 0$ MeV), Lattice QCD calculations reveal that the transition between hadronic matter and a Quark-Gluon Plasma (QGP) is a smooth crossover~\cite{Aoki:2006we}. QCD-based effective theories~\cite{Fu:2019hdw,Gao:2020fbl} predict that there is a first-order phase transition and a critical point (CP) at high $\mu_{B}$~\cite{Stephanov:1999zu,Ejiri:2008xt,Borderie:2019fii}. Mapping the QCD phase structure at high baryon density, namely the first-order phase transition boundary and the location of the CP, is the primary goal of the Beam Energy Scan (BES) program at the Relativistic Heavy-ion Collider (RHIC)~\cite{STAR:2010vob,Luo:2017faz}. Between 2010 and 2021, RHIC completed data collection for both phases of RHIC BES (BES-I and BES-II) in succession. During this period, the STAR experiment recorded data of Au+Au collisions at {\sNN} = 7.7 -- \SI{200}{GeV} in collider mode. In addition, the STAR detector operated in Fixed-Target (FXT) mode, collecting data of Au+Au collisions at {\sNN} = 3 -- \SI{7.7}{GeV}, which allows us to access the QCD phase diagram with $\mu_{B}$ up to approximately 750 MeV.

Light nuclei are loosely bound objects with binding energies on the order of a few MeV. Over the past half century, their production in heavy-ion collisions across a wide range of energies has been extensively studied both experimentally~\cite{E814:1994kon,E802:1999hit,E864:2000auv,FOPI:2010xrt,STAR:2010gyg,STAR:2011eej,ALICE:2015wav,NA49:2016qvu,STAR:2016ydv,ALICE:2017xrp,STAR:2019sjh,PhysRevLett.130.202301} and theoretically~\cite{Scheibl:1998tk,Oh:2009gx,Andronic:2010qu,Cleymans:2011pe,Shah:2015oha,Sun:2017xrx,Andronic:2017pug,Braun-Munzinger:2018hat,Shuryak:2020yrs,Donigus:2020ctf,Vovchenko:2020dmv,Zhao:2021dka}.
Two primary mechanisms proposed to explain their production in heavy-ion collisions are nucleon coalescence and thermal emission.
Based on the nucleon coalescence model, which combines dynamical models with the coalescence approach, predicts that the compound yield ratio $N_{p} \times N_{t} / N_{d}^2$ is sensitive to the neutron density fluctuations. This sensitivity makes it a valuable observable for probing first-order phase transitions and CP in the QCD phase diagram~\cite{Sun:2018jhg,Shuryak:2019ikv}.
In RHIC BES-I, the STAR experiment measured the production of deuterons~\cite{STAR:2019sjh} and tritons~\cite{PhysRevLett.130.202301} in Au+Au collisions at {\sNN} = 7.7 -- \SI{200}{GeV}. An enhancement in the yield ratio $N_{p} \times N_{t} / N_{d}^2$ relative to the coalescence baseline was observed in the 0-10\% central Au+Au collisions at {\sNN} = 19.6 and \SI{27}{GeV}, with a combined significance of 4.1$\sigma$~\cite{PhysRevLett.130.202301}. To determine whether these enhancements are related to a first-order phase transition or CP, dynamical modeling of heavy-ion collisions with a realistic equation of state is needed to compare with the experimental data. Additionally, the yield ratio $N_{p} \times N_{t} / N_{d}^2$ exhibits a scaling behavior that monotonically decreases with increasing charged-particle multiplicity ($dN/d\eta$), independent of collision energy and centrality~\cite{PhysRevLett.130.202301}. 
The observed decreasing trend and scaling behavior can be nicely explained by coalescence models, whereas the thermal model predicts an opposite trend. Thus, the systematic measurement of light nuclei production in heavy-ion collisions over a wide energy range serves as a valuable tool not only to probe the QCD phase structure but also to gain insight into the underlying production mechanism.

In this paper, we present the transverse momentum ($p_{T}$) spectra of protons ($p$), deuterons ($d$), tritons ($t$), $^{3}\mathrm{He}$, and $^{4}\mathrm{He}$ in FXT Au+Au collisions at {\sNN} = \SI{3}{GeV}. The analysis encompasses four centrality ranges (0-10\%, 10-20\%, 20-40\%, and 40-80\%) and spans from mid-rapidity to target rapidity. The proton spectra were obtained by subtracting the contributions from weak decays of hyperons. For light nuclei, no corrections for feed-down effects from excited states were applied, and we report their inclusive yields.
By fitting the $p_{T}$ spectra with a blast-wave model, we obtained the centrality and rapidity dependence of $dN/dy$ and $\langle p_{T}\rangle$ for various particles. Due to broad rapidity coverage, the $4 \pi$ yields can be obtained by integrating the particle $dN/dy$ from mid-rapidity to target rapidity. We further compare the measured yield and yield ratios of protons and light nuclei with calculations from various transport models, including the Jet AA Microscopic Transportation Model (JAM)~\cite{Nara:2019crj}, Simulating Many Accelerated Strongly-interacting Hadrons (SMASH)~\cite{SMASH:2016zqf}, Ultra-relativistic Quantum Molecular Dynamics (UrQMD)~\cite{Bass:1998ca}, and Parton Hadron Quantum Molecular Dynamics (PHQMD)~\cite{Glassel:2021rod} models. In these model calculations, except PHQMD, the light nuclei were produced from nucleon coalescence based on the formation probability from the Wigner function~\cite{Chen:2003qj,Zhao:2018lyf}.
Finally, we discuss the beam energy dependence of the coalescence parameters ($B_{2}$, $B_{3}$) and particle yield ratios ($d/p$, $t/p$, $N_{p} \times N_{t} / N_{d}^{2}$).

\section{Experiment and data analysis}
\subsection{Dataset and Event Selection}
The STAR Fixed-Target (FXT) program~\cite{STAR:2020dav,STAR:2021fge} was conducted to achieve lower center-of-mass energies and, thus higher baryon density. A gold foil target, with a thickness of \SI{250}{\micro\meter}, corresponding to a 1\% interaction probability~\cite{Sun:2008hg,STAR:2021fge}, was installed in the vacuum pipe at \SI{200.7}{cm} west of the nominal interaction point and \SI{2}{cm} down from the central beam-pipe axis of the STAR detector.
\begin{figure}[htb]\label{fig:PID}
\centering
    \includegraphics[width=0.85\columnwidth]{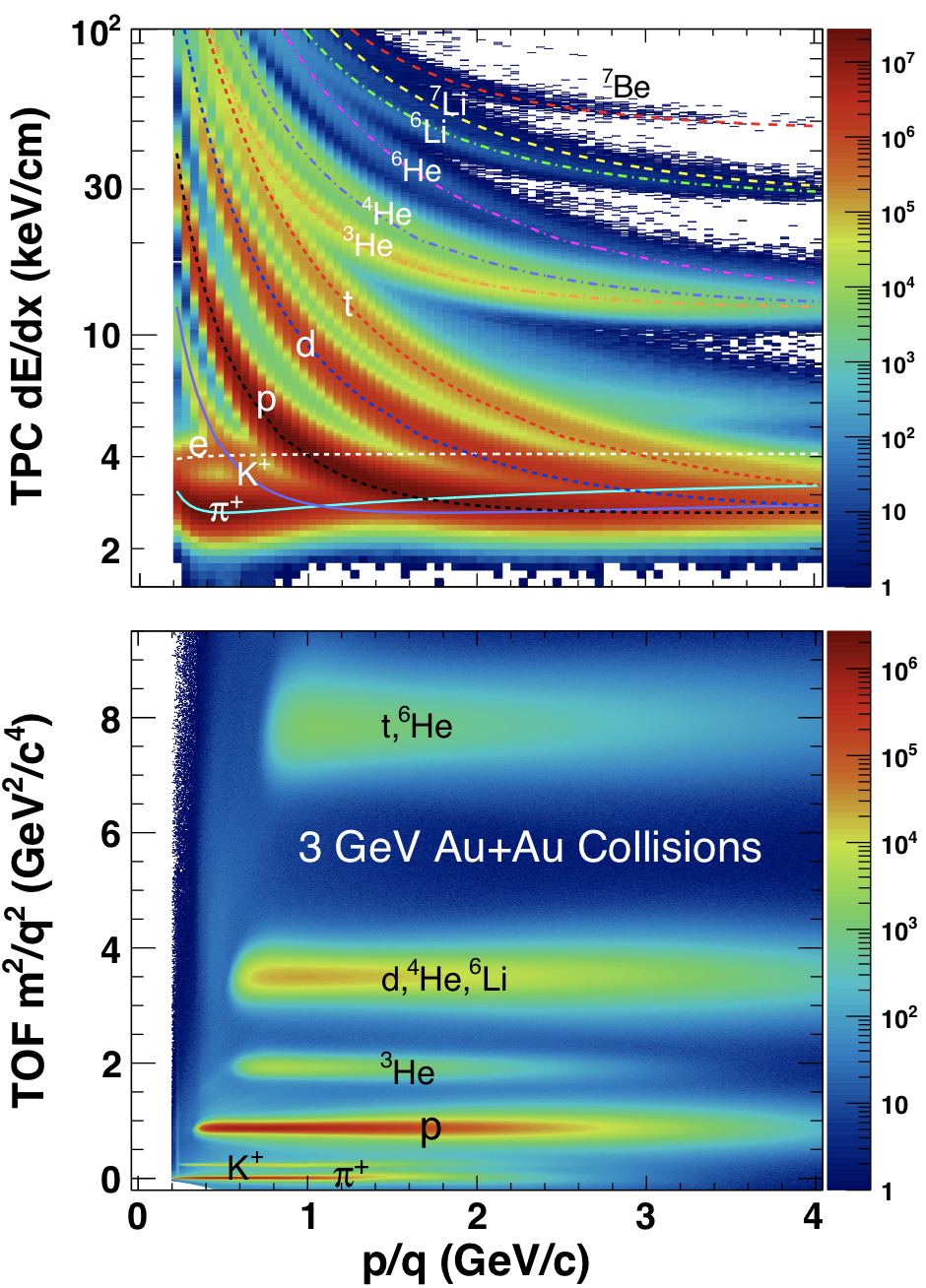}
	\caption{(Top) The $\langle dE/dx \rangle$ of charged tracks versus particle rigidity from Au+Au collisions at {\sNN} = \SI{3}{GeV}. The lines are Bichsel theoretical curves for the corresponding particles. (Bottom) The $m^{2}/q^{2}$ versus rigidity distribution of particle.}
\end{figure}
The experimental data were obtained by bombarding the gold (Au) target with an Au beam of \SI{3.85}{GeV/u}. The corresponding center-of-mass energy is $\sqrt{s_{\mathrm{NN}}}$ = \SI{3}{GeV}. 
The collision vertex of each event was required to have a $z$-coordinate (parallel to the beam axis) located within \SI{2}{cm} of the detector's fixed target position (set 198 $\leq \mathrm{V}_{z} \leq$ \SI{202}{cm} for this analysis), and a transverse $(x, y)$ position within a circle of radius \SI{2}{cm} centered \SI{2}{cm} from the beamline (set $\mathrm{V}_{x}^{2} + (\mathrm{V}_{y} + 2)^{2} < (\SI{2}{cm})^{2}$ for this analysis).
In total, about 260 million minimum-bias events were retained after applying offline event selection criteria with careful quality assurance. Minimum-bias events were selected by detecting simultaneous signals from the Beam-Beam Counter (BBC)~\cite{Bieser:2002ah} and Time of Flight (TOF)~\cite{Llope:2012zz} detector systems.
Collision centralities were determined by fitting the charged-particle multiplicity (referred to below as FXTMult), calculated from the raw number of tracks within the pseudo-rapidity range $-2 < \eta < 0$, using the Glauber model.
In the analysis, four centrality bins (0-10\%, 10-20\%, 20-40\%, and 40-80\%) were used. The FXTMult ranges and the mean values of the number of participating nucleons $\langle N_{\mathrm{Part}}\rangle$ for the corresponding centrality bins are shown in Table~\ref{tab:centrality}.
\begin{table}[h]
\centering
    \caption{Centrality definition and the corresponding mean value of $\langle N_{\mathrm{Part}}\rangle$ along with the statistical and systematic uncertainties in Au+Au collisions at $\sqrt{s_{\mathrm{NN}}}$ = \SI{3}{GeV}.}
    \begin{tabular}{lcc}
    \hline \hline
    Centrality  &   FXTMult     &   $\langle N_{\mathrm{Part}}\rangle$         \\
    \hline
    $0-10\%$    &   $195-119$   &   $310.7 \pm 0.1 \pm 8.3$    \\
    $10-20\%$   &   $118-86$    &   $224.2 \pm 0.1 \pm 8.0$    \\
    $20-40\%$   &   $85-41$     &   $135.0 \pm 0.1 \pm 5.3$    \\
    $40-80\%$   &   $40-5$      &   $39.7  \pm 0.1 \pm 1.9$    \\
    \hline \hline
    \end{tabular}
    \label{tab:centrality}
\end{table}

\begin{figure*}[htp]\label{fig:Phase Space}
\centering
    \subfloat{
        \includegraphics[width=1.6\columnwidth]{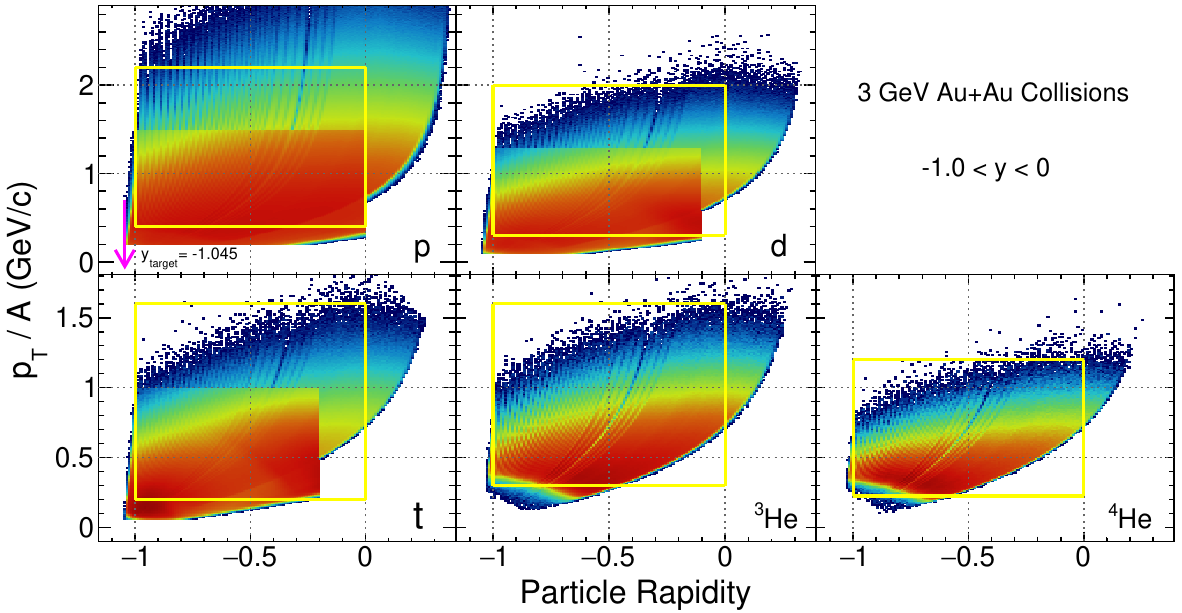}
    }
	\caption{Atomic mass number normalized transverse momentum ($p_T/A$) versus rapidity distributions for identified protons, deuterons, tritons, $^{3}\mathrm{He}$, and $^{4}\mathrm{He}$. For protons, deuterons, and tritons, the portion identified only by the TPC is superimposed. Yellow boxes indicate the regions for further analysis.}

\end{figure*}

\subsection{Track Selection and Particle Identification}\label{trackcuts}

In the analysis, particle identification was performed using the Time Projection Chamber (TPC)~\cite{Anderson:2003ur} and the Time of Flight (TOF)~\cite{Llope:2012zz} detectors. To ensure the track quality, at least 20 out of the maximum 45 possible hits (\nhitsfit) in the TPC were required to reconstruct a track. To prevent multiple counting of reconstructed tracks from the same particle, the number of hits used needed to exceed 52\% of the maximum possible fit points. 
In addition, the number of points (\nhitsdedx) used to calculate the energy loss ($dE/dx$) value was also required to be greater than 10. The distance of the closest approach (DCA) from the reconstructed track to the primary vertex was required to be less than \SI{3}{cm} for protons and \SI{1}{cm} for light nuclei, to suppress contamination from spallation in the beam pipe.
In the TOF measurement, additional track cuts were applied based on the hit positions of tracks at local Y and Z coordinates ($|$btofYLocal$| <$ \SI{1.8}{cm} and $|$btofZLocal$| <$ \SI{2.8}{cm}) to reduce the mismatch between TPC tracks and TOF hits.

The top panel of Fig.~\ref{fig:PID} shows the ionization energy loss of charged-particles ($dE/dx$) measured by the TPC versus particle rigidity ($p/q$), where $q$ is the particle charge. To identify the specified particles, the variables $n \sigma_{p}$ and $Z_{ln}$ were defined as: 
\begin{equation}
	n \sigma_{p} = \frac{1}{\sigma_{R}} \ln \frac{\langle dE/dx\rangle}{\langle dE/dx\rangle_{Bichsel}^{p}}
\end{equation}
\begin{equation}\label{ZLN}
    Z_{ln} =\ln \frac{\langle dE/dx\rangle}{\langle dE/dx\rangle_{Bichsel}^{ln}}
\end{equation}
where the $\langle dE/dx\rangle_{Bichsel}$ ($\langle dE/dx\rangle_{Bichsel}^{p}$ for protons, $\langle dE/dx\rangle_{Bichsel}^{ln}$ for light clusters / light nuclei) is the theoretical value of the energy loss obtained from the Bichsel function~\cite{Shao:2005iu}, which was represented by a dashed line in the figure, and the parameter $\sigma_{R}$ denotes the TPC $\ln \langle dE/dx\rangle $ resolution ($\sim 8 \%$).
At low momentum, the raw signal of protons was obtained by fitting a Gaussian function to the distribution of $n \sigma_{p}$~\cite{STAR:2021iop}, and the raw signal of light nuclei was obtained by fitting a Gaussian function to the $Z_{ln}$-distribution described in Eq.~\ref{ZLN}. At high momentum, the raw signal was extracted by using the mass squared ($m^{2}$) distributions from the TOF detector, in addition to the $n \sigma_{p}$ or $Z_{ln}$ information. The $m^{2}$ was calculated as: 
\begin{equation}
 {m}^{2}={p}^{2}\left(\frac{1}{\beta^2}-1\right) 
\end{equation}
where $p$ is the momentum of the particle, $\beta = L / ct$, and $L, c, t$ represent the track path length, the speed of light, and the time of flight, respectively.
Table~\ref{tab:PID} lists the rapidity ranges and transverse momentum ($p_{T}$) cutoffs for particle identification using TPC and TOF, the rapidity is calculated in the center-of-mass frame. 

The bottom panel of Fig.~\ref{fig:PID} shows the particle $m^{2}$ versus particle rigidity ($p/q$). One can observe that there are very clear bands in the $m^{2}$ distribution of different particles. The particle signals were extracted by fitting the $m^{2}$ distributions using a Student-t function~\cite{10.1093/biomet/83.4.891,STAR:2019sjh} with an exponential background tail. 
For particles in the low transverse momentum ($p_{T}$) region, particle identification generally relies on the $dE/dx$ information from the TPC. 
However, it was found that TPC information alone was insufficient to accurately identify deuterons in the rapidity range of $-0.1 < y < 0$ and tritons in the range of $-0.2 < y < 0$. 
Consequently, TOF information was also utilized in these specific rapidity ranges, even at low $p_{T}$.
Figure \ref{fig:Phase Space} shows the phase space coverage ($p_{T}$ versus rapidity) for each particle. The rapidity range measured for each particle in this analysis was -1.0 to 0, which is denoted by yellow boxes in Fig.~\ref{fig:Phase Space}. 

\begin{table}[h]
\centering
    \caption{The $p_{T}$ range (in GeV/c) of PID by TPC or TPC+TOF for different particles.}
    \begin{tabular}{lcc}
    \hline  \hline
    Particle                    & TPC               & TPC+TOF           \\
    \hline  
    proton                      & $p_{T} \leq 1.5$  & $p_{T} > 1.5$     \\
    deuteron$(-0.1<y<0)$        & ---               & $p_{T} > 0.6$     \\
    deuteron$(-1.0<y<-0.1)$     & $p_{T} \leq 2.6$  & $p_{T} > 2.6$     \\
    triton$(-0.2<y<0)$          & ---               & $p_{T} > 0.6$     \\
    triton$(-1.0<y<-0.2)$       & $p_{T} \leq 3.0$  & $p_{T} > 3.0$     \\
    $^{3}\mathrm{He}$           & ---               & $p_{T} \geq 0.9$  \\
    $^{4}\mathrm{He}$           & ---               & $p_{T} \geq 0.9$  \\
    \hline \hline
    \end{tabular}
    \label{tab:PID}
\end{table}

\begin{table*}[htb]
\centering
\caption{Weak decay feed-down fraction of protons (\%) at different centralities in Au+Au collisions at {\sNN} = \SI{3}{GeV}. The uncertainties represent statistical and systematic uncertainties, respectively.}
    \begin{tabular}{c|c|c|c|c}
    \hline \hline
    Rapidity  & 0-10\%                          &       10-20\%                 &   20-40\%                     &   40-80\%     \\  
    \hline
    -0.1 $<$ y $<$ 0    & 1.838 $\pm$ 0.017 $\pm$ 0.214 & 1.642 $\pm$ 0.013 $\pm$ 0.184 & 1.500 $\pm$ 0.006 $\pm$ 0.173 & 1.046 $\pm$ 0.002 $\pm$ 0.140  \\
    -0.2 $<$ y $<$ -0.1 & 1.984 $\pm$ 0.017 $\pm$ 0.198 & 1.774 $\pm$ 0.012 $\pm$ 0.182 & 1.511 $\pm$ 0.006 $\pm$ 0.167 & 1.016 $\pm$ 0.002 $\pm$ 0.103 \\
    -0.3 $<$ y $<$ -0.2 & 2.029 $\pm$ 0.015 $\pm$ 0.188 & 1.809 $\pm$ 0.011 $\pm$ 0.162 & 1.481 $\pm$ 0.005 $\pm$ 0.148 & 2.901 $\pm$ 0.004 $\pm$ 0.333 \\
    -0.4 $<$ y $<$ -0.3 & 1.895 $\pm$ 0.014 $\pm$ 0.197 & 1.637 $\pm$ 0.010 $\pm$ 0.147 & 1.490 $\pm$ 0.005 $\pm$ 0.155 & 0.969 $\pm$ 0.001 $\pm$ 0.084 \\
    -0.5 $<$ y $<$ -0.4 & 1.813 $\pm$ 0.013 $\pm$ 0.168 & 1.616 $\pm$ 0.010	$\pm$ 0.154 & 1.288 $\pm$ 0.005 $\pm$ 0.119 & 0.816 $\pm$ 0.001 $\pm$ 0.094 \\
    -0.6 $<$ y $<$ -0.5 & 1.685 $\pm$ 0.012 $\pm$ 0.155 & 1.456 $\pm$ 0.009 $\pm$ 0.127 & 1.182 $\pm$ 0.004 $\pm$ 0.119 & 0.718 $\pm$ 0.001 $\pm$ 0.093 \\
    -0.7 $<$ y $<$ -0.6 & 1.490 $\pm$ 0.010 $\pm$ 0.140 & 1.355 $\pm$ 0.009 $\pm$ 0.112 & 1.034 $\pm$ 0.004 $\pm$ 0.090 & 0.722 $\pm$ 0.001 $\pm$ 0.086 \\
    -0.8 $<$ y $<$ -0.7 & 1.299 $\pm$ 0.008 $\pm$ 0.117 & 1.121 $\pm$ 0.007 $\pm$ 0.088 & 0.854 $\pm$ 0.003 $\pm$ 0.080 & 0.569 $\pm$ 0.001 $\pm$ 0.103 \\
    -0.9 $<$ y $<$ -0.8 & 1.096 $\pm$ 0.007 $\pm$ 0.086 & 0.866 $\pm$ 0.005 $\pm$ 0.073 & 0.699 $\pm$ 0.003 $\pm$ 0.079 & 0.293 $\pm$ 0.001 $\pm$ 0.059 \\
    -1.0 $<$ y $<$ -0.9 & 1.141 $\pm$ 0.006 $\pm$ 0.102 & 0.883 $\pm$ 0.005 $\pm$ 0.100 & 0.687 $\pm$ 0.003 $\pm$ 0.065 & 0.329 $\pm$ 0.001 $\pm$ 0.072 \\
    \hline \hline  
    \end{tabular}
    \label{tab:feeddownvalue}
\end{table*}

\subsection{Efficiency Correction and Energy Loss Correction}
To obtain final particle spectra in each rapidity interval, efficiency corrections and energy loss corrections were applied. 
The TPC tracking acceptance and reconstruction efficiency were determined by the so-called embedding technique. 
In this technique, sampled Monte Carlo (MC) tracks, simulated within a given kinematic range using a GEANT model~\cite{STAR:2001mal,STAR:2019bjj} of the STAR detector and detector response simulators, are embedded into real events at the raw data level. This approach allows for assessing the quality and quantity of the reconstructed embedded tracks.
The acceptance and reconstruction efficiency is finally given by the ratio of the number of reconstructed tracks to the number of embedded MC tracks, 
as shown in Eq.~\ref{tpceff}:
\begin{equation}\label{tpceff}
    \varepsilon_{\mathrm{TPC}}(p_{T})=\frac{N_{\mathrm{rec.}}(p_{T})}{N_{\mathrm{emb.}}(p_{T})}
\end{equation}
where $N_{\mathrm{rec.}}$ and $N_{\mathrm{emb.}}$ are the number of reconstructed MC tracks satisfying the track quality cuts and the number of embedded MC tracks in a single $p_{T}$ bin, respectively. The TOF matching efficiency is defined as the ratio of the number of tracks matched to the TOF to the number of tracks identified by TPC. 

Low-momentum particles experience a substantial energy loss while traversing the detector material. Thus, it is necessary to correct the energy loss of these particles, especially the heavier ones.
The energy loss can be corrected with the embedding data by comparing the $p_{T}$ difference between the reconstructed and embedded MC tracks. The $p_{T}$-dependent correction factor was parametrized with Eq.~\ref{eq:eneloss}.
\begin{equation}\label{eq:eneloss}
    p^{rec.}_{T}-p^{MC}_{T}=p_{0}+p_{1}\left(1+\frac{p_{2}}{{\left(p^{rec.}_{T}\right)}^{2}}\right)^{p_{3}}
\end{equation}
where $p_{0}$, $p_{1}$, $p_{2}$, and $p_{3}$ are the fit parameters. For each particle, a set of fit parameters was obtained to estimate the $p_{T}$-dependent energy loss effect. These parameters were then utilized to correct the final particle $p_{T}$ values, and the uncertainty in the energy loss correction was found to be negligible.

A potential source of background contamination in the spectra and yield analysis are knockout-particles produced through interactions of high-energy particles with detectors materials or the beam pipe.  We completed a full GEANT simulation of the STAR detector with 1 million UrQMD Au+Au events at {\sNN} = \SI{3}{GeV} and found that knockout-particles constitute less that 2\% of the background contamination in the measured acceptance region. Therefore, no knockout correction was applied.
\begin{figure}[htb]
    \centering
    \includegraphics[width=0.8\linewidth]{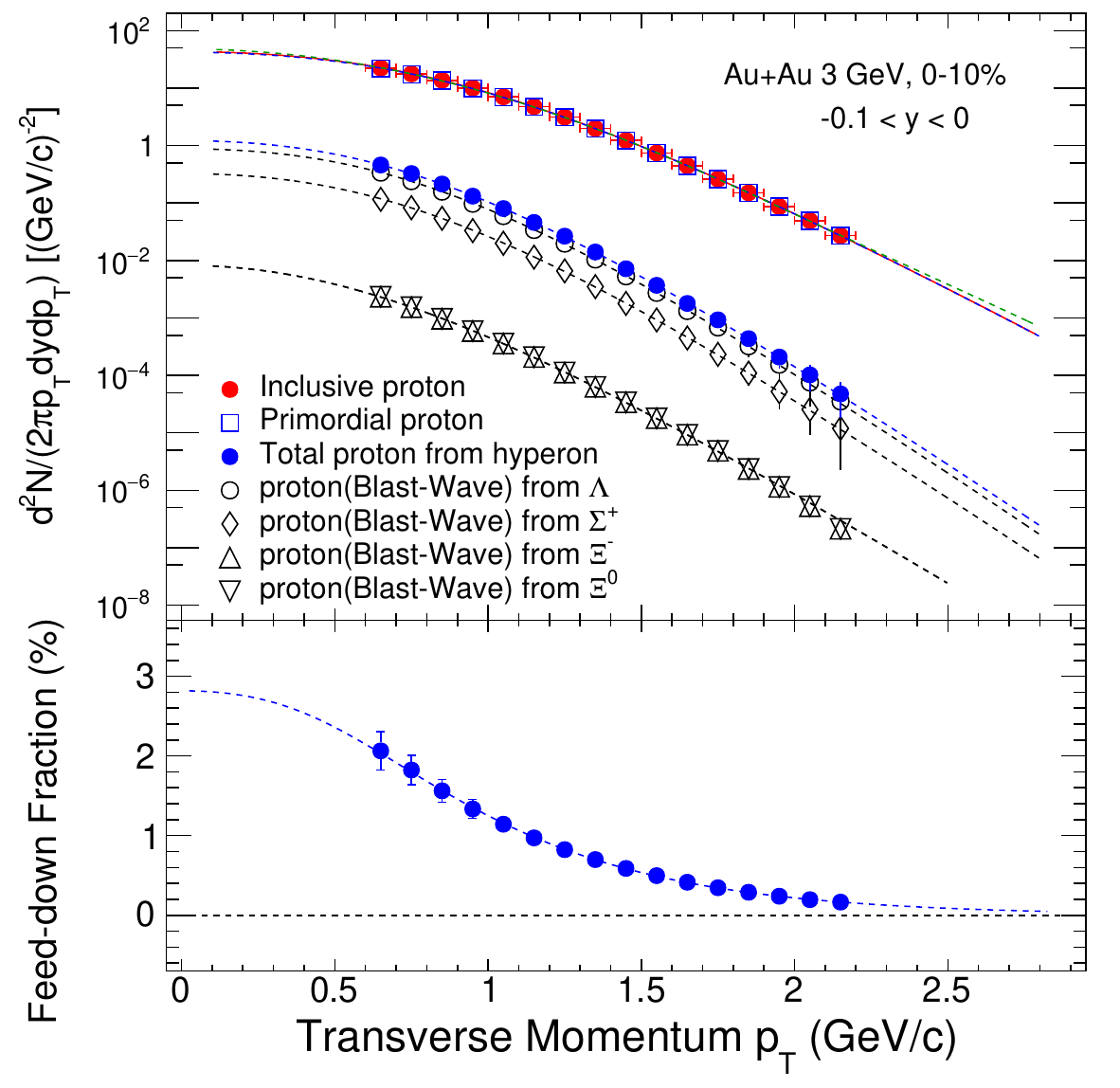}
	\caption{The $p_{T}$ dependence of the inclusive, primordial, and weak decay feed-down fractions of proton yields in Au+Au collisions at {\sNN} = \SI{3}{GeV}. The dashed/solid lines corresponding to the marker colors are the fitting results of the blast-wave model for the corresponding particles.}

    \label{fig:spefraction}
\end{figure}

\subsection{Weak decay Feed-down Correction for Protons}
In heavy-ion collisions, the weak decays of strange baryons, such as $\Lambda$ and $\Xi$ and their anti-particles, the contribute to the final yields of the (anti-)protons~\cite{PHENIX:2002svd,NA49:2010lhg,Oliinychenko:2020znl}. To obtain the primordial yields of (anti-)protons, it is necessary to subtract the contributions from weak decay. As reported in Ref.~\cite{PhysRevLett.130.202301}, the STAR experiment has published the energy dependence of the weak decay fractions for (anti-)protons in Au+Au collisions at {\sNN} = 7.7 -- \SI{200}{GeV}. 
Similarly to the previous analysis, for \SI{3}{GeV} the measured $p_{T}$ spectra of $\Lambda$ and $\Xi^{-}$~\cite{STAR:2021hyx} were used as inputs for the embedding to simulate the decay kinematics of hyperons and the $p_{T}$ spectra of the daughter protons. The main decay channels and their branching ratios (BR) are shown below~\cite{ParticleDataGroup:2018ovx}:
\begin{equation}\nonumber
\centering
\Lambda \longrightarrow p+\pi^{-}, \mathrm{BR} = 63.9 \%
\end{equation}
\vspace{-0.7cm}
\begin{equation}\nonumber
\centering
\Sigma^{+} \longrightarrow p+\pi^0, \mathrm{BR} = 51.57 \%
\end{equation}
\vspace{-0.7cm}
\begin{equation}\nonumber
\centering
\Xi^{-} \longrightarrow \Lambda+\pi^{-}, \mathrm{BR} = 99.887 \%
\end{equation}
\vspace{-0.7cm}
\begin{equation}\nonumber
\centering
\Xi^0 \longrightarrow \Lambda+\pi^0, \mathrm{BR} = 99.524 \%.
\end{equation}
The $p_{T}$ spectra of $\Sigma^{+}$ was obtained by multiplying the $\Lambda$ spectra by a factor of 0.224 ($\Sigma^{+}/\Lambda$ = 0.224 was estimated from the thermal model). Based on the ART~\cite{Yong:2022pyb} calculation, we assumed the spectra of $\Xi^{0}$ and $\Xi^{-}$ are the same and took 30\% of the $\Xi^{0}$ yield into the estimate of uncertainty.

\begin{figure}[htb]
    \centering
    \includegraphics[width=0.9\linewidth]{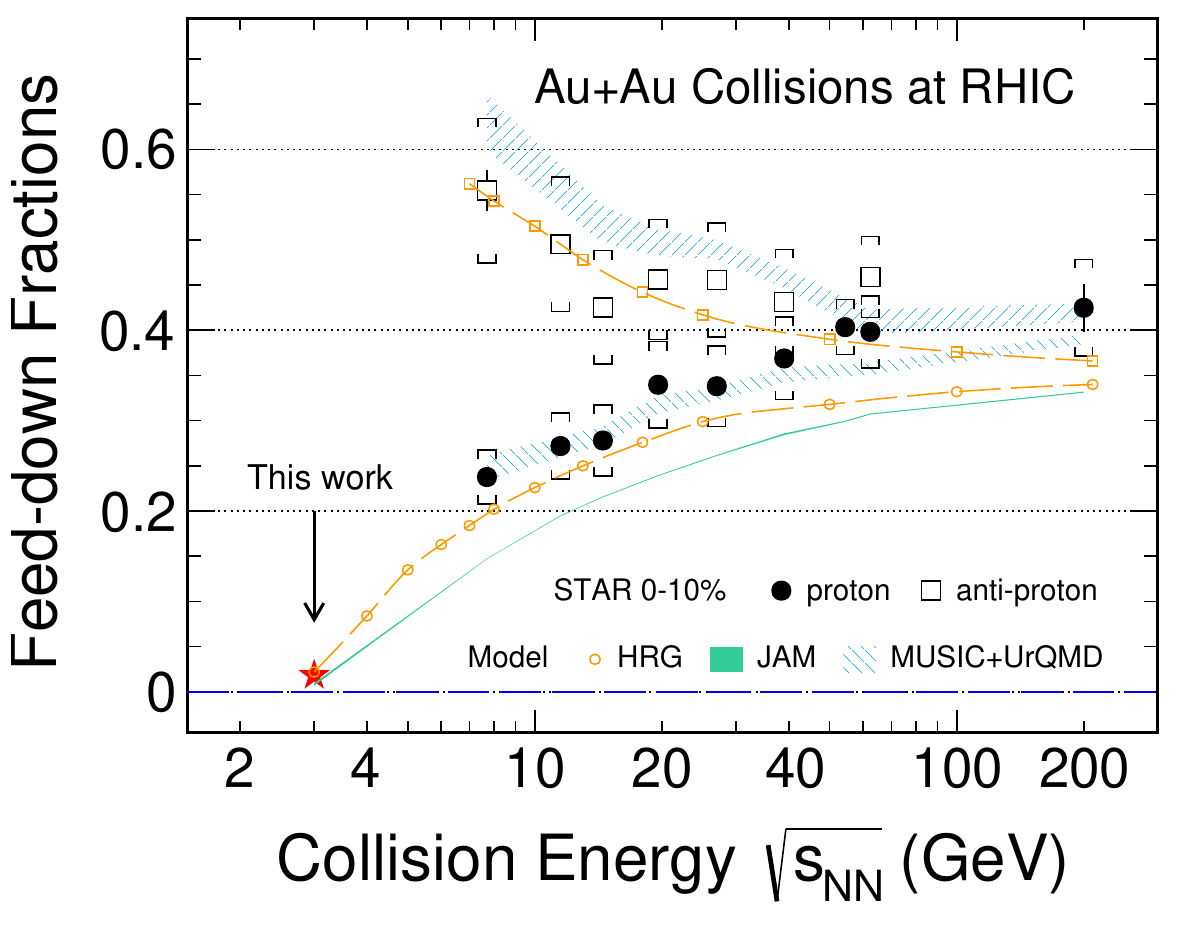}
	\caption{Energy dependence of the weak decay feed-down fraction of protons (filled circles) and anti-protons (open squares) in 0-10\% Au+Au collisions at RHIC~\cite{PhysRevLett.130.202301}, determined using a data-driven approach. The fraction for new STAR data at mid-rapidity ($-0.1 < y < 0$) at 3 GeV is marked with a red solid marker. Calculations from the HRG~\cite{Andronic:2005yp} (orange marker), JAM (dark-green band), and MUSIC+UrQMD~\cite{Zhao:2021dka} (dashed-blue area) models are plotted for comparison. The widths of the bands on the model results represent the statistical uncertainties.}
    \label{fig:fractionEne}
\end{figure}
\begin{figure*}[htb]
    \centering
    \includegraphics[width=0.87\linewidth]{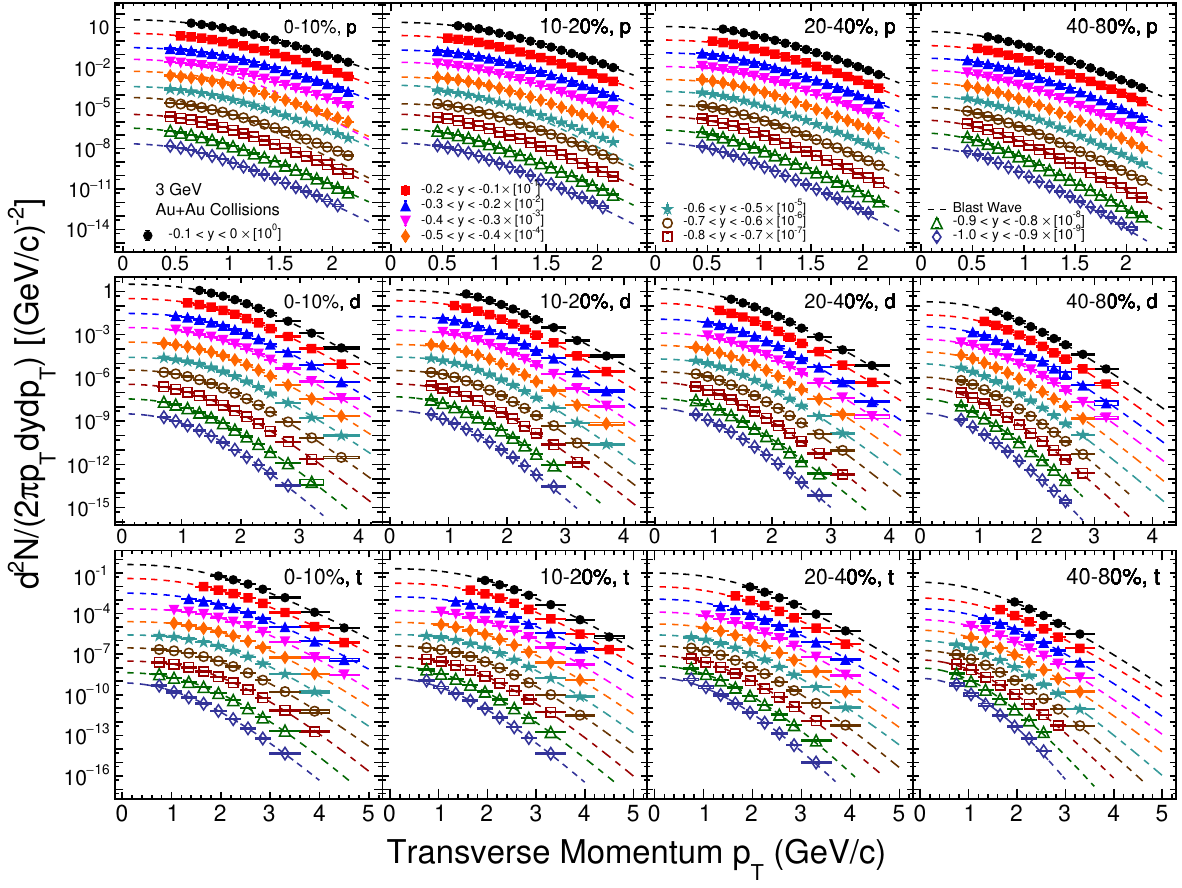}
	\caption{Transverse momentum spectra ($p_{T}$) of protons (Top), deuterons (Middle), and tritons (Bottom) from different rapidity ranges and centrality bins in Au+Au collisions at {\sNN} = \SI{3}{GeV}. For illustration purposes, the spectra are scaled by a factor from 1 at mid-rapidity to $10^{-9}$ at target rapidity. Systematic uncertainties are represented by boxes. The dotted lines are blast-wave model fits.}
    \label{fig:pdtspectra}
\end{figure*}

\begin{figure*}[htb]
    \centering
    \includegraphics[width=0.87\linewidth]{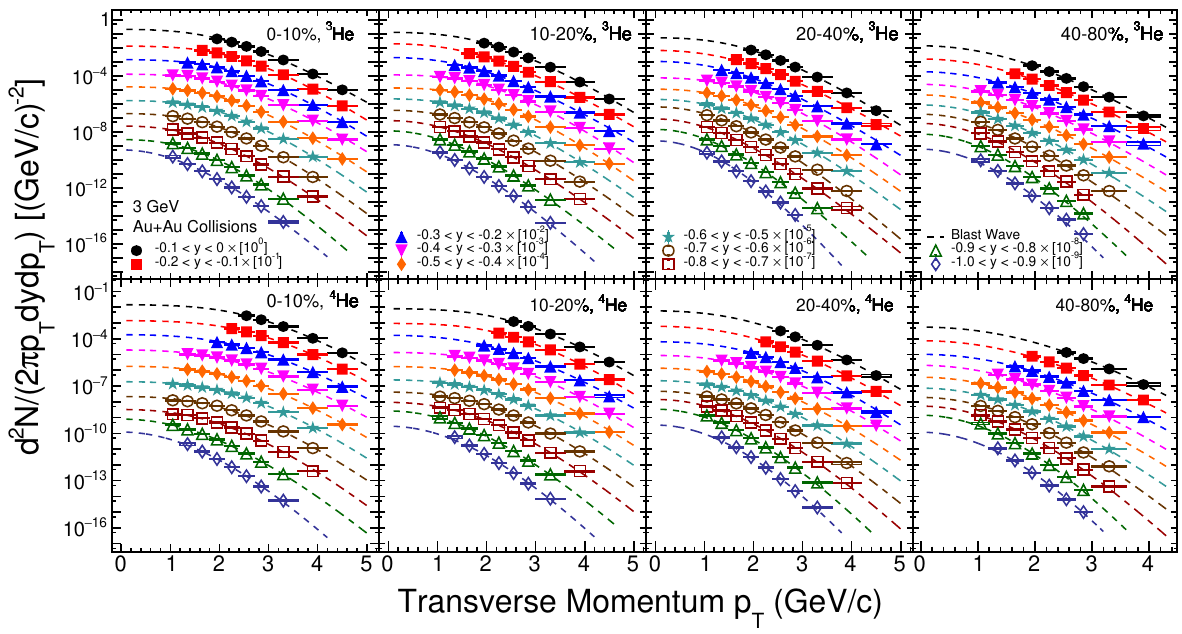}
	\caption{Similar to Fig.~\ref{fig:pdtspectra}, the transverse momentum spectra ($p_{T}$) of $^{3}\mathrm{He}$ (Top) and $^{4}\mathrm{He}$ (Bottom) from different rapidity ranges and centrality bins. Systematic uncertainties are represented by boxes. The dotted lines are blast-wave model fits.}
    \label{fig:he34spectra}
\end{figure*}
As shown in Fig.~\ref{fig:spefraction}, the top panel includes the $p_{T}$ spectra of the inclusive protons, the protons from the weak-decay of strange baryons, and the primordial protons at mid-rapidity in 0-10\% Au+Au collisions at $\sqrt{s_{\mathrm{NN}}}$ = \SI{3}{GeV}. The bottom panel shows the $p_{T}$ dependence of the fraction from the feed-down contributions.  Finally, the obtained weak decay feed-down fractions of protons for each centrality and rapidity window were listed in Table~\ref{tab:feeddownvalue}, with the maximum contribution is about 2\%. 
The statistical uncertainties of the weak decay fractions were obtained by adding the statistical uncertainties from different particle spectra in quadrature. The systematic uncertainty consists of two components added in quadrature: the systematic uncertainty of the primordial protons (5-7\%) and the difference between using the double $p_{T}$ exponential function (expressed in Eq.~\ref{eq:pT2}) and the blast-wave model (expressed in Eq.~\ref{eq:blastwave}) (6-9\%).

Figure~\ref{fig:fractionEne} shows the energy dependence of the weak decay feed-down fractions for protons and anti-protons at mid-rapidity in 0-10\% central Au+Au collisions. The filled circles and open squares denote the results of protons and anti-protons, respectively. While the weak decayed proton fractions decrease as collision energy decreases, the fractions of anti-protons show the opposite trend. At higher energies, the two fractions approach each other and reach saturation around $\sim$40\%. The weak decay fractions of (anti-)protons calculated from the hadron resonance gas (HRG)~\cite{Andronic:2005yp} and MUSIC+UrQMD~\cite{Zhao:2021dka} models show good agreement with the measured data, while the JAM model underestimates the feed-down fractions with respect to our results.
\begin{figure*}[htb]
	\centering
	\includegraphics[width=0.81\linewidth]{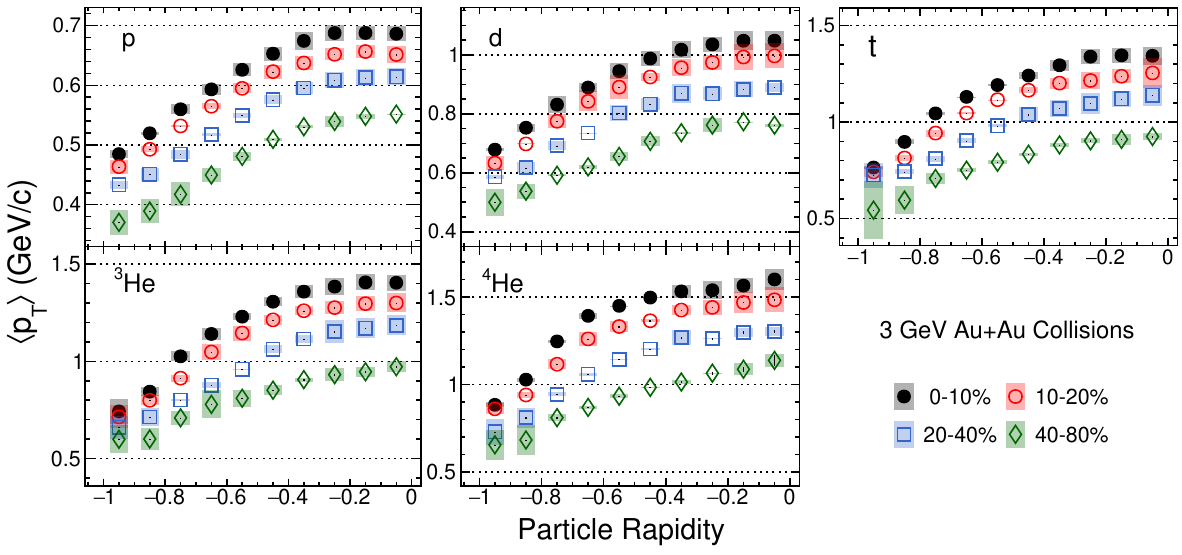}
	\caption{Collision centrality and particle rapidity dependence of the averaged transverse momentum $\langle p_{T} \rangle$ of protons and light nuclei from Au+Au collisions at {\sNN} = \SI{3}{GeV}. The boxes indicate systematic uncertainties. Statistical uncertainties are smaller than the markers and are not visible.}
     \label{fig:meanpT}                          
\end{figure*} 

\subsection{Systematic Uncertainty}

There are two dominant sources of systematic uncertainties for the $p_{T}$ spectra. The first arises from the variations in the track quality cuts mentioned in Sec.~\ref{trackcuts}. 
For each of the three track parameters, the default cutoffs were adjusted to two different values: \nhitsfit (15, 25), \nhitsdedx (8, 12), and DCA (2.4, 2.6 for protons and 0.8, 1.2 for light nuclei). Systematic uncertainties were then calculated for each set of modified cutoffs. The second source of uncertainty is related to the tracking efficiency determined from the embedding simulation, with a conservative estimate of $5\%$ applied to all particles.
The details of the systematic uncertainty are shown in Table \ref{tab:pTsys}. The ranges of the systematic uncertainties denotes the minimum and maximum values over the entire range of rapidity and $p_{T}$. In general, the minimum uncertainties come from the low $p_{T}$ for most of the rapidity windows, while the maximum values correspond to the high $p_{T}$ at the target rapidity.

\begin{table}[h]
\renewcommand{\arraystretch}{1.3}
\centering
    \caption{Systematic uncertainty of the particle $p_{T}$ spectra at all rapidity and centrality ranges.}
    \begin{tabular}{lccccc}
    \hline \hline
    Sources       & $p$      & $d$      & $t$       & $^{3}\mathrm{He}$ & $^{4} \mathrm{He}$ \\
    \hline
    \nhitsfit     & 3$-$5\%  & 2$-$4\%  & 3$-$5\%   & 2$-$3\%           & 1$-$5\%    \\
    \nhitsdedx    & 1$-$2\%  & 1$-$2\%  & 1$-$2\%   & 1$-$2\%           & 1$-$4\%    \\
    DCA           & 3$-$6\%  & 1$-$4\%  & 3$-$5\%   & 1$-$3\%           & 1$-$5\%    \\
    Cuts (total)  & 3$-$7\%  & 2$-$5\%  & 3$-$6\%   & 2$-$4\%           & 2$-$6\%    \\
    Tracking eff. & 5\%      & 5\%      & 5\%       & 5\%               & 5\%        \\
    \hline \hline
    \end{tabular}
    \label{tab:pTsys}
\end{table}

\begin{figure*}[htb]
    \centering
    \includegraphics[width=0.85\linewidth]{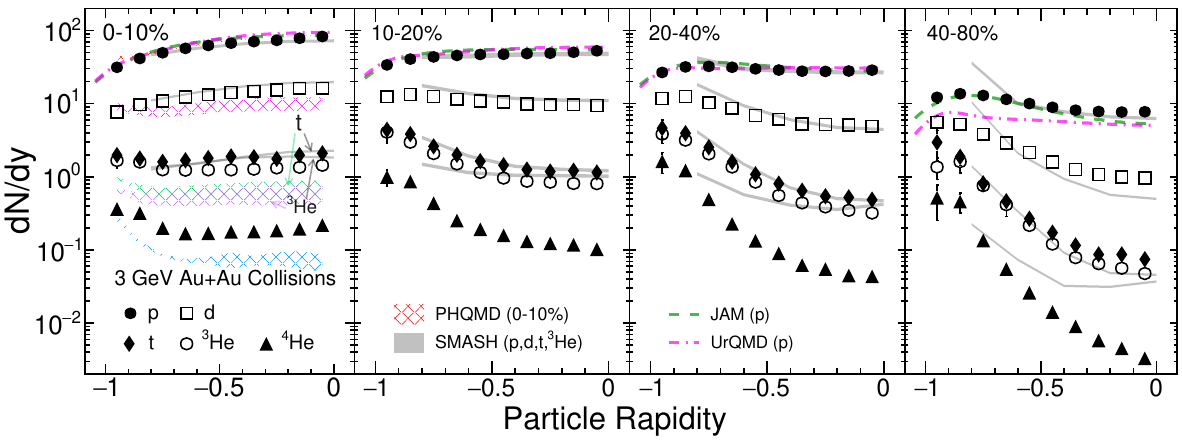}
	\caption{Collision centrality dependence of primordial protons and light nuclei $dN/dy$ from Au+Au collisions at {\sNN} = \SI{3}{GeV}. The vertical lines represent the orthogonal sum of statistical and systematic uncertainties. The gray bands and colored dotted lines show results from hadronic transport models: SMASH (for protons, deuterons, tritons, and $^{3}\mathrm{He}$), JAM (for protons), and UrQMD (for protons) for all centralities. The colored grid bands represent results from PHQMD calculations for protons, deuterons, tritons, $^{3}\mathrm{He}$, and $^{4}\mathrm{He}$ for the top 0-10\% central collisions.}

    \label{fig:dndy_fit}
\end{figure*}
\begin{table*}[htb]\scriptsize
\centering
\caption{Integral Yield ($dN/dy$) of inclusive and primordial protons at different centralities in Au+Au collisions at $\sqrt{s_\mathrm{NN}}$ = 3 GeV. The errors represent statistical and systematic uncertainties, respectively.}
\begin{tabular}{c|c|c|c|c}
\hline	\hline
Rapidity			&	0-10\%							&		10-20\%						& 	20-40\%							&	40-80\%		\\	
\hline
\multicolumn{5}{c}{Inclusive Proton}	\\
-0.1 $<$ y $<$ 0	& 84.2269 $\pm$ 0.0065 $\pm$ 6.5297 & 53.5612 $\pm$ 0.0055 $\pm$ 3.8242 & 29.0693 $\pm$ 0.0030 $\pm$ 2.1287 & 7.8225  $\pm$ 0.0013 $\pm$ 0.7345 \\
-0.2 $<$ y $<$ -0.1 & 79.7020 $\pm$ 0.0058 $\pm$ 5.2750 & 51.1497 $\pm$ 0.0048 $\pm$ 3.5403 & 28.2922 $\pm$ 0.0026 $\pm$ 2.0013 & 7.7403  $\pm$ 0.0011 $\pm$ 0.5458 \\
-0.3 $<$ y $<$ -0.2 & 74.6585 $\pm$ 0.0052 $\pm$ 4.5472 & 49.6261 $\pm$ 0.0044 $\pm$ 2.9821 & 27.9482 $\pm$ 0.0024 $\pm$ 1.8241 & 8.0325  $\pm$ 0.0010 $\pm$ 0.4771 \\
-0.4 $<$ y $<$ -0.3 & 71.6650 $\pm$ 0.0051 $\pm$ 4.8750 & 49.0187 $\pm$ 0.0044 $\pm$ 2.9275 & 28.1383 $\pm$ 0.0024 $\pm$ 1.9833 & 8.2011  $\pm$ 0.0011 $\pm$ 0.4955 \\
-0.5 $<$ y $<$ -0.4 & 68.1892 $\pm$ 0.0050 $\pm$ 4.2165 & 47.9539 $\pm$ 0.0044 $\pm$ 3.0252 & 29.1134 $\pm$ 0.0025 $\pm$ 1.8381 & 8.9535  $\pm$ 0.0011 $\pm$ 0.7249 \\
-0.6 $<$ y $<$ -0.5 & 63.3888 $\pm$ 0.0049 $\pm$ 3.8917 & 47.6113 $\pm$ 0.0044 $\pm$ 2.8138 & 30.2151 $\pm$ 0.0026 $\pm$ 2.1248 & 10.0751 $\pm$ 0.0012 $\pm$ 0.8930 \\
-0.7 $<$ y $<$ -0.6 & 57.7172 $\pm$ 0.0048 $\pm$ 3.6508 & 46.1672 $\pm$ 0.0044 $\pm$ 2.6254 & 31.7023 $\pm$ 0.0027 $\pm$ 1.9231 & 11.5032 $\pm$ 0.0014 $\pm$ 0.8742 \\
-0.8 $<$ y $<$ -0.7 & 50.1650 $\pm$ 0.0045 $\pm$ 3.0270 & 43.9809 $\pm$ 0.0045 $\pm$ 2.4251 & 32.5151 $\pm$ 0.0029 $\pm$ 2.0264 & 12.9956 $\pm$ 0.0015 $\pm$ 1.4232 \\
-0.9 $<$ y $<$ -0.8 & 41.8546 $\pm$ 0.0043 $\pm$ 2.3117 & 40.9181 $\pm$ 0.0045 $\pm$ 2.3713 & 31.9247 $\pm$ 0.0030 $\pm$ 2.3222 & 13.6821 $\pm$ 0.0017 $\pm$ 1.6381 \\
-1.0 $<$ y $<$ -0.9 & 31.7067 $\pm$ 0.0039 $\pm$ 1.9123 & 34.1028 $\pm$ 0.0043 $\pm$ 2.4771 & 26.8336 $\pm$ 0.0028 $\pm$ 1.6951 & 12.2254 $\pm$ 0.0017 $\pm$ 1.5859 \\
\multicolumn{5}{c}{Primordial Proton}   \\
-0.1 $<$ y $<$ 0    & 82.6797 $\pm$ 0.0069 $\pm$ 7.1551 & 52.6816 $\pm$ 0.0058 $\pm$ 4.5478 & 28.6342 $\pm$ 0.0031 $\pm$ 2.5641 & 7.7406  $\pm$ 0.0014 $\pm$ 0.7364 \\	 
-0.2 $<$ y $<$ -0.1 & 78.1212 $\pm$ 0.0061 $\pm$ 5.8114 & 50.2430 $\pm$ 0.0050 $\pm$ 3.8371 & 27.8656 $\pm$ 0.0027 $\pm$ 2.3687 & 7.6617  $\pm$ 0.0012 $\pm$ 0.5569 \\
-0.3 $<$ y $<$ -0.2 & 73.1433 $\pm$ 0.0054 $\pm$ 5.1103 & 48.7280 $\pm$ 0.0046 $\pm$ 3.2221 & 27.5364 $\pm$ 0.0025 $\pm$ 2.0731 & 7.9401  $\pm$ 0.0011 $\pm$ 0.5067 \\
-0.4 $<$ y $<$ -0.3 & 70.3076 $\pm$ 0.0053 $\pm$ 5.5114 & 48.2161 $\pm$ 0.0046 $\pm$ 3.2354 & 27.7211 $\pm$ 0.0025 $\pm$ 2.1184 & 8.1217  $\pm$ 0.0011 $\pm$ 0.5011 \\
-0.5 $<$ y $<$ -0.4 & 66.9528 $\pm$ 0.0052 $\pm$ 4.6377 & 47.1806 $\pm$ 0.0045 $\pm$ 3.3778 & 28.7408 $\pm$ 0.0026 $\pm$ 1.9352 & 8.8804  $\pm$ 0.0012 $\pm$ 0.7318 \\
-0.6 $<$ y $<$ -0.5 & 62.3203 $\pm$ 0.0051 $\pm$ 4.2731 & 46.9185 $\pm$ 0.0046 $\pm$ 3.0108 & 29.8603 $\pm$ 0.0027 $\pm$ 2.1564 & 10.0022 $\pm$ 0.0013 $\pm$ 0.9445 \\
-0.7 $<$ y $<$ -0.6 & 56.8581 $\pm$ 0.0049 $\pm$ 3.9454 & 45.5422 $\pm$ 0.0046 $\pm$ 2.7241 & 31.3767 $\pm$ 0.0028 $\pm$ 1.9461 & 11.4186 $\pm$ 0.0014 $\pm$ 1.0531	\\
-0.8 $<$ y $<$ -0.7 & 49.5232 $\pm$ 0.0047 $\pm$ 3.2879 & 43.4985 $\pm$ 0.0046 $\pm$ 2.4453 & 32.2542 $\pm$ 0.0029 $\pm$ 2.2320 & 12.9221 $\pm$ 0.0016 $\pm$ 1.8662	\\
-0.9 $<$ y $<$ -0.8 & 41.4016 $\pm$ 0.0044 $\pm$ 2.3341 & 40.5701 $\pm$ 0.0046 $\pm$ 2.4841 & 31.7371 $\pm$ 0.0030 $\pm$ 2.7653 & 13.6422 $\pm$ 0.0017 $\pm$ 2.1961	\\
-1.0 $<$ y $<$ -0.9 & 31.3449 $\pm$ 0.0034 $\pm$ 2.0611 & 33.8040 $\pm$ 0.0039 $\pm$ 2.9454 & 26.6531 $\pm$ 0.0026 $\pm$ 1.8582 & 12.1841 $\pm$ 0.0015 $\pm$ 2.1424	\\
\hline	\hline
\end{tabular}
\label{tab:proton_dndy}
\end{table*}

\begin{table*}[htb]\scriptsize
\centering
\caption{Integral Yield ($dN/dy$) of light nuclei at different centralities in Au+Au collisions at $\sqrt{s_\mathrm{NN}}$ = 3 GeV. The errors represent statistical and systematic uncertainties, respectively.}
\begin{tabular}{c|c|c|c|c}
\hline	\hline
Rapidity			&	0-10\%							 &		10-20\%						  & 	20-40\%						   &	40-80\%		\\	
\hline
\multicolumn{5}{c}{Deuteron}    \\
-0.1 $<$ y $<$ 0	& 16.2055 $\pm$ 0.0035 $\pm$ 2.2718 & 9.3904  $\pm$ 0.0029 $\pm$ 1.0265  & 4.8936  $\pm$ 0.0017 $\pm$ 0.3123 & 0.9668 $\pm$ 0.0008 $\pm$ 0.0586 \\
-0.2 $<$ y $<$ -0.1 & 16.1856 $\pm$ 0.0030 $\pm$ 1.4321 & 9.6358  $\pm$ 0.0024 $\pm$ 1.1131  & 5.1030  $\pm$ 0.0014 $\pm$ 0.4626 & 0.9962 $\pm$ 0.0006 $\pm$ 0.0546 \\
-0.3 $<$ y $<$ -0.2 & 15.2651 $\pm$ 0.0025 $\pm$ 1.1290 & 9.6575  $\pm$ 0.0021 $\pm$ 0.7286  & 5.1802  $\pm$ 0.0012 $\pm$ 0.3152 & 1.0728 $\pm$ 0.0005 $\pm$ 0.0989 \\
-0.4 $<$ y $<$ -0.3 & 14.7571 $\pm$ 0.0025 $\pm$ 1.1883 & 9.8211  $\pm$ 0.0022 $\pm$ 0.7626  & 5.2948  $\pm$ 0.0012 $\pm$ 0.4964 & 1.2552 $\pm$ 0.0004 $\pm$ 0.0696 \\
-0.5 $<$ y $<$ -0.4 & 14.1393 $\pm$ 0.0023 $\pm$ 1.0007 & 10.5301 $\pm$ 0.0020 $\pm$ 0.6487  & 6.0936  $\pm$ 0.0011 $\pm$ 0.4200 & 1.5903 $\pm$ 0.0005 $\pm$ 0.1167 \\
-0.6 $<$ y $<$ -0.5 & 13.2526 $\pm$ 0.0022 $\pm$ 0.9648 & 10.8063 $\pm$ 0.0021 $\pm$ 0.8709  & 6.8961  $\pm$ 0.0012 $\pm$ 0.5749 & 2.1439 $\pm$ 0.0006 $\pm$ 0.1449 \\
-0.7 $<$ y $<$ -0.6 & 12.2633 $\pm$ 0.0022 $\pm$ 0.7779 & 11.4487 $\pm$ 0.0022 $\pm$ 0.8635  & 8.6326  $\pm$ 0.0015 $\pm$ 0.4736 & 2.8985 $\pm$ 0.0007 $\pm$ 0.1725 \\
-0.8 $<$ y $<$ -0.7 & 10.8655 $\pm$ 0.0021 $\pm$ 0.9661 & 12.4667 $\pm$ 0.0024 $\pm$ 0.9404  & 10.2733 $\pm$ 0.0016 $\pm$ 1.2942 & 3.8083 $\pm$ 0.0009 $\pm$ 0.2365 \\
-0.9 $<$ y $<$ -0.8 & 9.6343  $\pm$ 0.0021 $\pm$ 0.6206 & 13.3361 $\pm$ 0.0026 $\pm$ 0.7320  & 12.5062 $\pm$ 0.0020 $\pm$ 1.0884 & 5.2252 $\pm$ 0.0011 $\pm$ 0.7009 \\
-1.0 $<$ y $<$ -0.9 & 7.6776  $\pm$ 0.0020 $\pm$ 0.4384 & 12.3905 $\pm$ 0.0027 $\pm$ 1.0454  & 11.7238 $\pm$ 0.0020 $\pm$ 0.6558 & 5.5553 $\pm$ 0.0013 $\pm$ 1.2751 \\ 
\multicolumn{5}{c}{Triton} \\
-0.1 $<$ y $<$ 0	& 2.0913 $\pm$ 0.0015 $\pm$ 0.2191 & 1.1445 $\pm$ 0.0013 $\pm$ 0.2144 & 0.4856 $\pm$ 0.0007 $\pm$ 0.0599 & 0.0746 $\pm$ 0.0003 $\pm$ 0.0151 \\
-0.2 $<$ y $<$ -0.1 & 1.9739 $\pm$ 0.0012 $\pm$ 0.1842 & 1.2105 $\pm$ 0.0010 $\pm$ 0.1540 & 0.5220 $\pm$ 0.0005 $\pm$ 0.0558 & 0.0867 $\pm$ 0.0002 $\pm$ 0.0144 \\
-0.3 $<$ y $<$ -0.2 & 1.7539 $\pm$ 0.0009 $\pm$ 0.1375 & 1.1792 $\pm$ 0.0008 $\pm$ 0.1381 & 0.5414 $\pm$ 0.0004 $\pm$ 0.0712 & 0.0867 $\pm$ 0.0002 $\pm$ 0.0066 \\
-0.4 $<$ y $<$ -0.3 & 1.8515 $\pm$ 0.0008 $\pm$ 0.1254 & 1.2523 $\pm$ 0.0007 $\pm$ 0.0880 & 0.6438 $\pm$ 0.0004 $\pm$ 0.0439 & 0.1153 $\pm$ 0.0002 $\pm$ 0.0068 \\
-0.5 $<$ y $<$ -0.4 & 1.8933 $\pm$ 0.0008 $\pm$ 0.1152 & 1.4409 $\pm$ 0.0008 $\pm$ 0.0875 & 0.8055 $\pm$ 0.0004 $\pm$ 0.0775 & 0.1721 $\pm$ 0.0002 $\pm$ 0.0159 \\
-0.6 $<$ y $<$ -0.5 & 1.8016 $\pm$ 0.0007 $\pm$ 0.0992 & 1.6316 $\pm$ 0.0007 $\pm$ 0.0901 & 1.0800 $\pm$ 0.0004 $\pm$ 0.0686 & 0.2673 $\pm$ 0.0002 $\pm$ 0.0159 \\
-0.7 $<$ y $<$ -0.6 & 1.6791 $\pm$ 0.0007 $\pm$ 0.0920 & 1.9951 $\pm$ 0.0008 $\pm$ 0.1103 & 1.6042 $\pm$ 0.0006 $\pm$ 0.0974 & 0.4626 $\pm$ 0.0003 $\pm$ 0.0280 \\
-0.8 $<$ y $<$ -0.7 & 1.5980 $\pm$ 0.0007 $\pm$ 0.0888 & 2.6041 $\pm$ 0.0010 $\pm$ 0.1678 & 2.5242 $\pm$ 0.0007 $\pm$ 0.1573 & 0.8110 $\pm$ 0.0003 $\pm$ 0.0826 \\
-0.9 $<$ y $<$ -0.8 & 1.7754 $\pm$ 0.0008 $\pm$ 0.1069 & 3.8254 $\pm$ 0.0013 $\pm$ 0.2850 & 3.9358 $\pm$ 0.0009 $\pm$ 0.2622 & 1.8718 $\pm$ 0.0006 $\pm$ 0.4538 \\
-1.0 $<$ y $<$ -0.9 & 1.9749 $\pm$ 0.0009 $\pm$ 0.1246 & 4.3823 $\pm$ 0.0014 $\pm$ 0.5853 & 4.6199 $\pm$ 0.0010 $\pm$ 1.3601 & 2.9489 $\pm$ 0.0008 $\pm$ 1.3143 \\ 
\multicolumn{5}{c}{$^{3}\mathrm{He}$} \\
-0.1 $<$ y $<$ 0	& 1.4361 $\pm$ 0.0012 $\pm$ 0.1254 & 0.8050 $\pm$ 0.0011 $\pm$ 0.0717 & 0.3194 $\pm$ 0.0006 $\pm$ 0.0329 & 0.0476 $\pm$ 0.0003 $\pm$ 0.0066 \\
-0.2 $<$ y $<$ -0.1 & 1.3468 $\pm$ 0.0009 $\pm$ 0.1217 & 0.8099 $\pm$ 0.0008 $\pm$ 0.0444 & 0.3481 $\pm$ 0.0004 $\pm$ 0.0431 & 0.0563 $\pm$ 0.0002 $\pm$ 0.0067 \\
-0.3 $<$ y $<$ -0.2 & 1.3205 $\pm$ 0.0008 $\pm$ 0.0967 & 0.8434 $\pm$ 0.0007 $\pm$ 0.0643 & 0.3873 $\pm$ 0.0004 $\pm$ 0.0453 & 0.0648 $\pm$ 0.0001 $\pm$ 0.0083 \\
-0.4 $<$ y $<$ -0.3 & 1.2696 $\pm$ 0.0008 $\pm$ 0.0779 & 0.8677 $\pm$ 0.0006 $\pm$ 0.0513 & 0.4395 $\pm$ 0.0003 $\pm$ 0.0372 & 0.0785 $\pm$ 0.0001 $\pm$ 0.0051 \\
-0.5 $<$ y $<$ -0.4 & 1.2405 $\pm$ 0.0007 $\pm$ 0.0695 & 0.9597 $\pm$ 0.0006 $\pm$ 0.0662 & 0.5549 $\pm$ 0.0004 $\pm$ 0.0362 & 0.1203 $\pm$ 0.0002 $\pm$ 0.0095 \\
-0.6 $<$ y $<$ -0.5 & 1.2434 $\pm$ 0.0007 $\pm$ 0.0736 & 1.1465 $\pm$ 0.0007 $\pm$ 0.0657 & 0.8450 $\pm$ 0.0005 $\pm$ 0.0468 & 0.2159 $\pm$ 0.0002 $\pm$ 0.0032 \\
-0.7 $<$ y $<$ -0.6 & 1.2290 $\pm$ 0.0007 $\pm$ 0.0919 & 1.4973 $\pm$ 0.0009 $\pm$ 0.1317 & 1.3583 $\pm$ 0.0007 $\pm$ 0.0967 & 0.4154 $\pm$ 0.0003 $\pm$ 0.1300 \\
-0.8 $<$ y $<$ -0.7 & 1.2451 $\pm$ 0.0008 $\pm$ 0.1163 & 2.0762 $\pm$ 0.0011 $\pm$ 0.1447 & 2.0529 $\pm$ 0.0009 $\pm$ 0.1236 & 0.7560 $\pm$ 0.0005 $\pm$ 0.1247 \\
-0.9 $<$ y $<$ -0.8 & 1.5867 $\pm$ 0.0011 $\pm$ 0.1462 & 2.9643 $\pm$ 0.0015 $\pm$ 0.2644 & 3.1660 $\pm$ 0.0012 $\pm$ 0.4133 & 1.6094 $\pm$ 0.0009 $\pm$ 0.4884 \\
-1.0 $<$ y $<$ -0.9 & 1.6674 $\pm$ 0.0012 $\pm$ 0.3469 & 4.0648 $\pm$ 0.0021 $\pm$ 1.2223 & 3.8249 $\pm$ 0.0015 $\pm$ 0.9857 & 1.3608 $\pm$ 0.0009 $\pm$ 0.4378 \\ 
\multicolumn{5}{c}{$^{4}\mathrm{He}$} \\
-0.1 $<$ y $<$ 0	& 0.2187 $\pm$ 0.0006 $\pm$ 0.0207 & 0.1023 $\pm$ 0.0005 $\pm$ 0.0179 & 0.0435 $\pm$ 0.0003 $\pm$ 0.0025 & 0.0033 $\pm$ 0.0001 $\pm$ 0.0002 \\
-0.2 $<$ y $<$ -0.1 & 0.1943 $\pm$ 0.0005 $\pm$ 0.0162 & 0.1179 $\pm$ 0.0004 $\pm$ 0.0157 & 0.0447 $\pm$ 0.0002 $\pm$ 0.0037 & 0.0045 $\pm$ 0.0001 $\pm$ 0.0008 \\
-0.3 $<$ y $<$ -0.2 & 0.1842 $\pm$ 0.0004 $\pm$ 0.0196 & 0.1229 $\pm$ 0.0003 $\pm$ 0.0114 & 0.0548 $\pm$ 0.0002 $\pm$ 0.0031 & 0.0058 $\pm$ 0.0001 $\pm$ 0.0003 \\
-0.4 $<$ y $<$ -0.3 & 0.1778 $\pm$ 0.0003 $\pm$ 0.0102 & 0.1313 $\pm$ 0.0003 $\pm$ 0.0083 & 0.0620 $\pm$ 0.0001 $\pm$ 0.0054 & 0.0090 $\pm$ 0.0001 $\pm$ 0.0005 \\
-0.5 $<$ y $<$ -0.4 & 0.1767 $\pm$ 0.0003 $\pm$ 0.0106 & 0.1582 $\pm$ 0.0003 $\pm$ 0.0087 & 0.0898 $\pm$ 0.0001 $\pm$ 0.0050 & 0.0142 $\pm$ 0.0001 $\pm$ 0.0008 \\
-0.6 $<$ y $<$ -0.5 & 0.1693 $\pm$ 0.0002 $\pm$ 0.0103 & 0.1916 $\pm$ 0.0003 $\pm$ 0.0135 & 0.1335 $\pm$ 0.0002 $\pm$ 0.0073 & 0.0260 $\pm$ 0.0001 $\pm$ 0.0016 \\
-0.7 $<$ y $<$ -0.6 & 0.1680 $\pm$ 0.0003 $\pm$ 0.0117 & 0.2530 $\pm$ 0.0003 $\pm$ 0.0189 & 0.2271 $\pm$ 0.0002 $\pm$ 0.0127 & 0.0548 $\pm$ 0.0001 $\pm$ 0.0030 \\
-0.8 $<$ y $<$ -0.7 & 0.1993 $\pm$ 0.0003 $\pm$ 0.0131 & 0.4351 $\pm$ 0.0005 $\pm$ 0.0333 & 0.4917 $\pm$ 0.0004 $\pm$ 0.0287 & 0.1348 $\pm$ 0.0002 $\pm$ 0.0121 \\
-0.9 $<$ y $<$ -0.8 & 0.3182 $\pm$ 0.0004 $\pm$ 0.0290 & 0.8622 $\pm$ 0.0007 $\pm$ 0.0929 & 1.2183 $\pm$ 0.0008 $\pm$ 0.2014 & 0.4572 $\pm$ 0.0004 $\pm$ 0.1385 \\
-1.0 $<$ y $<$ -0.9 & 0.3631 $\pm$ 0.0006 $\pm$ 0.0280 & 0.9943 $\pm$ 0.0011 $\pm$ 0.2552 & 1.6292 $\pm$ 0.0014 $\pm$ 0.5169 & 0.5158 $\pm$ 0.0008 $\pm$ 0.2623 \\ 
\hline	\hline
\end{tabular}
\label{tab:LN_dndy}
\end{table*}

The $p_{T}$ integral yield ($dN/dy$) for various particles was obtained by adding the yields in the measured $p_{T}$ region and the unmeasured $p_{T}$ range, which was extrapolated from the blast-wave function~\cite{Schnedermann:1993ws}, 
which can be expressed in Eq.~\ref{eq:blastwave}:

\begin{equation}\label{eq:blastwave}  
\centering
\begin{split}
    \frac{1}{2 \pi p_{T}} \frac{d^{2} N}{d p_{T} d y} \propto \int_{0}^{R} {r d r m_{T}} I_{0}\left(\frac{p_{T} \sinh \rho(r)}{T_{kin}}\right) \\ 
    \times K_{1}\left(\frac{m_{T} \cosh \rho(r)}{T_{kin}}\right)                              
\end{split}
\end{equation}

where $m_{T}$ is the transverse mass of the particle, $I_{0}$ and $K_{1}$ are the modified Bessel functions, $T_{kin}$ is the kinetic freeze-out temperature, and $\rho(r) = \tanh ^{-1} \beta_{T}$ is the velocity profile. The transverse radial flow velocity $\beta_{T}$ in the region $0 \leq r \leq R$ can be expressed as $\beta_{T}=\beta_{S}(r / R)^{n}$, where $\beta_{S}$ is the surface velocity, $r/R$ is the relative radial extent of the thermal source, and the exponent $n$ reflects the form of the flow velocity profile (fixed $n = 1$ in this analysis as proposed in Ref.~\cite{Schnedermann:1993ws,Braun-Munzinger:2018hat,Andronic:2018vqh,STAR:2019sjh}). 

The main source of systematic uncertainty for $dN/dy$ originated from the extrapolation of the unmeasured yield at low $p_{T}$ region. This uncertainty was estimated by fitting the $p_{T}$ spectra using the double $p_{T}$ exponential function, as expressed in Eq.~\ref{eq:pT2} 
\begin{equation}\label{eq:pT2}
	\frac{1}{2 \pi p_{T}} \frac{d^{2} N}{d p_{T} d y} \propto p_{0} \exp \left(\frac{-p_{T}^{2}}{p_{1}}\right) + p_{2} \exp \left(\frac{-p_{T}^{2}}{p_{3}}\right)
\end{equation}
 and comparing the corresponding extrapolated yields to the default ones obtained from the blast-wave model. The systematic uncertainties from the extrapolation at different centralities are approximately 3-6\% for protons and increases for light nuclei with a maximum contribution to 18\% for $^{4}\mathrm{He}$. The final systematic uncertainties were calculated by summing the uncertainties from extrapolation and tracking efficiency in quadrature. The total uncertainties are about 6-8\% for protons, 6-12\% for deuteron, 6-11\% for triton, 7-11\% for $^{3}\mathrm{He}$, and 6-20\% for $^{4}\mathrm{He}$. 

For the systematic uncertainty for compound yield ratios, the default yield ratios were obtained by fitting the spectra with the blast-wave model. Other functions, such as the double $p_{T}$ exponential function mentioned above in Eq.~\ref{eq:pT2}, in addition to the Boltzmann function expressed in Eq.~\ref{eq:Boltz},
\begin{equation}\label{eq:Boltz}
	\frac{1}{2 \pi m_{T}} \frac{d^{2} N}{d m_{T} d y} \propto p_{0} m_{\mathrm{T}} \exp \left(\frac{-m_{\mathrm{T}}}{p_{1}}\right) 
\end{equation}
the Levy function expressed in Eq.~\ref{eq:levy},
\begin{equation}\label{eq:levy}
	\frac{1}{2 \pi m_{T}} \frac{d^{2} N}{d m_{T} d y} \propto p_{0}\left(1+\frac{m_{\mathrm{T}}-m_{\mathrm{0}}}{p_{1}p_{2}}\right)^{-p_{1}}
\end{equation}
and the $m_{T}$ exponential function expressed in Eq.~\ref{eq:mTexp}
\begin{equation}\label{eq:mTexp}
	\frac{1}{2 \pi m_{T}} \frac{d^{2} N}{d m_{T} d y} \propto p_{0} \exp \left(\frac{-m_{\mathrm{T}}}{p_{1}}\right)
\end{equation}
were also applied to calculate the yield ratios. The differences between these results and the default value were the main source of systematic uncertainty. The total systematic uncertainty for the ratio ($N_{p} \times N_{t} / N_{d}^{2}$) was about 2-15\% for different rapidity and centrality bins. The systematic uncertainty increased to 25\% when considering $^{3}\mathrm{He}$, and $^{4}\mathrm{He}$ in the yield ratio.

\section{Results and Discussions}
\subsection{Transverse Momentum Spectra}
Figures \ref{fig:pdtspectra} and~\ref{fig:he34spectra} show the transverse momentum spectra ($p_{T}$) for primordial protons, deuterons, tritons, $^{3}\mathrm{He}$, and $^{4}\mathrm{He}$ in 0-10\%, 10-20\%, 20-40\%, and 40-80\% central Au+Au collisions at $\sqrt{s_{\mathrm{NN}}}$ = \SI{3}{GeV}. The results are presented for various rapidity windows with a bin width of 0.1. For illustration purposes, the data points were scaled by a factor from 1 at mid-rapidity to $10^{-9}$ at target rapidity. 
The dotted lines represent the fits using the blast-wave model. 

\subsection{Averaged Transverse Momentum ($\langle p_{T}\rangle$)}
The averaged transverse momentum, $\langle p_{T}\rangle$, was calculated from the measured $p_{T}$ range and extrapolated to the unmeasured region with individual blast-wave model fits.
The rapidity dependence of $\langle p_{T}\rangle$ for $p, d, t, ^{3}\mathrm{He}$, and $^{4}\mathrm{He}$ in 0-10\%, 10-20\%, 20-40\%, and 40-80\% central Au+Au collisions at $\sqrt{s_{\mathrm{NN}}}$ = \SI{3}{GeV} are shown in Fig.~\ref{fig:meanpT}. For each particle, $\langle p_{T}\rangle$ shows a trend of monotonically decreasing from mid-rapidity to target rapidity and from central to peripheral collisions.

The systematic uncertainty for the $\langle p_{T}\rangle$ of particles was estimated in the same way as for $dN/dy$. The total systematic uncertainties of the $\langle p_{T}\rangle$ are 1-3\% for protons, 3-9\% for deuterons, 5-13\% for tritons, 3-12\% for $^{3}\mathrm{He}$, and 3-12\% for $^{4}\mathrm{He}$.

\begin{table*}[htb]
\centering
    \caption{The percentage (\%) of the measured $p_{T}$ range for primordial protons and light nuclei relative to the total $dN/dy$ in 0-10\% and 40-80\% Au+Au collisions at $\sqrt{s_{\mathrm{NN}}}$ = \SI{3}{GeV}. The errors are the combined statistical and systematic uncertainties.}
    \begin{tabular}{c|cc|cc|cc|cc|cc}
    \hline \hline
    \multirow{2}{*}{Rapidity} &	\multicolumn{2}{c|}{Proton} & \multicolumn{2}{c|}{Deuteron}		   & \multicolumn{2}{c|}{Triton}		   & \multicolumn{2}{c|}{$^{3}\mathrm{He}$} & \multicolumn{2}{c}{$^{4}\mathrm{He}$}	\\ 
	\cline{2-11}
							  &	\multicolumn{1}{c}{0-10\%} & 40-80\%  & \multicolumn{1}{c}{0-10\%} & 40-80\% & \multicolumn{1}{c}{0-10\%} & 40-80\% & \multicolumn{1}{c}{0-10\%} & 40-80\%  & \multicolumn{1}{c}{0-10\%} & 40-80\%		\\
    \hline
	-0.1 $<$ y $<$ 0    &	\multicolumn{1}{c}{$55 \pm 4$} & $39 \pm 4$ & \multicolumn{1}{c}{$36 \pm 5$} & $14 \pm 1$ & \multicolumn{1}{c}{$23 \pm 2$} & $6  \pm 2$  & \multicolumn{1}{c}{$26 \pm 2$} & $7  \pm 1$  & \multicolumn{1}{c}{$15 \pm 1$} & $4  \pm 1$	\\		
	-0.2 $<$ y $<$ -0.1 &	\multicolumn{1}{c}{$66 \pm 4$} & $51 \pm 4$ & \multicolumn{1}{c}{$50 \pm 5$} & $26 \pm 1$ & \multicolumn{1}{c}{$38 \pm 4$} & $12 \pm 2$  & \multicolumn{1}{c}{$42 \pm 4$} & $14 \pm 2$  & \multicolumn{1}{c}{$25 \pm 2$} & $12 \pm 2$	\\
	-0.3 $<$ y $<$ -0.2 &	\multicolumn{1}{c}{$77 \pm 5$} & $63 \pm 4$ & \multicolumn{1}{c}{$65 \pm 5$} & $42 \pm 4$ & \multicolumn{1}{c}{$56 \pm 4$} & $25 \pm 2$  & \multicolumn{1}{c}{$57 \pm 4$} & $27 \pm 3$  & \multicolumn{1}{c}{$37 \pm 4$} & $21 \pm 1$	\\
	-0.4 $<$ y $<$ -0.3 &	\multicolumn{1}{c}{$77 \pm 5$} & $64 \pm 4$ & \multicolumn{1}{c}{$64 \pm 5$} & $59 \pm 3$ & \multicolumn{1}{c}{$74 \pm 5$} & $43 \pm 3$  & \multicolumn{1}{c}{$74 \pm 5$} & $45 \pm 3$  & \multicolumn{1}{c}{$66 \pm 4$} & $33 \pm 2$	\\
	-0.5 $<$ y $<$ -0.4 &	\multicolumn{1}{c}{$75 \pm 5$} & $61 \pm 5$ & \multicolumn{1}{c}{$77 \pm 6$} & $57 \pm 4$ & \multicolumn{1}{c}{$69 \pm 4$} & $39 \pm 4$  & \multicolumn{1}{c}{$73 \pm 4$} & $41 \pm 3$  & \multicolumn{1}{c}{$64 \pm 4$} & $51 \pm 3$	\\
	-0.6 $<$ y $<$ -0.5 &	\multicolumn{1}{c}{$73 \pm 5$} & $57 \pm 5$ & \multicolumn{1}{c}{$75 \pm 6$} & $51 \pm 3$ & \multicolumn{1}{c}{$84 \pm 5$} & $63 \pm 4$  & \multicolumn{1}{c}{$68 \pm 4$} & $39 \pm 6$  & \multicolumn{1}{c}{$77 \pm 5$} & $47 \pm 3$	\\
	-0.7 $<$ y $<$ -0.6 &	\multicolumn{1}{c}{$70 \pm 4$} & $52 \pm 4$ & \multicolumn{1}{c}{$72 \pm 5$} & $47 \pm 3$ & \multicolumn{1}{c}{$82 \pm 5$} & $60 \pm 4$  & \multicolumn{1}{c}{$64 \pm 5$} & $37 \pm 11$ & \multicolumn{1}{c}{$72 \pm 5$} & $42 \pm 2$	\\
	-0.8 $<$ y $<$ -0.7 &	\multicolumn{1}{c}{$67 \pm 4$} & $47 \pm 5$ & \multicolumn{1}{c}{$68 \pm 6$} & $45 \pm 3$ & \multicolumn{1}{c}{$79 \pm 4$} & $58 \pm 6$  & \multicolumn{1}{c}{$55 \pm 5$} & $29 \pm 5$	& \multicolumn{1}{c}{$66 \pm 4$} & $38 \pm 3$	\\
	-0.9 $<$ y $<$ -0.8 &	\multicolumn{1}{c}{$62 \pm 3$} & $42 \pm 5$ & \multicolumn{1}{c}{$61 \pm 4$} & $37 \pm 5$ & \multicolumn{1}{c}{$71 \pm 4$} & $44 \pm 11$ & \multicolumn{1}{c}{$39 \pm 4$} & $16 \pm 5$ 	& \multicolumn{1}{c}{$52 \pm 5$} & $26 \pm 8$	\\
	-1.0 $<$ y $<$ -0.9 &	\multicolumn{1}{c}{$57 \pm 3$} & $39 \pm 5$ & \multicolumn{1}{c}{$54 \pm 3$} & $32 \pm 7$ & \multicolumn{1}{c}{$63 \pm 4$} & $38 \pm 17$ & \multicolumn{1}{c}{$32 \pm 6$} & $17 \pm 5$ 	& \multicolumn{1}{c}{$24 \pm 2$} & $7  \pm 4$	\\
    \hline \hline
    \end{tabular}
    \label{tab:dndypercentage}
\end{table*}

\begin{table*}[htb] 
\centering                                                                                                                                                                              
    \caption{$4\pi$ yield for primordial protons and light nuclei. The errors represent statistical and systematic (from measurements and fitting, respectively) uncertainties.} 
    \begin{tabular}{lcccc}                                                                                                                                                              
    \hline \hline                                                                                                                                                                       
    Centrality              & $0-10 \%$                            & $10-20 \%$                           & $20-40 \%$                          & $40-80 \%$ \\                         
    \hline                                                                                                                                                                              
    proton                  & $134.78 \pm 0.10 \pm 8.64 \pm 0.05$  & $105.47 \pm 0.09 \pm 6.37 \pm 1.33$  & $66.82 \pm 0.05 \pm 4.41 \pm 3.17$  & $23.38 \pm 0.03 \pm 2.16 \pm 1.53$    \\ 
    deuteron                & $29.03 \pm 0.05 \pm 2.09 \pm 0.17$   & $27.34 \pm 0.05 \pm 1.77 \pm 1.84$   & $19.33 \pm 0.03 \pm 1.20 \pm 3.29$  & $6.66 \pm 0.01 \pm 0.45 \pm 1.28$     \\
    triton                  & $4.70 \pm 0.02 \pm 0.26 \pm 0.40$    & $5.15 \pm 0.02 \pm 0.39 \pm 0.81$    & $4.32 \pm 0.01 \pm 0.36 \pm 0.94$   & $1.86 \pm 0.01 \pm 0.21 \pm 0.47$     \\
    $^{3}\mathrm{He}$       & $3.55 \pm 0.02 \pm 0.24 \pm 0.41$    & $4.28 \pm 0.02 \pm 0.31 \pm 0.53$    & $3.57 \pm 0.01 \pm 0.29 \pm 0.65$   & $1.20 \pm 0.01 \pm 0.17 \pm 0.25$     \\
    $^{4}\mathrm{He}$       & $0.56 \pm 0.01 \pm 0.03 \pm 0.07$    & $0.84 \pm 0.01 \pm 0.08 \pm 0.17$    & $1.06 \pm 0.01 \pm 0.08 \pm 0.26$   & $0.33 \pm 0.01 \pm 0.04 \pm 0.08$     \\
    \hline \hline                                                                                                                                                                       
    \end{tabular}                                                                                                                                                                       
    \label{tab:4piyield}                                                                                                                                                                
\end{table*}                                                                                                                                                                            

\begin{figure*}[htb]
    \centering
    \includegraphics[width=0.85\linewidth]{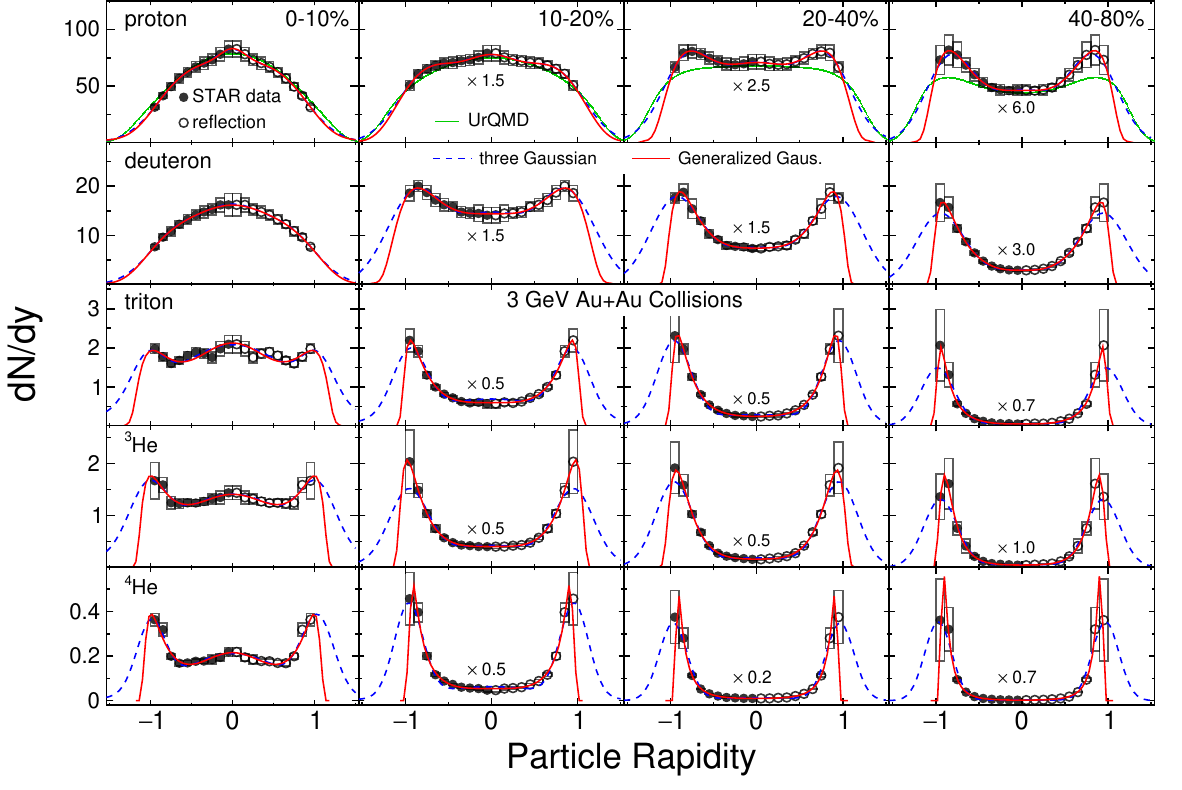}
	\caption{Particle rapidity dependence of protons and light nuclei $dN/dy$ from Au+Au collisions at {\sNN} = \SI{3}{GeV}. For illustrative purposes, the maximum values of non-central collisions are scaled to the values of the most central collision for the same particle. The blue and red lines represent fits to the distributions using three-Gaussians and generalized Gaussian functions, respectively. The green lines represent the proton distribution from the UrQMD model, with values scaled by the indicated factor to match the experiment at mid-rapidity.}

    \label{fig:dndy_fit_rap}
\end{figure*}

\subsection{$dN/dy$ and $4\pi$ Yields of Particles}
Figure~\ref{fig:dndy_fit} shows the rapidity dependence of the $p_{T}$ integrated yield ($dN/dy$) for primordial protons and light nuclei in 0-10\%, 10-20\%, 20-40\%, and 40-80\% central Au+Au collisions at $\sqrt{s_{\mathrm{NN}}}$ = \SI{3}{GeV}. In each panel, different markers represent the distributions for different particles. Due to the interplay between baryon stopping and spectators, the $dN/dy$ of protons and deuterons decrease from mid-rapidity to target rapidity in the 0-10\% most central collisions, while in peripheral collisions, the values of $dN/dy$ are peaked near the target rapidity. For tritons, $^{3}\mathrm{He}$, and $^{4}\mathrm{He}$, the peak structures at target rapidity are increasingly prominent as we move from central to peripheral collisions, primarily due to the fragmentation of the spectators~\cite{Bondorf:1995ua}. The measured $dN/dy$ for various particles were listed in Table~\ref{tab:proton_dndy} and Table~\ref{tab:LN_dndy}. At the same time, considering the significant contribution of the extrapolated $p_{T}$ range to the total yield extraction, Table~\ref{tab:dndypercentage} lists the percentage of dN/dy for the measured $p_{T}$ range to the total $dN/dy$ at all rapidity. The results for central and peripheral collisions provide a percentage range under different collision centralities.
Calculations of protons and light nuclei $dN/dy$ distributions using the hadronic transport models (JAM, PHQMD, SMASH, and UrQMD) were compared with the experimental data. The rapidity distributions of protons $dN/dy$ in 0-10\%, 10-20\%, and 20-40\% centrality bins can be well described by these models. For the SMASH model's calculations of light nuclei ($d$, $t$, and $^{3}\mathrm{He}$), the Wigner function~\cite{Chen:2003qj,Zhao:2018lyf,Zhao:2021dka} was used to compute their formation probability. For better visualization, the statistical uncertainties are expanded by a factor of 5 to set the band width. It was found that the rapidity distributions of $d$ and $t$ were well described by the SMASH model in central and mid-central collisions. 
In contrast, the PHQMD model forms clusters dynamically through attractive interactions and identifies them using an advanced Minimum Spanning Tree (aMST) method in coordinate space~\cite{Glassel:2021rod,Bratkovskaya:2022vqi,Coci:2023daq}, with the MST radius set to $r = 4$ fm. 
The uncertainties in this model arise from different finite size corrections~\cite{Coci:2023daq}. One can see in Fig.~\ref{fig:dndy_fit} (colored grid bands), that for central collisions, the PHQMD model results for proton rapidity distributions are consistent with data, but the yields of d, t, $^{3}\mathrm{He}$, and $^{4}\mathrm{He}$ are all underpredicted at mid-rapidity.

To obtain the yields for the full phase space (4$\pi$ yields), the $dN/dy$ distributions were fitted by the three-Gaussians function and the modified generalized Gaussian function~\cite{Moulin:1999mp,2011Geodesics,2012On}, the latter of whic of which can be expressed in Eq.~\ref{eq:GG1G}.

\onecolumngrid
\lipsum[0]
\begin{eqnarray}\label{eq:GG1G}
\centering
    Fitfunc. \! = \! p_{0} \! \cdot \! \left(\frac{e^{-\frac{1}{2}\left(\frac{\log \left(1-\frac{k}{p}(x-x_{1})\right.}{k}\right)^{2}}}{\sqrt{2 \pi} \cdot(p-k \cdot(x-x_{1}))} \! + \! \frac{e^{-\frac{1}{2}\left(\frac{\log \left(1+\frac{k}{p}(x+x_{1})\right.}{k}\right)^{2}}}{\sqrt{2 \pi} \cdot(p+k \cdot(x+x_{1}))}\right) \! + \! p_{4} \! \cdot \! e^{-\frac{1}{2} \cdot\left(\frac{x-p_{5}}{p_{6}}\right)^{2}}
\end{eqnarray}
\twocolumngrid
where the parts in parentheses are the variant of the standard form of the generalized Gaussian function: $p (x; \alpha, \beta)=\frac{\alpha}{2 \beta \Gamma(1 / \alpha)} e^{-(|x| / \beta)^{\alpha}}, \alpha, \beta>0$. This distribution is symmetric about $x=0$ and exhibits a sharp peak at $x = 0$. 
The parameters $k$, $p$, and $x_{1}$ in the formula are given in the following forms:         
\begin{equation}                                                
A=p_3^2+\sqrt{p_3^4+4 p_3^2}+2
\end{equation}
\begin{equation}
k=\sqrt{\log \left[\left(\frac{A}{2}\right)^{\frac{1}{3}}+\left(\frac{2}{A}\right)^{\frac{1}{3}}-1\right]}
\end{equation}                                                  
\begin{equation}                                                
p=p_2 \cdot k \cdot \frac{e^{-\frac{1}{2} k^2}}{\sqrt{e^{k^2}-1}}           
\end{equation}                                                  
\begin{equation}                                                
x_{1}=\frac{p_1 \cdot k \cdot e^{k^2}+p\left(1-e^{k^2}\right)}{k \cdot e^{k^2}}
\end{equation}                                                  
with the above functions, the sharp peak will occur at $x = \pm p_{1}$. Here $p_{2}$ is the standard deviation, and $p_{3}$ represents skewness.
The term outside the parentheses is a Gaussian function with an expectation value $p_{5}=0$.

The 4$\pi$ yield was obtained by summing the true value of the measured region with the fitted value of the unmeasured region. As shown by the red and blue lines in Fig.~\ref{fig:dndy_fit_rap}, they represent the fitting results of the generalized Gaussian and three-Gaussians functions, respectively. The average of the two fits outside the target rapidity region ($1.0<|y|<2.0$) was used as the yield value for the extrapolation region.

The systematic uncertainty for the 4$\pi$ yield, in the measured rapidity region ($-1.0 < y < 0$), is the sum of the measured $dN/dy$ systematic errors. For the unmeasured region, the discrepancy between the two fits described above was considered as the systematic uncertainty.
Finally, the systematic uncertainties of the 4$\pi$ yields are 6-11\% for protons, 7-20\% for deuterons, 11-28\% for tritons, 13-25\% for $^{3}\mathrm{He}$, and 14-28\% for $^{4}\mathrm{He}$. Table~\ref{tab:4piyield} lists the values, statistical, and systematic uncertainties for each particle across all centrality bins. 

\begin{figure}[H]
    \centering
    \includegraphics[width=0.85\linewidth]{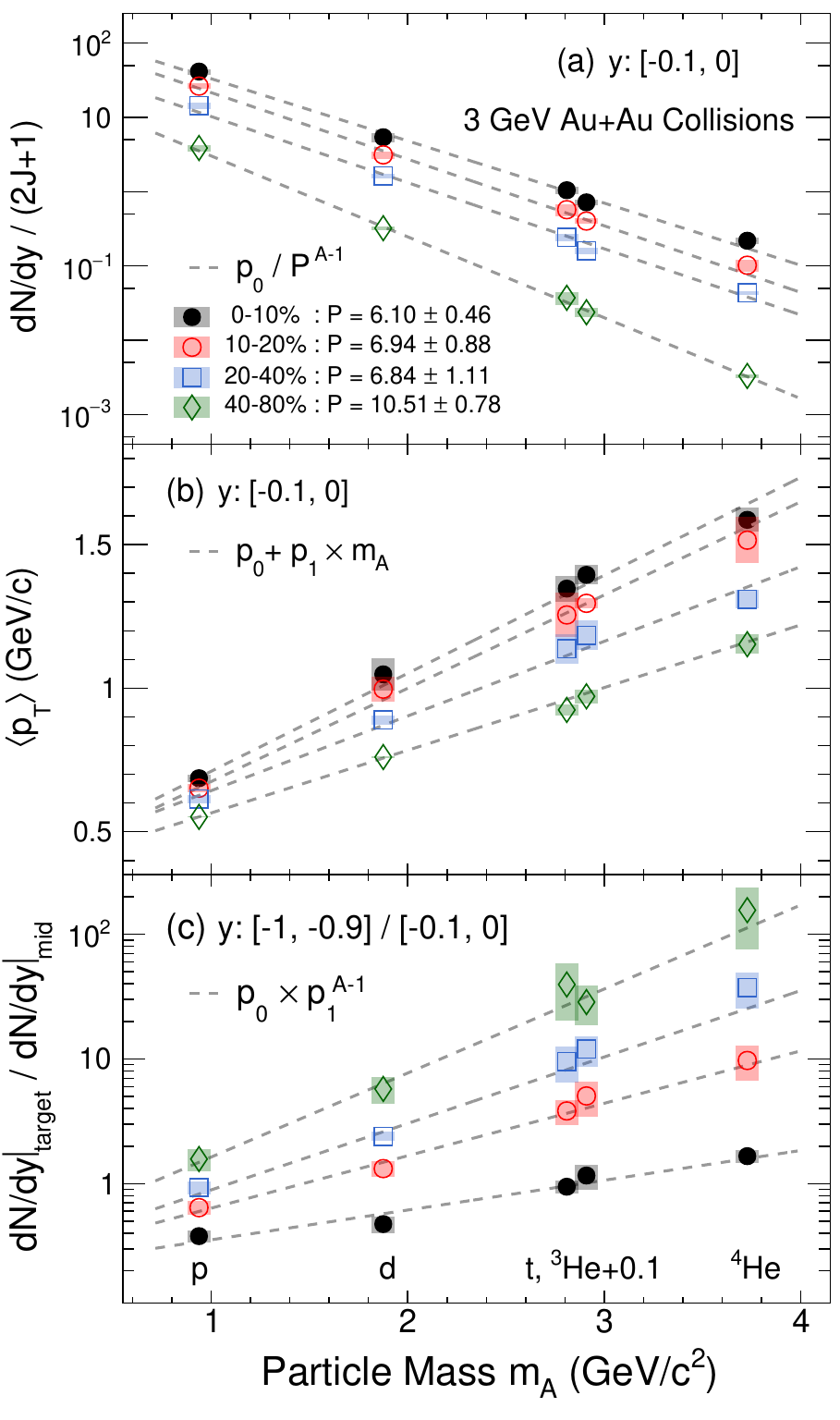}
	\caption{(a) Mid-rapidity particle yields $dN/dy$, (b) mean transverse momentum $\langle p_{T} \rangle$, and (c) ratio of $dN/dy$ measured at target rapidity ($-1.0 < y < -0.9$) to mid-rapidity ($-0.1 < y < 0$) as a function of collision centrality and particle mass $m_{A}$ (where $A$ is the mass number) from Au+Au collisions at $\sqrt{s_{\mathrm{NN}}}$ = \SI{3}{GeV}. For clarity, the mass of $^{3}\mathrm{He}$ is shifted by 0.1 GeV/$c^2$. The boxes represent the quadratic sum of the measured statistical uncertainties and the extrapolation uncertainties. Dashed lines are fitting results of an exponential function to (a) the yields and (c) the yield ratios, and a first-order polynomial to (b) the $\langle p_{T} \rangle$, respectively.}
    \label{fig:4pimass}
\end{figure}

\begin{figure*}[htb]
    \centering
    \includegraphics[width=0.85\linewidth]{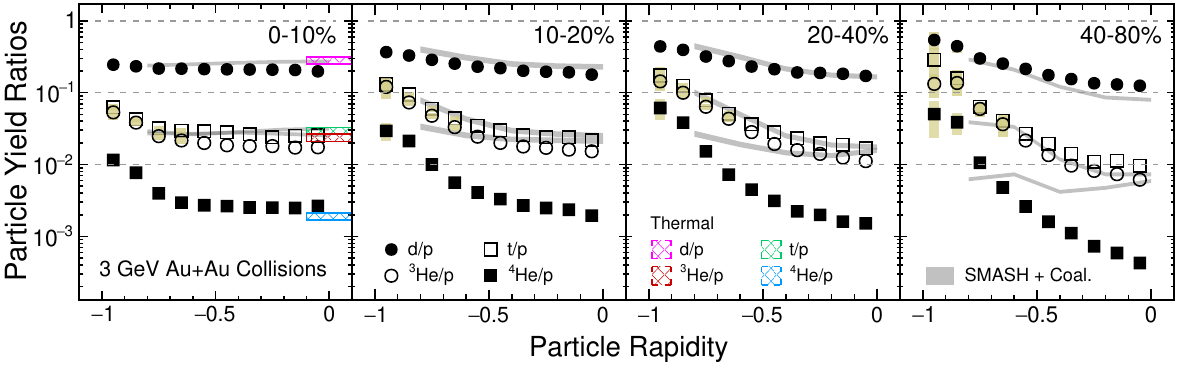}
	\caption{Rapidity dependence of particle ratios $d/p$, $t/p$, $^{3}\mathrm{He}/p$, and $^{4}\mathrm{He}/p$ for 0-10\%, 10-20\%, 20-40\%, and 40-80\% centralities in Au+Au collisions at $\sqrt{s_{\mathrm{NN}}}$ = \SI{3}{GeV}. The vertical yellow areas represent systematic uncertainties. Results of SMASH + Coalescence and thermal model calculations are shown as solid gray bands and hatched colored bands (mid-rapidity only), respectively. For demonstration purposes, a band of uniform width is applied to the thermal model results, and the statistical uncertainties of the SMASH model are expanded by a factor of 5 to set the width of the band.}
    \label{fig:particleratio}
\end{figure*}

\begin{figure}[htb]
    \centering
    \includegraphics[width=0.87\linewidth]{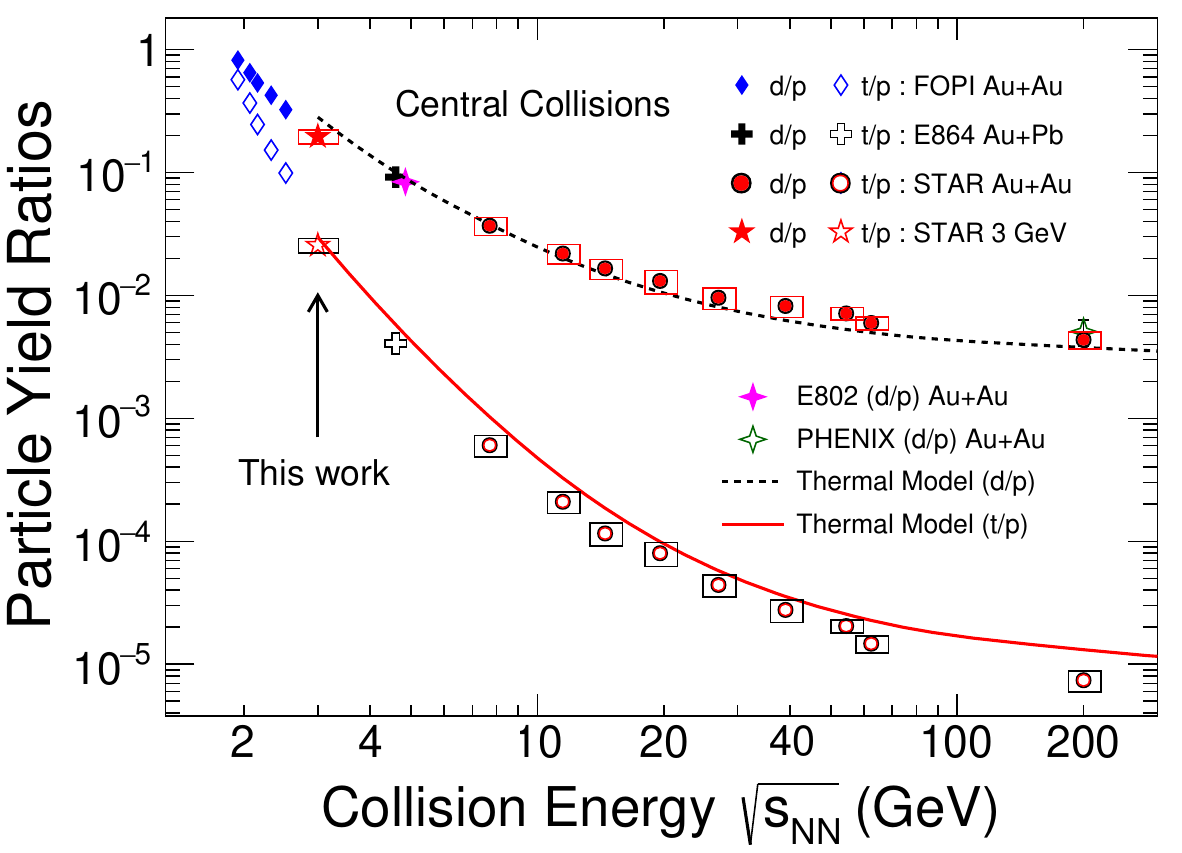}
	\caption{Energy dependence of $d/p$ (filled circles) and $t/p$ (open circles) yield ratios. Statistical and systematic uncertainties are shown as vertical lines and boxes, respectively. The experimental results come from STAR (0-10\%)~\cite{STAR:2019sjh,PhysRevLett.130.202301}, FOPI (impact parameter $b_{0}<0.15$)~\cite{FOPI:2010xrt,Shuryak:2020yrs}, E864 (0-10\%)~\cite{E864:2000auv}, and E802 (0-12\%)~\cite{E802:1999hit}. Results from the thermal model~\cite{Vovchenko:2020dmv} are displayed as lines.}
    \label{fig:parratioEne}
\end{figure}
Figure~\ref{fig:4pimass} shows the mid-rapidity $dN/dy$, the $dN/dy$ ratio of target rapidity ($-1.0 < y < -0.9$) to mid-rapidity ($-0.1 < y < 0$), and $\langle p_{T}\rangle$ as a function of particle mass for four centralities. In Fig.~\ref{fig:4pimass} (a), the $dN/dy$ values for different particles were scaled by their corresponding spin degeneracy factor ($2\mathrm{J}+1$)~\cite{Andronic:2005yp,E864:1999zqh}. The mass dependence of the scaled $dN/dy$ values can be well fitted by an exponential function, $p_0 / P^{A-1}$. Here, $P$ is the penalty factor, which is related to the Boltzmann factor~$e^{\left(m_{N}-\mu_{B}\right) / T}$~\cite{Braun-Munzinger:1994zkz,E864:1999zqh,Steinheimer:2012gq,NA49:2016qvu} used in the statistical production of light nuclei. The penalty factors are about 6.1 $\pm$ 0.5 and 10.5 $\pm$ 0.8 for the most central and peripheral Au+Au collisions at $\sqrt{s_{\mathrm{NN}}}$ = \SI{3}{GeV}, respectively. Figure~\ref{fig:4pimass} (b) shows that $\langle p_{T}\rangle$ increases linearly with increasing mass of the particle and exhibits centrality dependence. This indicates the light nuclei participate in the collective expansion of the fireball, which is stronger in central collisions than in peripheral collisions.
Figure~\ref{fig:4pimass} (c) shows the ratio of the $dN/dy$ values measured at target rapidity to mid-rapidity as a function of particle mass. The ratio reveals the relative contributions of nuclear fragmentation to the yields of light nuclei at different collision centralities and particle mass. The ratio increases exponentially with increasing particle mass, with this upward trend being more pronounced in peripheral collisions compared to central collisions. It indicates that, from light to heavy nuclei, the proportion of contributions originating from the nuclear fragments increases. 

\subsection{Particle Ratio}
Figure \ref{fig:particleratio} shows the rapidity and centrality dependence of the light nuclei to proton yield ratios ($d/p$, $t/p$, $^{3}\mathrm{He}/p$, and $^{4}\mathrm{He}/p$) in Au+Au collisions at $\sqrt{s_{\mathrm{NN}}}$ = \SI{3}{GeV}. Based on the conclusions from the previous analysis of spectator contributions, it is observed that these particle ratios monotonically decrease from target to mid-rapidity and exhibit a stronger rapidity dependence in peripheral collisions than in central collisions. The particle ratios calculated from the SMASH model are shown for comparison and denoted by the gray bands. In addition, using the parameters $T = 85$ MeV and $\mu_{B} = 728$ MeV~\cite{Vovchenko:2015idt}, we estimated the mid-rapidity particle yield ratios in 0-10\% central collisions from the thermal model without excited nuclei contributions~\cite{Vovchenko:2015idt,Vovchenko:2020dmv}. 
Figure~\ref{fig:parratioEne} shows the energy dependence of the mid-rapidity ($-0.1 < y < 0$) $d/p$ and $t/p$ yield ratios in central Au+Au collisions. Both the $d/p$ and $t/p$ ratios at \SI{3}{GeV} follow the world trend of energy dependence observed by the STAR Beam Energy Scan I~\cite{STAR:2019sjh,PhysRevLett.130.202301}, FOPI~\cite{FOPI:2010xrt,Shuryak:2020yrs}, and AGS~\cite{E802:1999hit,E864:2000auv} experiments. The thermal model successfully predicts energy-dependent trends in $d/p$ and $t/p$. At 3 GeV, the $d/p$ and $t/p$ ratios show increases of 3.4$\sigma$ and 1.2$\sigma$, respectively. 

\begin{figure}[htb]
    \centering
    \includegraphics[width=0.97\linewidth]{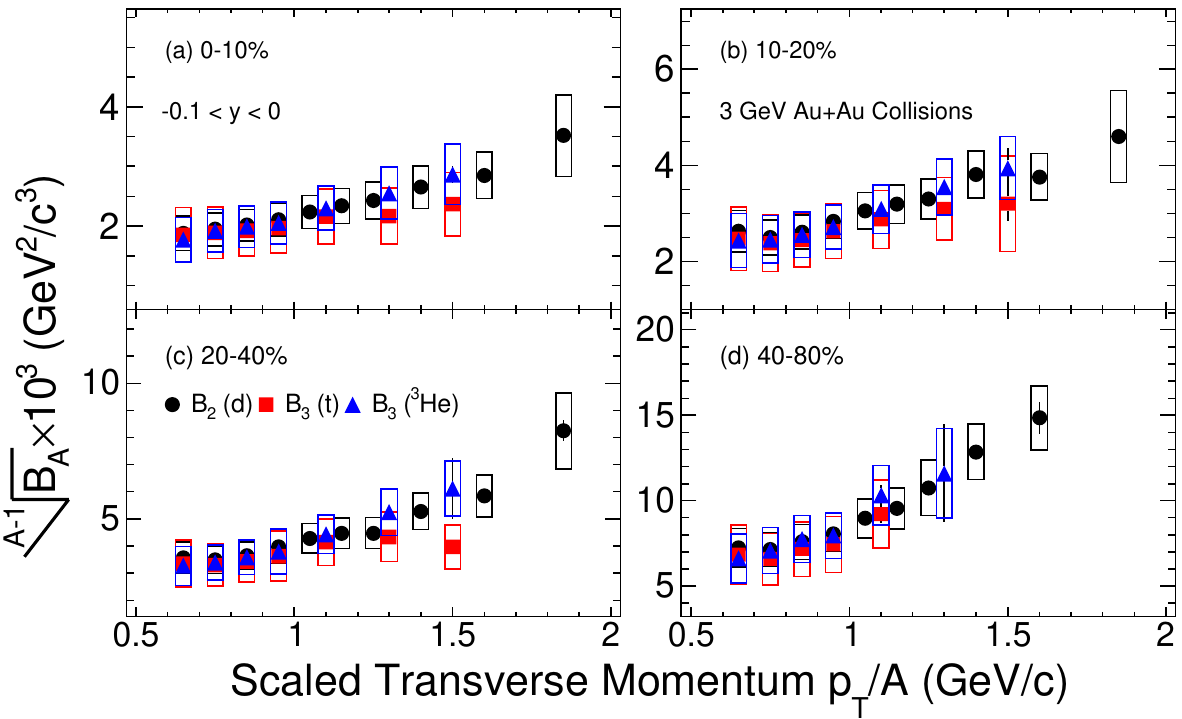}
    \caption{Coalescence parameters $B_{2}(d)$, $\sqrt{B_{3}}(t)$, and $\sqrt{B_{3}}(^{3}\mathrm{He})$ as a function of $p_{T}/A$ for different centrality bins in Au+Au collisions at $\sqrt{s_{\mathrm{NN}}}$ = \SI{3}{GeV}. The boxes represent systematic uncertainties.}
    \label{fig:ba_pT}
\end{figure} 

\begin{figure*}[htb]
    \centering                                          
    \includegraphics[width=0.84\linewidth]{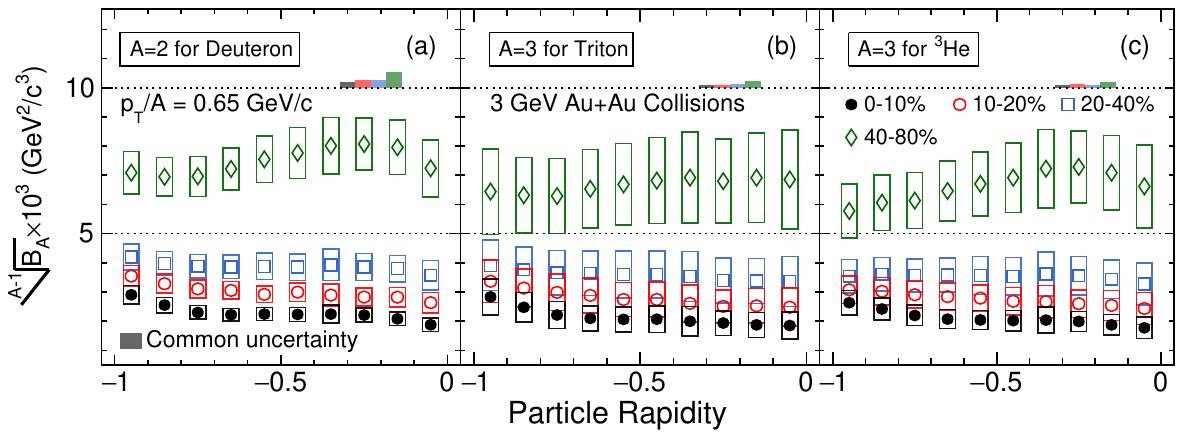}
	\caption{Rapidity dependence of the coalescence parameters $B_{2}(d)$, $B_{3}(t)$, and $B_{3}(^{3}\mathrm{He})$ at $p_{T}/A$ = \SI{0.65}{GeV/c} for different centrality bins in Au+Au collisions at $\sqrt{s_{\mathrm{NN}}}$ = \SI{3}{GeV}. The boxes represent systematic uncertainties. The colored bands represent common uncertainties in the rapidity dependence.}

    \label{fig:ba_rap}
\end{figure*} 
             
\subsection{Coalescence Parameter}
In the coalescence model~\cite{Butler:1963pp,Yu:2018kvh}, light nuclei are formed via the coalescence of their constituents (protons and neutrons), thus the relation between the momentum spectra of light nuclei, protons, and neutrons is given by:

\begin{equation}
\centering
\begin{split}
    E_{A} \frac{d^{3} N_{A}}{d^{3} p_{A}}=B_{A}\left(E_{p} \frac{d^{3} N_{p}}{d^{3} p_{p}}\right)^{Z}\left(E_{n} \frac{d^{3} N_{n}}{d^{3} p_{n}}\right)^{A-Z}\\
	\left.\approx (1.3)^{A-Z}B_{A}\left(E_{p} \frac{d^{3} N_{p}}{d^{3} p_{p}}\right)^{A}\right|_{p_{p}=p_{n}=\frac{p_{A}}{A}}
\end{split}
\end{equation}
where $A$ is the mass number of the light nuclei, and $Z$ is the number of protons in it. We assume the same $p_{T}$, rapidity, and centrality dependence between protons and neutrons. The neutron spectrum is derived by scaling the proton spectrum with a factor of $n/p=1.3 \pm 0.1$. This scaling factor is estimated from the $t/^{3}\mathrm{He}$ ratio~\cite{Kolesnikov:2007ps} measured at \SI{3}{GeV}, and this ratio is consistent with the value of 1.28 calculated by the thermal model.
$B_{A} \propto(1 / V_{\mathrm{eff}})^{(\mathrm{A}-1)}$ denotes the coalescence parameter, which is to reflect the coalescence probability of light nuclei with the mass number $A$.
$V_{\mathrm{eff}}$ is the effective volume of the nucleon emission source. 

Figure~\ref{fig:ba_pT} describes the scaled transverse momentum ($p_{T}/A$) dependence of the coalescence parameters for $B_{2}(d)$, $B_{3}(t)$, and $B_{3}(^{3}\mathrm{He})$ at mid-rapidity in 0-10\%, 10-20\%, 20-40\%, and 40-80\% centrality bins. Within the  uncertainties, $^{A-1}\sqrt{B_{A}}$ of $d$, $t$, and $^{3}\mathrm{He}$ are consistent, and their values increase with increasing $p_{T}$. This increasing trend can be explained by the presence of collective flow, and the fact that the length of homogeneity becomes smaller at higher transverse momentum~\cite{Scheibl:1998tk}. As discussed in Ref.~\cite{Scheibl:1998tk}, there is a perfect scaling relation between the length of homogeneity and the effective volume. Figure \ref{fig:ba_rap} presents the rapidity and centrality dependence of the coalescence parameters: $B_{2}$ for deuterons, $\sqrt{B_{3}}$ for tritons, and $^{3}\mathrm{He}$ at $p_{T}/A$ = \SI{0.65}{GeV/c}. As shown by the colored bands, the common uncertainties in rapidity dependence were removed from the final systematic uncertainties. Both $B_{2}$ and $\sqrt{B_{3}}$ increase from central to peripheral collisions and, as mentioned before, these behaviors can be interpreted as indicating that the effective source volume decreases from central to peripheral collisions.     

\begin{figure}[htb]
    \centering
    \includegraphics[width=0.88\linewidth]{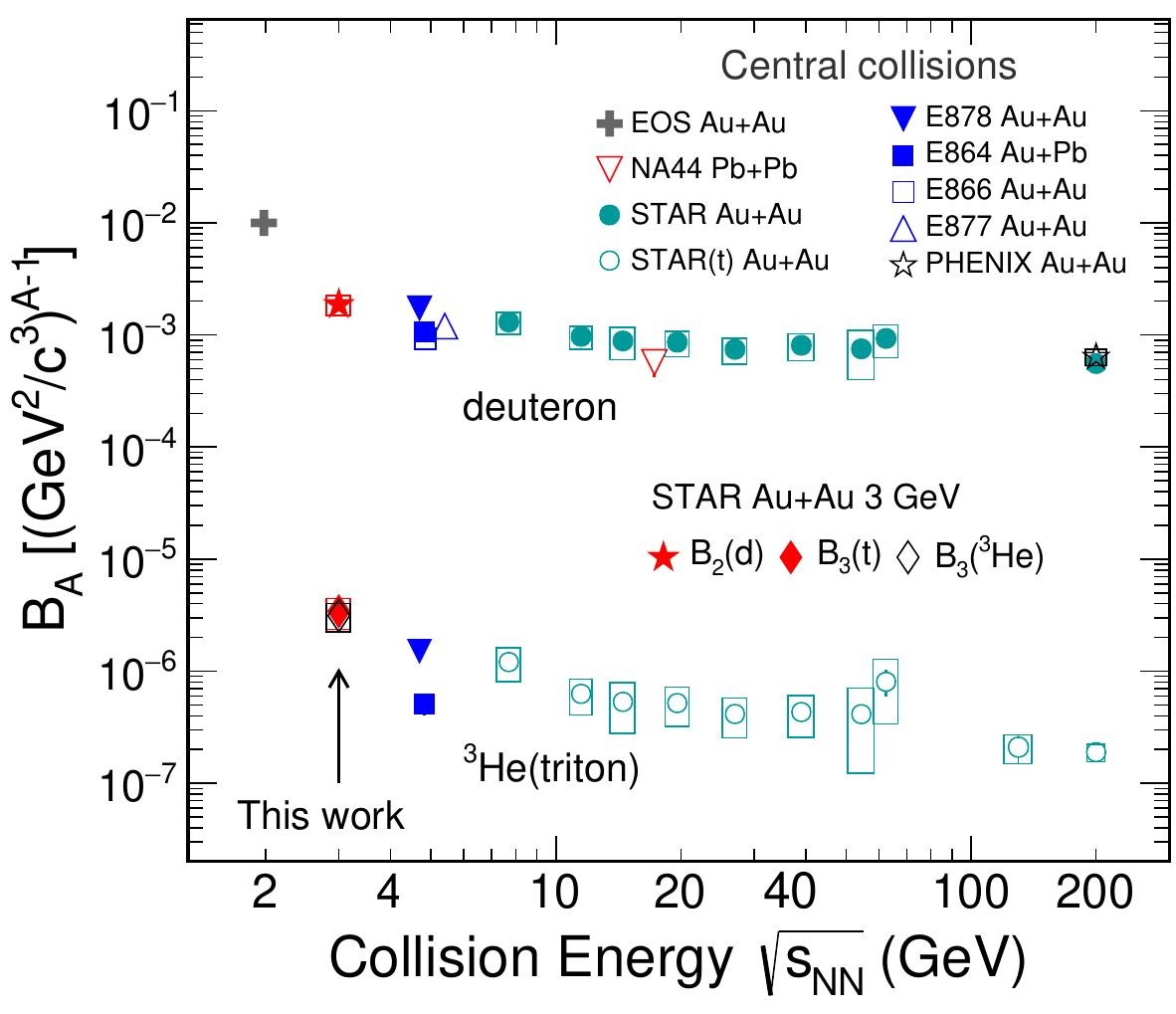}
	\caption{Energy dependence of the coalescence parameters for $B_{2}(d)$ and $B_{3}(t, ^{3}\mathrm{He})$ in central collisions. The vertical lines indicate statistical uncertainties. The boxes indicate systematic uncertainties. For comparison, results from EOS~\cite{NA49:2016qvu}, NA44 (0-10\%)~\cite{Bearden:2002ta}, AGS (0-10\%)~\cite{E878:1998vna,E864:2000auv}, PHENIX (0-20\%)~\cite{PHENIX:2004vqi}, and STAR BES-I (0-10\%)~\cite{STAR:2001pbk,PhysRevLett.130.202301} are also shown.}

    \label{fig:ba_ene}       
\end{figure}

Figure~\ref{fig:ba_ene} shows the energy dependence of the coalescence parameters in central heavy-ion collisions. The experimental data include the measurements from the EOS~\cite{NA49:2016qvu}, NA44~\cite{Bearden:2002ta}, AGS~\cite{E878:1998vna,E864:2000auv}, PHENIX~\cite{PHENIX:2004vqi}, and STAR BES-I~\cite{STAR:2001pbk,PhysRevLett.130.202301} experiments. At mid-rapidity, $B_{2}(d)$, $B_{3}(t)$, and $B_{3}(^{3}\mathrm{He})$ with transverse momentum $p_{T}/A$ = \SI{0.65}{GeV/c} at $\sqrt{s_{\mathrm{NN}}}$ = \SI{3}{GeV} follow the world trend, and there is a clear upward trend towards low energy, which implies that the overall effective volume of the nucleon emitting source ($V_{\mathrm{eff}}$) decreases with decreasing collision energies.
             
\begin{figure*}[htb]
	\centering
	\includegraphics[width=0.79\linewidth]{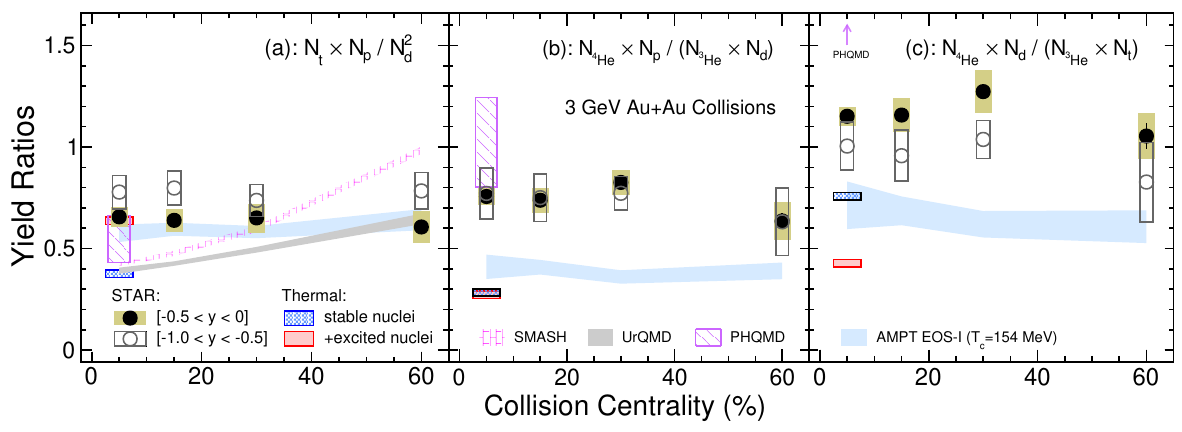}
	\caption{Centrality and rapidity dependence of the yield ratios $N_{p} \times N_{t} / N_{d}^{2}$, $N_{^{4}\mathrm{He}} \times N_{p} / \left(N_{^{3}\mathrm{He}} \times N_{d}\right)$, and $N_{^{4}\mathrm{He}} \times N_{d} / \left(N_{^{3}\mathrm{He}} \times N_{t}\right)$ in Au+Au collisions at $\sqrt{s_{\mathrm{NN}}}$ = \SI{3}{GeV}. Solid and open markers indicate ratios from mid-rapidity ($-0.5 < y < 0$) and target rapidity ($-1 < y < -0.5$), respectively. Statistical uncertainties are smaller than the size of the markers, and systematic uncertainties are shown by boxes. Results from hadronic transport models UrQMD, SMASH, and AMPT EoS-I are shown as colored bands. For the top 0-10\% central collisions, results from thermal and PHQMD models are also shown.}

     \label{fig:YieldRatio}             
\end{figure*}

\subsection{The Compound Yield Ratio of Light Nuclei}
Based on the coalescence model~\cite{Sun:2018jhg}, the compound yield ratio of tritons, deuterons, and protons ($N_{t} \times N_{p}/N_{d}^{2}$), was predicted to be sensitive to the neutron density fluctuations. Thus, it can be used to probe the signatures of the QCD critical point and/or the first-order phase transition in heavy-ion collisions. The STAR experiment has reported the centrality and the beam energy dependence of this yield ratio in Au+Au collisions at {\sNN} = 7.7 -- \SI{200}{GeV}~\cite{PhysRevLett.130.202301}. The measured yield ratio $N_{p} \times N_{t} / N_{d}^{2}$ monotonically decreases with increasing charged-particle multiplicity and shows scaling behavior regardless of the energy and centrality. As the yield of light nuclei primarily depends on the available phase space for nucleons and the spatial extent of the light nuclei, this scaling behavior of the yield ratio with charged-particle multiplicity, can be explained by the interplay between the finite size of the light nuclei and the overall system size within the framework of the coalescence model~\cite{PhysRevLett.130.202301,Csernai:1986qf,Scheibl:1998tk,Zhao:2021dka}. 

Figure~\ref{fig:YieldRatio} shows the centrality dependence of the yield ratios: $N_{t} \times N_{p} / N_{d}^{2}$, $N_{^{4}\mathrm{He}}\times N_{p}/N_{^{3}\mathrm{He}}\times N_{d}$, and $N_{^{4}\mathrm{He}} \times N_{d} /\left(N_{^{3}{\mathrm{He}}} \times N_{t}\right)$. In each panel, the black solid and open circles denote the results from mid-rapidity and target rapidity, respectively. 
Panels (a), (b), and (c) show that the experimental results exhibit almost no centrality dependence.
Calculations from various models were applied to compare with the data. In panel (a), the results from the SMASH and UrQMD models~\cite{Sun:2022cxp} show a monotonically increasing trends from central to peripheral collisions. The result from the thermal model calculation, which includes the decay from the excited nuclei to light nuclei (red band), is consistent with the experimental result in central collisions. 
In recent AMPT calculations~\cite{Sun:2022cxp}, implementing a first-order phase transition, also align well with the centrality dependence observed in the data. 
Panel (b) depicts that the $N_{^{4}\mathrm{He}} \times N_{p} /\left(N_{^{3}{\mathrm{He}}} \times N_{d}\right)$ shows no centrality dependence, and the AMPT model underestimates the experimental values. In most central collisions, the results of the thermal model are lower than the experimental data whether or not the contribution of excited state decay was considered.
In contrast to panel (b), the $N_{^{4}\mathrm{He}} \times N_{d} /\left(N_{^{3}{\mathrm{He}}} \times N_{t}\right)$ displayed in panel (c) demonstrates that the value from the thermal model is lower than the experimental result in central collisions, while the value from the PHQMD model is much higher. The AMPT model shows a relatively flat trend across different centralities, with ratios consistently lower than the experimental values.

\begin{figure}[htb]
    \centering
    \includegraphics[width=0.9\linewidth]{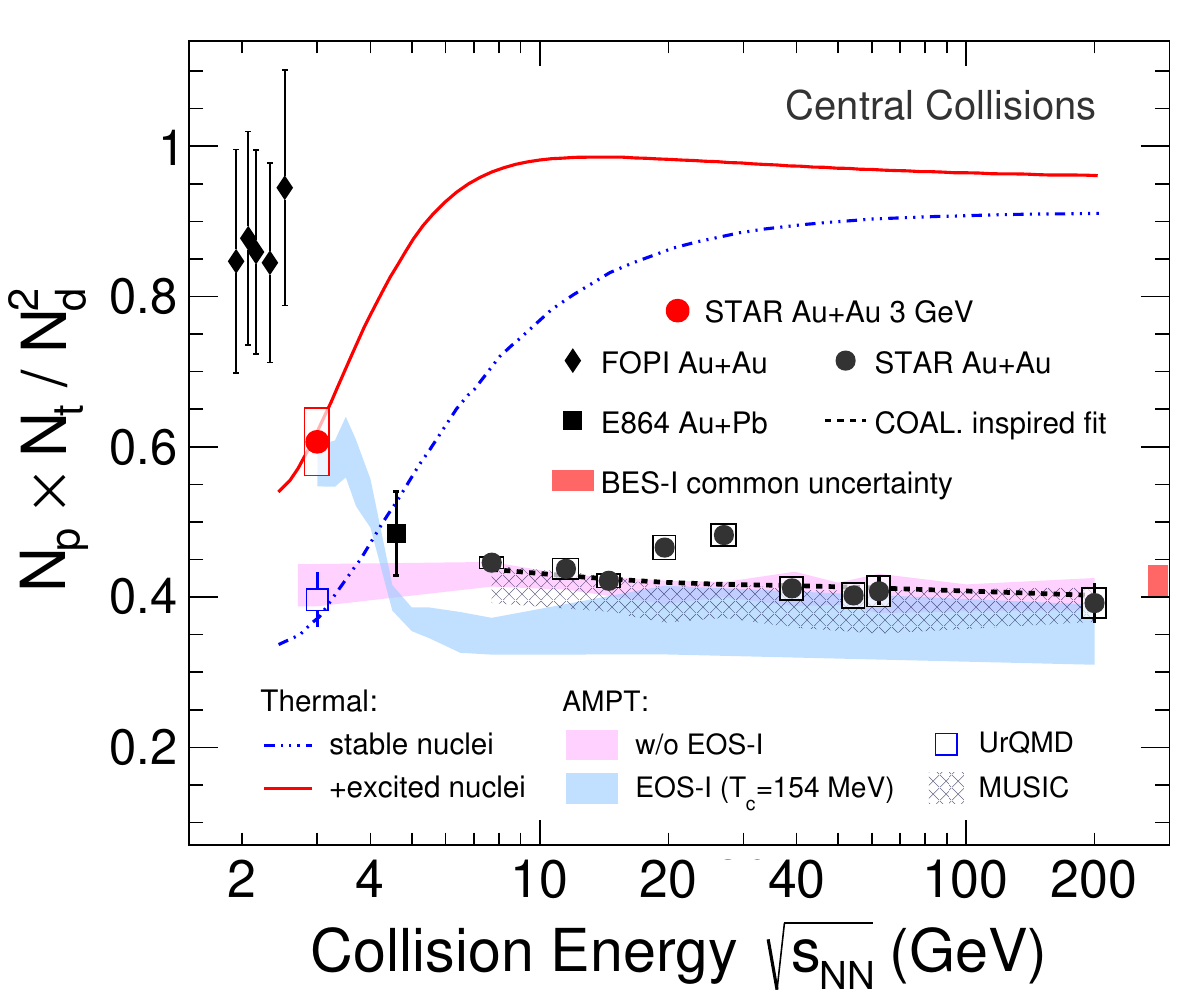}
	\caption{Energy dependence of the ratio $N_{p} \times N_{t} / N_{d}^{2}$ in the most central 0-10\% Au+Au collisions. Vertical lines and boxes represent statistical and systematic uncertainties, respectively. The experimental results come from STAR (0-10\%), E864 (0-10\%), and FOPI (impact parameter $b_{0}<0.15$). The red band at the right side of the plot indicates the common uncertainty ($\sim$4.2\%) in the BES-I result. Colored bands denote the ratios from MUSIC+UrQMD and AMPT (without and with EOS-I) hybrid model calculations. The ratio from the UrQMD model is shown by an open square. Dashed black lines represent the coalescence baselines obtained from a coalescence-inspired fit~\cite{Sun:2018mqq}. Solid red and dotted blue lines represent thermal model results.}
    \label{fig:yieldratio_ene}
\end{figure}

The energy dependence of $N_{t} \times N_{p} / N_{d}^{2}$ in central heavy-ion collisions at mid-rapidity ($-0.5<y<0$) was presented in Fig.~\ref{fig:yieldratio_ene}. The colored bands and gray grid represent the calculations obtained from the hadronic transport model AMPT~\cite{Sun:2022cxp} and the hybrid model MUSIC+UrQMD~\cite{Zhao:2021dka}, respectively. The black dashed line corresponds to the coalescence baseline obtained by fitting the charged-particle multiplicity dependence of the yield ratio from STAR BES-I data~\cite{PhysRevLett.130.202301}. Enhancements of the yield ratio relative to the coalescence baseline in 0-10\% central Au+Au collisions at \sNN\ = 19.6 and 27GeV are observed. At lower energies, the experimental results from the E864~\cite{E864:2000auv}, STAR, and FOPI~\cite{FOPI:2010xrt} experiments align with the world trend of the energy dependence, showing a monotonically increasing trend with decreasing energies. The thermal model, whether considering contributions from excited nuclear state decays or not, shows a monotonically increasing trend with increasing collision energies, eventually reaching saturation at energies around a few tens of GeV. While the thermal model can describe the yield ratio at \sNN\ = 3 GeV when including the decay of excited nuclear states, this agreement is most likely coincidental. Coalescence-based calculations of central Au+Au collisions at \sNN\ = \SI{3}{GeV} from UrQMD and AMPT models without considering the excited nuclear states decays show smaller values than the \SI{3}{GeV} data. Furthermore, this yield ratio can be also reproduced by the AMPT model when employing a first-order phase transition. Those detailed comparisons between experimental data and various model calculations demonstrate that the production of light nuclei at mid-rapidity in Au+Au collisions at RHIC energies (\sNN\ = 7.7 - \SI{200}{GeV}) can be effectively explained by nucleon coalescence models. Additionally, it has been observed that the thermal model fails to describe the overall trend of the energy dependence of the yield ratios, and at lower energies, the yields of light nuclei receive significant contributions from the decays of excited nuclear states. 

\section{Summary}
We report the comprehensive measurements of protons ($p$) and light nuclei ($d$, $t$, $^{3}\mathrm{He}$, and $^{4}\mathrm{He}$) production from mid-rapidity to target rapidity in Au+Au collisions at $\sqrt{s_{\mathrm{NN}}}$ = \SI{3}{GeV} by the STAR experiment. The transverse momentum $p_{T}$ spectra, integrated yield ($dN/dy$), mean transverse momentum ($\langle p_{T}\rangle$), particle yield ratios ($d/p$, $t/p$,$^{3}\mathrm{He}/p$, $^{4}\mathrm{He}/p$), and coalescence parameters ($B_2$, $B_3$) are presented as a function of rapidity and collision centrality. The 4$\pi$ yields are extracted based on the measured rapidity distributions of $dN/dy$. It is observed that the mid-rapidity $dN/dy$ of protons and light nuclei can be well described by the exponential dependence of the particle mass via the Boltzmann factor $e^{\left(m_{N}-\mu_{B}\right) / T}$~\cite{E864:1999zqh,Nagle:1996vp,DasGupta:1981xx}. The contributions from spectator fragmentations become more pronounced in peripheral collisions. 
The compound yield ratio $N_{p} \times N_{t} / N_{d}^{2}$ shows no centrality dependence for both mid-rapidity and target rapidity. Calculations of $N_{p} \times N_{t} / N_{d}^{2}$ from coalescence-based UrQMD and SMASH transport models show significant increasing trends from central to peripheral collisions, which fail to describe the experimental data~\cite{Sun:2018mqq,Zhao:2021dka}. Furthermore, the increasing trend of the yield ratio $N_p \times N_t / N_d^2$ at energies below \SI{4}{GeV}, which cannot be explained by thermal model or transport model calculations, suggests the presence of additional physics beyond the scope of these models. A recent AMPT calculation~\cite{Sun:2022cxp} incorporating a first-order phase transition successfully reproduces this increasing trend at low energies. The systematic measurements of the production of protons and light nuclei at \SI{3}{GeV} provide valuable insights into the production dynamics of light nuclei and may potentially contribute to our understanding of the QCD phase structure at high baryon density.

\section{Acknowledgements}
We thank Drs. J. Aichelin,  E. Bratkovskaya, C. Ko, J. Steinheimer, K. Sun, and W. Zhao for fruitful discussions about the production mechanism of light nuclei. 
We thank the RHIC Operations Group and RCF at BNL, the NERSC Center at LBNL, and the Open Science Grid consortium for providing resources and support.  This work was supported in part by the Office of Nuclear Physics within the U.S. DOE Office of Science, the U.S. National Science Foundation, National Natural Science Foundation of China, Chinese Academy of Science, the Ministry of Science and Technology of China and the Chinese Ministry of Education, the Higher Education Sprout Project by Ministry of Education at NCKU, the National Research Foundation of Korea, Czech Science Foundation and Ministry of Education, Youth and Sports of the Czech Republic, Hungarian National Research, Development and Innovation Office, New National Excellency Programme of the Hungarian Ministry of Human Capacities, Department of Atomic Energy and Department of Science and Technology of the Government of India, the National Science Centre and WUT ID-UB of Poland, the Ministry of Science, Education and Sports of the Republic of Croatia, German Bundesministerium f\"ur Bildung, Wissenschaft, Forschung and Technologie (BMBF), Helmholtz Association, Ministry of Education, Culture, Sports, Science, and Technology (MEXT), Japan Society for the Promotion of Science (JSPS) and Agencia Nacional de Investigaci\'on y Desarrollo (ANID) of Chile.

\bibliographystyle{apsrev} 
\bibliography{draft_LN_reference} 

\begin{thebibliography}{90}
\expandafter\ifx\csname natexlab\endcsname\relax\def\natexlab#1{#1}\fi
\expandafter\ifx\csname bibnamefont\endcsname\relax
  \def\bibnamefont#1{#1}\fi
\expandafter\ifx\csname bibfnamefont\endcsname\relax
  \def\bibfnamefont#1{#1}\fi
\expandafter\ifx\csname citenamefont\endcsname\relax
  \def\citenamefont#1{#1}\fi
\expandafter\ifx\csname url\endcsname\relax
  \def\url#1{\texttt{#1}}\fi
\expandafter\ifx\csname urlprefix\endcsname\relax\def\urlprefix{URL }\fi
\providecommand{\bibinfo}[2]{#2}
\providecommand{\eprint}[2][]{\url{#2}}

\bibitem[{\citenamefont{Arsene et~al.}(2005)}]{BRAHMS:2004adc}
\bibinfo{author}{\bibfnamefont{I.}~\bibnamefont{Arsene}} \bibnamefont{et~al.}
  (\bibinfo{collaboration}{BRAHMS Collaboration}), \bibinfo{journal}{Nucl.
  Phys. A} \textbf{\bibinfo{volume}{757}}, \bibinfo{pages}{1}
  (\bibinfo{year}{2005}), \eprint{nucl-ex/0410020}.

\bibitem[{\citenamefont{Adcox et~al.}(2005)}]{PHENIX:2004vcz}
\bibinfo{author}{\bibfnamefont{K.}~\bibnamefont{Adcox}} \bibnamefont{et~al.}
  (\bibinfo{collaboration}{PHENIX Collaboration}), \bibinfo{journal}{Nucl.
  Phys. A} \textbf{\bibinfo{volume}{757}}, \bibinfo{pages}{184}
  (\bibinfo{year}{2005}), \eprint{nucl-ex/0410003}.

\bibitem[{\citenamefont{Back et~al.}(2005)}]{PHOBOS:2004zne}
\bibinfo{author}{\bibfnamefont{B.~B.} \bibnamefont{Back}} \bibnamefont{et~al.}
  (\bibinfo{collaboration}{PHOBOS Collaboration}), \bibinfo{journal}{Nucl.
  Phys. A} \textbf{\bibinfo{volume}{757}}, \bibinfo{pages}{28}
  (\bibinfo{year}{2005}), \eprint{nucl-ex/0410022}.

\bibitem[{\citenamefont{Adams et~al.}(2005)}]{STAR:2005gfr}
\bibinfo{author}{\bibfnamefont{J.}~\bibnamefont{Adams}} \bibnamefont{et~al.}
  (\bibinfo{collaboration}{STAR Collaboration}), \bibinfo{journal}{Nucl. Phys.
  A} \textbf{\bibinfo{volume}{757}}, \bibinfo{pages}{102}
  (\bibinfo{year}{2005}), \eprint{nucl-ex/0501009}.

\bibitem[{\citenamefont{ALICE-Collaboration}(2022)}]{ALICE:2022wpn}
\bibinfo{author}{\bibnamefont{ALICE-Collaboration}}
  (\bibinfo{collaboration}{ALICE Collaboration}), \emph{\bibinfo{title}{{The
  ALICE experiment -- A journey through QCD}}} (\bibinfo{year}{2022}),
  \eprint{2211.04384}.

\bibitem[{\citenamefont{Chen et~al.}(2018)\citenamefont{Chen, Keane, Ma, Tang,
  and Xu}}]{Chen:2018tnh}
\bibinfo{author}{\bibfnamefont{J.}~\bibnamefont{Chen}},
  \bibinfo{author}{\bibfnamefont{D.}~\bibnamefont{Keane}},
  \bibinfo{author}{\bibfnamefont{Y.-G.} \bibnamefont{Ma}},
  \bibinfo{author}{\bibfnamefont{A.}~\bibnamefont{Tang}}, \bibnamefont{and}
  \bibinfo{author}{\bibfnamefont{Z.}~\bibnamefont{Xu}}, \bibinfo{journal}{Phys.
  Rept.} \textbf{\bibinfo{volume}{760}}, \bibinfo{pages}{1}
  (\bibinfo{year}{2018}), \eprint{1808.09619}.

\bibitem[{\citenamefont{Luo et~al.}(2022)\citenamefont{Luo, Wang, Xu, and
  Zhuang}}]{Luo:2022mtp}
\bibinfo{editor}{\bibfnamefont{X.}~\bibnamefont{Luo}},
  \bibinfo{editor}{\bibfnamefont{Q.}~\bibnamefont{Wang}},
  \bibinfo{editor}{\bibfnamefont{N.}~\bibnamefont{Xu}}, \bibnamefont{and}
  \bibinfo{editor}{\bibfnamefont{P.}~\bibnamefont{Zhuang}}, eds.,
  \emph{\bibinfo{title}{{Properties of QCD Matter at High Baryon Density}}}
  (\bibinfo{publisher}{Springer}, \bibinfo{year}{2022}), ISBN
  \bibinfo{isbn}{978-981-19444-0-6, 978-981-19444-1-3}.

\bibitem[{\citenamefont{Chen et~al.}(2024)}]{Chen:2024zwk}
\bibinfo{author}{\bibfnamefont{J.}~\bibnamefont{Chen}} \bibnamefont{et~al.}
  (\bibinfo{year}{2024}), \eprint{2407.02935}.

\bibitem[{\citenamefont{Aoki et~al.}(2006)\citenamefont{Aoki, Endrodi, Fodor,
  Katz, and Szabo}}]{Aoki:2006we}
\bibinfo{author}{\bibfnamefont{Y.}~\bibnamefont{Aoki}},
  \bibinfo{author}{\bibfnamefont{G.}~\bibnamefont{Endrodi}},
  \bibinfo{author}{\bibfnamefont{Z.}~\bibnamefont{Fodor}},
  \bibinfo{author}{\bibfnamefont{S.~D.} \bibnamefont{Katz}}, \bibnamefont{and}
  \bibinfo{author}{\bibfnamefont{K.~K.} \bibnamefont{Szabo}},
  \bibinfo{journal}{Nature} \textbf{\bibinfo{volume}{443}},
  \bibinfo{pages}{675} (\bibinfo{year}{2006}), \eprint{hep-lat/0611014}.

\bibitem[{\citenamefont{Fu et~al.}(2020)\citenamefont{Fu, Pawlowski, and
  Rennecke}}]{Fu:2019hdw}
\bibinfo{author}{\bibfnamefont{W.-J.} \bibnamefont{Fu}},
  \bibinfo{author}{\bibfnamefont{J.~M.} \bibnamefont{Pawlowski}},
  \bibnamefont{and} \bibinfo{author}{\bibfnamefont{F.}~\bibnamefont{Rennecke}},
  \bibinfo{journal}{Phys. Rev. D} \textbf{\bibinfo{volume}{101}},
  \bibinfo{pages}{054032} (\bibinfo{year}{2020}), \eprint{1909.02991}.

\bibitem[{\citenamefont{Gao and Pawlowski}(2021)}]{Gao:2020fbl}
\bibinfo{author}{\bibfnamefont{F.}~\bibnamefont{Gao}} \bibnamefont{and}
  \bibinfo{author}{\bibfnamefont{J.~M.} \bibnamefont{Pawlowski}},
  \bibinfo{journal}{Phys. Lett. B} \textbf{\bibinfo{volume}{820}},
  \bibinfo{pages}{136584} (\bibinfo{year}{2021}), \eprint{2010.13705}.

\bibitem[{\citenamefont{Stephanov et~al.}(1999)\citenamefont{Stephanov,
  Rajagopal, and Shuryak}}]{Stephanov:1999zu}
\bibinfo{author}{\bibfnamefont{M.~A.} \bibnamefont{Stephanov}},
  \bibinfo{author}{\bibfnamefont{K.}~\bibnamefont{Rajagopal}},
  \bibnamefont{and} \bibinfo{author}{\bibfnamefont{E.~V.}
  \bibnamefont{Shuryak}}, \bibinfo{journal}{Phys. Rev. D}
  \textbf{\bibinfo{volume}{60}}, \bibinfo{pages}{114028}
  (\bibinfo{year}{1999}), \eprint{hep-ph/9903292}.

\bibitem[{\citenamefont{Ejiri}(2008)}]{Ejiri:2008xt}
\bibinfo{author}{\bibfnamefont{S.}~\bibnamefont{Ejiri}},
  \bibinfo{journal}{Phys. Rev. D} \textbf{\bibinfo{volume}{78}},
  \bibinfo{pages}{074507} (\bibinfo{year}{2008}), \eprint{0804.3227}.

\bibitem[{\citenamefont{Borderie and Frankland}(2019)}]{Borderie:2019fii}
\bibinfo{author}{\bibfnamefont{B.}~\bibnamefont{Borderie}} \bibnamefont{and}
  \bibinfo{author}{\bibfnamefont{J.~D.} \bibnamefont{Frankland}},
  \bibinfo{journal}{Prog. Part. Nucl. Phys.} \textbf{\bibinfo{volume}{105}},
  \bibinfo{pages}{82} (\bibinfo{year}{2019}), \eprint{1903.02881}.

\bibitem[{\citenamefont{Aggarwal et~al.}(2010)}]{STAR:2010vob}
\bibinfo{author}{\bibfnamefont{M.~M.} \bibnamefont{Aggarwal}}
  \bibnamefont{et~al.} (\bibinfo{collaboration}{STAR Collaboration})
  (\bibinfo{year}{2010}), \eprint{1007.2613}.

\bibitem[{\citenamefont{Luo and Xu}(2017)}]{Luo:2017faz}
\bibinfo{author}{\bibfnamefont{X.}~\bibnamefont{Luo}} \bibnamefont{and}
  \bibinfo{author}{\bibfnamefont{N.}~\bibnamefont{Xu}}, \bibinfo{journal}{Nucl.
  Sci. Tech.} \textbf{\bibinfo{volume}{28}}, \bibinfo{pages}{112}
  (\bibinfo{year}{2017}), \eprint{1701.02105}.

\bibitem[{\citenamefont{Barrette et~al.}(1994)}]{E814:1994kon}
\bibinfo{author}{\bibfnamefont{J.}~\bibnamefont{Barrette}} \bibnamefont{et~al.}
  (\bibinfo{collaboration}{E814 Collaboration}), \bibinfo{journal}{Phys. Rev.
  C} \textbf{\bibinfo{volume}{50}}, \bibinfo{pages}{1077}
  (\bibinfo{year}{1994}).

\bibitem[{\citenamefont{Ahle et~al.}(1999)}]{E802:1999hit}
\bibinfo{author}{\bibfnamefont{L.}~\bibnamefont{Ahle}} \bibnamefont{et~al.}
  (\bibinfo{collaboration}{E802 Collaboration}), \bibinfo{journal}{Phys. Rev.
  C} \textbf{\bibinfo{volume}{60}}, \bibinfo{pages}{064901}
  (\bibinfo{year}{1999}).

\bibitem[{\citenamefont{Armstrong et~al.}(2000)}]{E864:2000auv}
\bibinfo{author}{\bibfnamefont{T.~A.} \bibnamefont{Armstrong}}
  \bibnamefont{et~al.} (\bibinfo{collaboration}{E864 Collaboration}),
  \bibinfo{journal}{Phys. Rev. C} \textbf{\bibinfo{volume}{61}},
  \bibinfo{pages}{064908} (\bibinfo{year}{2000}), \eprint{nucl-ex/0003009}.

\bibitem[{\citenamefont{Reisdorf et~al.}(2010)}]{FOPI:2010xrt}
\bibinfo{author}{\bibfnamefont{W.}~\bibnamefont{Reisdorf}} \bibnamefont{et~al.}
  (\bibinfo{collaboration}{FOPI Collaboration}), \bibinfo{journal}{Nucl. Phys.
  A} \textbf{\bibinfo{volume}{848}}, \bibinfo{pages}{366}
  (\bibinfo{year}{2010}), \eprint{1005.3418}.

\bibitem[{\citenamefont{Abelev et~al.}(2010)}]{STAR:2010gyg}
\bibinfo{author}{\bibfnamefont{B.~I.} \bibnamefont{Abelev}}
  \bibnamefont{et~al.} (\bibinfo{collaboration}{STAR}),
  \bibinfo{journal}{Science} \textbf{\bibinfo{volume}{328}},
  \bibinfo{pages}{58} (\bibinfo{year}{2010}), \eprint{1003.2030}.

\bibitem[{\citenamefont{Agakishiev et~al.}(2011)}]{STAR:2011eej}
\bibinfo{author}{\bibfnamefont{H.}~\bibnamefont{Agakishiev}}
  \bibnamefont{et~al.} (\bibinfo{collaboration}{STAR Collaboration}),
  \bibinfo{journal}{Nature} \textbf{\bibinfo{volume}{473}},
  \bibinfo{pages}{353} (\bibinfo{year}{2011}), \bibinfo{note}{[Erratum: Nature
  475, 412 (2011)]}, \eprint{1103.3312}.

\bibitem[{\citenamefont{Adam et~al.}(2016)}]{ALICE:2015wav}
\bibinfo{author}{\bibfnamefont{J.}~\bibnamefont{Adam}} \bibnamefont{et~al.}
  (\bibinfo{collaboration}{ALICE Collaboration}), \bibinfo{journal}{Phys. Rev.
  C} \textbf{\bibinfo{volume}{93}}, \bibinfo{pages}{024917}
  (\bibinfo{year}{2016}), \eprint{1506.08951}.

\bibitem[{\citenamefont{Anticic et~al.}(2016)}]{NA49:2016qvu}
\bibinfo{author}{\bibfnamefont{T.}~\bibnamefont{Anticic}} \bibnamefont{et~al.}
  (\bibinfo{collaboration}{NA49 Collaboration}), \bibinfo{journal}{Phys. Rev.
  C} \textbf{\bibinfo{volume}{94}}, \bibinfo{pages}{044906}
  (\bibinfo{year}{2016}), \eprint{1606.04234}.

\bibitem[{\citenamefont{Adamczyk et~al.}(2016)}]{STAR:2016ydv}
\bibinfo{author}{\bibfnamefont{L.}~\bibnamefont{Adamczyk}} \bibnamefont{et~al.}
  (\bibinfo{collaboration}{STAR Collaboration}), \bibinfo{journal}{Phys. Rev.
  C} \textbf{\bibinfo{volume}{94}}, \bibinfo{pages}{034908}
  (\bibinfo{year}{2016}), \eprint{1601.07052}.

\bibitem[{\citenamefont{Acharya et~al.}(2018)}]{ALICE:2017xrp}
\bibinfo{author}{\bibfnamefont{S.}~\bibnamefont{Acharya}} \bibnamefont{et~al.}
  (\bibinfo{collaboration}{ALICE Collaboration}), \bibinfo{journal}{Phys. Rev.
  C} \textbf{\bibinfo{volume}{97}}, \bibinfo{pages}{024615}
  (\bibinfo{year}{2018}), \eprint{1709.08522}.

\bibitem[{\citenamefont{Adam et~al.}(2019)}]{STAR:2019sjh}
\bibinfo{author}{\bibfnamefont{J.}~\bibnamefont{Adam}} \bibnamefont{et~al.}
  (\bibinfo{collaboration}{STAR Collaboration}), \bibinfo{journal}{Phys. Rev.
  C} \textbf{\bibinfo{volume}{99}}, \bibinfo{pages}{064905}
  (\bibinfo{year}{2019}), \eprint{1903.11778}.

\bibitem[{\citenamefont{Abdulhamid et~al.}(2023)}]{PhysRevLett.130.202301}
\bibinfo{author}{\bibfnamefont{M.~I.} \bibnamefont{Abdulhamid}}
  \bibnamefont{et~al.} (\bibinfo{collaboration}{STAR Collaboration}),
  \bibinfo{journal}{Phys. Rev. Lett.} \textbf{\bibinfo{volume}{130}},
  \bibinfo{pages}{202301} (\bibinfo{year}{2023}), \eprint{2209.08058}.

\bibitem[{\citenamefont{Scheibl and Heinz}(1999)}]{Scheibl:1998tk}
\bibinfo{author}{\bibfnamefont{R.}~\bibnamefont{Scheibl}} \bibnamefont{and}
  \bibinfo{author}{\bibfnamefont{U.~W.} \bibnamefont{Heinz}},
  \bibinfo{journal}{Phys. Rev. C} \textbf{\bibinfo{volume}{59}},
  \bibinfo{pages}{1585} (\bibinfo{year}{1999}), \eprint{nucl-th/9809092}.

\bibitem[{\citenamefont{Oh et~al.}(2009)\citenamefont{Oh, Lin, and
  Ko}}]{Oh:2009gx}
\bibinfo{author}{\bibfnamefont{Y.}~\bibnamefont{Oh}},
  \bibinfo{author}{\bibfnamefont{Z.-W.} \bibnamefont{Lin}}, \bibnamefont{and}
  \bibinfo{author}{\bibfnamefont{C.~M.} \bibnamefont{Ko}},
  \bibinfo{journal}{Phys. Rev. C} \textbf{\bibinfo{volume}{80}},
  \bibinfo{pages}{064902} (\bibinfo{year}{2009}), \eprint{0910.1977}.

\bibitem[{\citenamefont{Andronic et~al.}(2011)\citenamefont{Andronic,
  Braun-Munzinger, Stachel, and Stocker}}]{Andronic:2010qu}
\bibinfo{author}{\bibfnamefont{A.}~\bibnamefont{Andronic}},
  \bibinfo{author}{\bibfnamefont{P.}~\bibnamefont{Braun-Munzinger}},
  \bibinfo{author}{\bibfnamefont{J.}~\bibnamefont{Stachel}}, \bibnamefont{and}
  \bibinfo{author}{\bibfnamefont{H.}~\bibnamefont{Stocker}},
  \bibinfo{journal}{Phys. Lett. B} \textbf{\bibinfo{volume}{697}},
  \bibinfo{pages}{203} (\bibinfo{year}{2011}), \eprint{1010.2995}.

\bibitem[{\citenamefont{Cleymans et~al.}(2011)\citenamefont{Cleymans, Kabana,
  Kraus, Oeschler, Redlich, and Sharma}}]{Cleymans:2011pe}
\bibinfo{author}{\bibfnamefont{J.}~\bibnamefont{Cleymans}},
  \bibinfo{author}{\bibfnamefont{S.}~\bibnamefont{Kabana}},
  \bibinfo{author}{\bibfnamefont{I.}~\bibnamefont{Kraus}},
  \bibinfo{author}{\bibfnamefont{H.}~\bibnamefont{Oeschler}},
  \bibinfo{author}{\bibfnamefont{K.}~\bibnamefont{Redlich}}, \bibnamefont{and}
  \bibinfo{author}{\bibfnamefont{N.}~\bibnamefont{Sharma}},
  \bibinfo{journal}{Phys. Rev. C} \textbf{\bibinfo{volume}{84}},
  \bibinfo{pages}{054916} (\bibinfo{year}{2011}), \eprint{1105.3719}.

\bibitem[{\citenamefont{Shah et~al.}(2016)\citenamefont{Shah, Ma, Chen, and
  Zhang}}]{Shah:2015oha}
\bibinfo{author}{\bibfnamefont{N.}~\bibnamefont{Shah}},
  \bibinfo{author}{\bibfnamefont{Y.~G.} \bibnamefont{Ma}},
  \bibinfo{author}{\bibfnamefont{J.~H.} \bibnamefont{Chen}}, \bibnamefont{and}
  \bibinfo{author}{\bibfnamefont{S.}~\bibnamefont{Zhang}},
  \bibinfo{journal}{Phys. Lett. B} \textbf{\bibinfo{volume}{754}},
  \bibinfo{pages}{6} (\bibinfo{year}{2016}), \eprint{1511.08266}.

\bibitem[{\citenamefont{Sun et~al.}(2017)\citenamefont{Sun, Chen, Ko, and
  Xu}}]{Sun:2017xrx}
\bibinfo{author}{\bibfnamefont{K.-J.} \bibnamefont{Sun}},
  \bibinfo{author}{\bibfnamefont{L.-W.} \bibnamefont{Chen}},
  \bibinfo{author}{\bibfnamefont{C.~M.} \bibnamefont{Ko}}, \bibnamefont{and}
  \bibinfo{author}{\bibfnamefont{Z.}~\bibnamefont{Xu}}, \bibinfo{journal}{Phys.
  Lett. B} \textbf{\bibinfo{volume}{774}}, \bibinfo{pages}{103}
  (\bibinfo{year}{2017}), \eprint{1702.07620}.

\bibitem[{\citenamefont{Andronic et~al.}(2018)\citenamefont{Andronic,
  Braun-Munzinger, Redlich, and Stachel}}]{Andronic:2017pug}
\bibinfo{author}{\bibfnamefont{A.}~\bibnamefont{Andronic}},
  \bibinfo{author}{\bibfnamefont{P.}~\bibnamefont{Braun-Munzinger}},
  \bibinfo{author}{\bibfnamefont{K.}~\bibnamefont{Redlich}}, \bibnamefont{and}
  \bibinfo{author}{\bibfnamefont{J.}~\bibnamefont{Stachel}},
  \bibinfo{journal}{Nature} \textbf{\bibinfo{volume}{561}},
  \bibinfo{pages}{321} (\bibinfo{year}{2018}), \eprint{1710.09425}.

\bibitem[{\citenamefont{Braun-Munzinger and
  D\"onigus}(2019)}]{Braun-Munzinger:2018hat}
\bibinfo{author}{\bibfnamefont{P.}~\bibnamefont{Braun-Munzinger}}
  \bibnamefont{and}
  \bibinfo{author}{\bibfnamefont{B.}~\bibnamefont{D\"onigus}},
  \bibinfo{journal}{Nucl. Phys. A} \textbf{\bibinfo{volume}{987}},
  \bibinfo{pages}{144} (\bibinfo{year}{2019}), \eprint{1809.04681}.

\bibitem[{\citenamefont{Shuryak and
  Torres-Rincon}(2020{\natexlab{a}})}]{Shuryak:2020yrs}
\bibinfo{author}{\bibfnamefont{E.}~\bibnamefont{Shuryak}} \bibnamefont{and}
  \bibinfo{author}{\bibfnamefont{J.~M.} \bibnamefont{Torres-Rincon}},
  \bibinfo{journal}{Eur. Phys. J. A} \textbf{\bibinfo{volume}{56}},
  \bibinfo{pages}{241} (\bibinfo{year}{2020}{\natexlab{a}}),
  \eprint{2005.14216}.

\bibitem[{\citenamefont{D\"onigus}(2020)}]{Donigus:2020ctf}
\bibinfo{author}{\bibfnamefont{B.}~\bibnamefont{D\"onigus}},
  \bibinfo{journal}{Int. J. Mod. Phys. E} \textbf{\bibinfo{volume}{29}},
  \bibinfo{pages}{2040001} (\bibinfo{year}{2020}), \eprint{2004.10544}.

\bibitem[{\citenamefont{Vovchenko et~al.}(2020)\citenamefont{Vovchenko,
  D\"onigus, Kardan, Lorenz, and Stoecker}}]{Vovchenko:2020dmv}
\bibinfo{author}{\bibfnamefont{V.}~\bibnamefont{Vovchenko}},
  \bibinfo{author}{\bibfnamefont{B.}~\bibnamefont{D\"onigus}},
  \bibinfo{author}{\bibfnamefont{B.}~\bibnamefont{Kardan}},
  \bibinfo{author}{\bibfnamefont{M.}~\bibnamefont{Lorenz}}, \bibnamefont{and}
  \bibinfo{author}{\bibfnamefont{H.}~\bibnamefont{Stoecker}},
  \bibinfo{journal}{Phys. Lett.} \textbf{\bibinfo{volume}{B}},
  \bibinfo{pages}{135746} (\bibinfo{year}{2020}), \eprint{2004.04411}.

\bibitem[{\citenamefont{Zhao et~al.}(2021)\citenamefont{Zhao, Sun, Ko, and
  Luo}}]{Zhao:2021dka}
\bibinfo{author}{\bibfnamefont{W.}~\bibnamefont{Zhao}},
  \bibinfo{author}{\bibfnamefont{K.-j.} \bibnamefont{Sun}},
  \bibinfo{author}{\bibfnamefont{C.~M.} \bibnamefont{Ko}}, \bibnamefont{and}
  \bibinfo{author}{\bibfnamefont{X.}~\bibnamefont{Luo}},
  \bibinfo{journal}{Phys. Lett. B} \textbf{\bibinfo{volume}{820}},
  \bibinfo{pages}{136571} (\bibinfo{year}{2021}), \eprint{2105.14204}.

\bibitem[{\citenamefont{Sun et~al.}(2018)\citenamefont{Sun, Chen, Ko, Pu, and
  Xu}}]{Sun:2018jhg}
\bibinfo{author}{\bibfnamefont{K.-J.} \bibnamefont{Sun}},
  \bibinfo{author}{\bibfnamefont{L.-W.} \bibnamefont{Chen}},
  \bibinfo{author}{\bibfnamefont{C.~M.} \bibnamefont{Ko}},
  \bibinfo{author}{\bibfnamefont{J.}~\bibnamefont{Pu}}, \bibnamefont{and}
  \bibinfo{author}{\bibfnamefont{Z.}~\bibnamefont{Xu}}, \bibinfo{journal}{Phys.
  Lett. B} \textbf{\bibinfo{volume}{781}}, \bibinfo{pages}{499}
  (\bibinfo{year}{2018}), \eprint{1801.09382}.

\bibitem[{\citenamefont{Shuryak and
  Torres-Rincon}(2020{\natexlab{b}})}]{Shuryak:2019ikv}
\bibinfo{author}{\bibfnamefont{E.}~\bibnamefont{Shuryak}} \bibnamefont{and}
  \bibinfo{author}{\bibfnamefont{J.~M.} \bibnamefont{Torres-Rincon}},
  \bibinfo{journal}{Phys. Rev. C} \textbf{\bibinfo{volume}{101}},
  \bibinfo{pages}{034914} (\bibinfo{year}{2020}{\natexlab{b}}),
  \eprint{1910.08119}.

\bibitem[{\citenamefont{Nara}(2019)}]{Nara:2019crj}
\bibinfo{author}{\bibfnamefont{Y.}~\bibnamefont{Nara}}, \bibinfo{journal}{EPJ
  Web Conf.} \textbf{\bibinfo{volume}{208}}, \bibinfo{pages}{11004}
  (\bibinfo{year}{2019}).

\bibitem[{\citenamefont{Weil et~al.}(2016)}]{SMASH:2016zqf}
\bibinfo{author}{\bibfnamefont{J.}~\bibnamefont{Weil}} \bibnamefont{et~al.}
  (\bibinfo{collaboration}{SMASH Collaboration}), \bibinfo{journal}{Phys. Rev.
  C} \textbf{\bibinfo{volume}{94}}, \bibinfo{pages}{054905}
  (\bibinfo{year}{2016}), \eprint{1606.06642}.

\bibitem[{\citenamefont{Bass et~al.}(1998)}]{Bass:1998ca}
\bibinfo{author}{\bibfnamefont{S.~A.} \bibnamefont{Bass}} \bibnamefont{et~al.},
  \bibinfo{journal}{Prog. Part. Nucl. Phys.} \textbf{\bibinfo{volume}{41}},
  \bibinfo{pages}{255} (\bibinfo{year}{1998}), \eprint{nucl-th/9803035}.

\bibitem[{\citenamefont{Gl\"a\ss{}el et~al.}(2022)\citenamefont{Gl\"a\ss{}el,
  Kireyeu, Voronyuk, Aichelin, Blume, Bratkovskaya, Coci, Kolesnikov, and
  Winn}}]{Glassel:2021rod}
\bibinfo{author}{\bibfnamefont{S.}~\bibnamefont{Gl\"a\ss{}el}},
  \bibinfo{author}{\bibfnamefont{V.}~\bibnamefont{Kireyeu}},
  \bibinfo{author}{\bibfnamefont{V.}~\bibnamefont{Voronyuk}},
  \bibinfo{author}{\bibfnamefont{J.}~\bibnamefont{Aichelin}},
  \bibinfo{author}{\bibfnamefont{C.}~\bibnamefont{Blume}},
  \bibinfo{author}{\bibfnamefont{E.}~\bibnamefont{Bratkovskaya}},
  \bibinfo{author}{\bibfnamefont{G.}~\bibnamefont{Coci}},
  \bibinfo{author}{\bibfnamefont{V.}~\bibnamefont{Kolesnikov}},
  \bibnamefont{and} \bibinfo{author}{\bibfnamefont{M.}~\bibnamefont{Winn}},
  \bibinfo{journal}{Phys. Rev. C} \textbf{\bibinfo{volume}{105}},
  \bibinfo{pages}{014908} (\bibinfo{year}{2022}), \eprint{2106.14839}.

\bibitem[{\citenamefont{Chen et~al.}(2003)\citenamefont{Chen, Ko, and
  Li}}]{Chen:2003qj}
\bibinfo{author}{\bibfnamefont{L.-W.} \bibnamefont{Chen}},
  \bibinfo{author}{\bibfnamefont{C.~M.} \bibnamefont{Ko}}, \bibnamefont{and}
  \bibinfo{author}{\bibfnamefont{B.-A.} \bibnamefont{Li}},
  \bibinfo{journal}{Phys. Rev. C} \textbf{\bibinfo{volume}{68}},
  \bibinfo{pages}{017601} (\bibinfo{year}{2003}), \eprint{nucl-th/0302068}.

\bibitem[{\citenamefont{Zhao et~al.}(2018)\citenamefont{Zhao, Zhu, Zheng, Ko,
  and Song}}]{Zhao:2018lyf}
\bibinfo{author}{\bibfnamefont{W.}~\bibnamefont{Zhao}},
  \bibinfo{author}{\bibfnamefont{L.}~\bibnamefont{Zhu}},
  \bibinfo{author}{\bibfnamefont{H.}~\bibnamefont{Zheng}},
  \bibinfo{author}{\bibfnamefont{C.~M.} \bibnamefont{Ko}}, \bibnamefont{and}
  \bibinfo{author}{\bibfnamefont{H.}~\bibnamefont{Song}},
  \bibinfo{journal}{Phys. Rev. C} \textbf{\bibinfo{volume}{98}},
  \bibinfo{pages}{054905} (\bibinfo{year}{2018}), \eprint{1807.02813}.

\bibitem[{\citenamefont{Adam et~al.}(2021)}]{STAR:2020dav}
\bibinfo{author}{\bibfnamefont{J.}~\bibnamefont{Adam}} \bibnamefont{et~al.}
  (\bibinfo{collaboration}{STAR Collaboration}), \bibinfo{journal}{Phys. Rev.
  C} \textbf{\bibinfo{volume}{103}}, \bibinfo{pages}{034908}
  (\bibinfo{year}{2021}), \eprint{2007.14005}.

\bibitem[{\citenamefont{Abdallah et~al.}(2022{\natexlab{a}})}]{STAR:2021fge}
\bibinfo{author}{\bibfnamefont{M.~S.} \bibnamefont{Abdallah}}
  \bibnamefont{et~al.} (\bibinfo{collaboration}{STAR Collaboration}),
  \bibinfo{journal}{Phys. Rev. Lett.} \textbf{\bibinfo{volume}{128}},
  \bibinfo{pages}{202303} (\bibinfo{year}{2022}{\natexlab{a}}),
  \eprint{2112.00240}.

\bibitem[{\citenamefont{Sun et~al.}(2008)}]{Sun:2008hg}
\bibinfo{author}{\bibfnamefont{Y.~J.} \bibnamefont{Sun}} \bibnamefont{et~al.},
  \bibinfo{journal}{Nucl. Instrum. Meth. A} \textbf{\bibinfo{volume}{593}},
  \bibinfo{pages}{307} (\bibinfo{year}{2008}), \eprint{0805.2459}.

\bibitem[{\citenamefont{Bieser et~al.}(2003)}]{Bieser:2002ah}
\bibinfo{author}{\bibfnamefont{F.~S.} \bibnamefont{Bieser}}
  \bibnamefont{et~al.}, \bibinfo{journal}{Nucl. Instrum. Meth. A}
  \textbf{\bibinfo{volume}{499}}, \bibinfo{pages}{766} (\bibinfo{year}{2003}).

\bibitem[{\citenamefont{Llope}(2012)}]{Llope:2012zz}
\bibinfo{author}{\bibfnamefont{W.~J.} \bibnamefont{Llope}}
  (\bibinfo{collaboration}{STAR Collaboration}), \bibinfo{journal}{Nucl.
  Instrum. Meth. A} \textbf{\bibinfo{volume}{661}}, \bibinfo{pages}{S110}
  (\bibinfo{year}{2012}).

\bibitem[{\citenamefont{Anderson et~al.}(2003)}]{Anderson:2003ur}
\bibinfo{author}{\bibfnamefont{M.}~\bibnamefont{Anderson}}
  \bibnamefont{et~al.}, \bibinfo{journal}{Nucl. Instrum. Meth. A}
  \textbf{\bibinfo{volume}{499}}, \bibinfo{pages}{659} (\bibinfo{year}{2003}),
  \eprint{nucl-ex/0301015}.

\bibitem[{\citenamefont{Shao et~al.}(2006)\citenamefont{Shao, Barannikova,
  Dong, Fisyak, Ruan, Sorensen, and Xu}}]{Shao:2005iu}
\bibinfo{author}{\bibfnamefont{M.}~\bibnamefont{Shao}},
  \bibinfo{author}{\bibfnamefont{O.~Y.} \bibnamefont{Barannikova}},
  \bibinfo{author}{\bibfnamefont{X.}~\bibnamefont{Dong}},
  \bibinfo{author}{\bibfnamefont{Y.}~\bibnamefont{Fisyak}},
  \bibinfo{author}{\bibfnamefont{L.}~\bibnamefont{Ruan}},
  \bibinfo{author}{\bibfnamefont{P.}~\bibnamefont{Sorensen}}, \bibnamefont{and}
  \bibinfo{author}{\bibfnamefont{Z.}~\bibnamefont{Xu}}, \bibinfo{journal}{Nucl.
  Instrum. Meth. A} \textbf{\bibinfo{volume}{558}}, \bibinfo{pages}{419}
  (\bibinfo{year}{2006}), \eprint{nucl-ex/0505026}.

\bibitem[{\citenamefont{Abdallah et~al.}(2021)}]{STAR:2021iop}
\bibinfo{author}{\bibfnamefont{M.}~\bibnamefont{Abdallah}} \bibnamefont{et~al.}
  (\bibinfo{collaboration}{STAR}), \bibinfo{journal}{Phys. Rev. C}
  \textbf{\bibinfo{volume}{104}}, \bibinfo{pages}{024902}
  (\bibinfo{year}{2021}), \eprint{2101.12413}.

\bibitem[{\citenamefont{PFANZAGL and SHEYNIN}(1996)}]{10.1093/biomet/83.4.891}
\bibinfo{author}{\bibfnamefont{J.}~\bibnamefont{PFANZAGL}} \bibnamefont{and}
  \bibinfo{author}{\bibfnamefont{O.}~\bibnamefont{SHEYNIN}},
  \bibinfo{journal}{Biometrika} \textbf{\bibinfo{volume}{83}},
  \bibinfo{pages}{891} (\bibinfo{year}{1996}), ISSN \bibinfo{issn}{0006-3444},
  \urlprefix\url{https://doi.org/10.1093/biomet/83.4.891}.

\bibitem[{\citenamefont{Adler et~al.}(2001{\natexlab{a}})}]{STAR:2001mal}
\bibinfo{author}{\bibfnamefont{C.}~\bibnamefont{Adler}} \bibnamefont{et~al.}
  (\bibinfo{collaboration}{STAR Collaboration}), \bibinfo{journal}{Phys. Rev.
  Lett.} \textbf{\bibinfo{volume}{87}}, \bibinfo{pages}{262302}
  (\bibinfo{year}{2001}{\natexlab{a}}), \eprint{nucl-ex/0110009}.

\bibitem[{\citenamefont{Adam et~al.}(2020)}]{STAR:2019bjj}
\bibinfo{author}{\bibfnamefont{J.}~\bibnamefont{Adam}} \bibnamefont{et~al.}
  (\bibinfo{collaboration}{STAR}), \bibinfo{journal}{Phys. Rev. C}
  \textbf{\bibinfo{volume}{102}}, \bibinfo{pages}{034909}
  (\bibinfo{year}{2020}), \eprint{1906.03732}.

\bibitem[{\citenamefont{Adcox et~al.}(2002)}]{PHENIX:2002svd}
\bibinfo{author}{\bibfnamefont{K.}~\bibnamefont{Adcox}} \bibnamefont{et~al.}
  (\bibinfo{collaboration}{PHENIX Collaboration}), \bibinfo{journal}{Phys. Rev.
  Lett.} \textbf{\bibinfo{volume}{89}}, \bibinfo{pages}{092302}
  (\bibinfo{year}{2002}), \eprint{nucl-ex/0204007}.

\bibitem[{\citenamefont{Anticic et~al.}(2011)}]{NA49:2010lhg}
\bibinfo{author}{\bibfnamefont{T.}~\bibnamefont{Anticic}} \bibnamefont{et~al.}
  (\bibinfo{collaboration}{NA49 Collaboration}), \bibinfo{journal}{Phys. Rev.
  C} \textbf{\bibinfo{volume}{83}}, \bibinfo{pages}{014901}
  (\bibinfo{year}{2011}), \eprint{1009.1747}.

\bibitem[{\citenamefont{Oliinychenko et~al.}(2021)\citenamefont{Oliinychenko,
  Shen, and Koch}}]{Oliinychenko:2020znl}
\bibinfo{author}{\bibfnamefont{D.}~\bibnamefont{Oliinychenko}},
  \bibinfo{author}{\bibfnamefont{C.}~\bibnamefont{Shen}}, \bibnamefont{and}
  \bibinfo{author}{\bibfnamefont{V.}~\bibnamefont{Koch}}
  (\bibinfo{collaboration}{SMASH Collaboration}), \bibinfo{journal}{Phys. Rev.
  C} \textbf{\bibinfo{volume}{103}}, \bibinfo{pages}{034913}
  (\bibinfo{year}{2021}), \eprint{2009.01915}.

\bibitem[{\citenamefont{Abdallah et~al.}(2022{\natexlab{b}})}]{STAR:2021hyx}
\bibinfo{author}{\bibfnamefont{M.~S.} \bibnamefont{Abdallah}}
  \bibnamefont{et~al.} (\bibinfo{collaboration}{STAR Collaboration}),
  \bibinfo{journal}{Phys. Lett. B} \textbf{\bibinfo{volume}{831}},
  \bibinfo{pages}{137152} (\bibinfo{year}{2022}{\natexlab{b}}),
  \eprint{2108.00924}.

\bibitem[{\citenamefont{Tanabashi et~al.}(2018)}]{ParticleDataGroup:2018ovx}
\bibinfo{author}{\bibfnamefont{M.}~\bibnamefont{Tanabashi}}
  \bibnamefont{et~al.} (\bibinfo{collaboration}{Particle Data Group}),
  \bibinfo{journal}{Phys. Rev. D} \textbf{\bibinfo{volume}{98}},
  \bibinfo{pages}{030001} (\bibinfo{year}{2018}).

\bibitem[{\citenamefont{Yong et~al.}(2022)\citenamefont{Yong, Li, Xiao, and
  Lin}}]{Yong:2022pyb}
\bibinfo{author}{\bibfnamefont{G.-C.} \bibnamefont{Yong}},
  \bibinfo{author}{\bibfnamefont{B.-A.} \bibnamefont{Li}},
  \bibinfo{author}{\bibfnamefont{Z.-G.} \bibnamefont{Xiao}}, \bibnamefont{and}
  \bibinfo{author}{\bibfnamefont{Z.-W.} \bibnamefont{Lin}},
  \bibinfo{journal}{Phys. Rev. C} \textbf{\bibinfo{volume}{106}},
  \bibinfo{pages}{024902} (\bibinfo{year}{2022}), \eprint{2206.10766}.

\bibitem[{\citenamefont{Andronic et~al.}(2006)\citenamefont{Andronic,
  Braun-Munzinger, and Stachel}}]{Andronic:2005yp}
\bibinfo{author}{\bibfnamefont{A.}~\bibnamefont{Andronic}},
  \bibinfo{author}{\bibfnamefont{P.}~\bibnamefont{Braun-Munzinger}},
  \bibnamefont{and} \bibinfo{author}{\bibfnamefont{J.}~\bibnamefont{Stachel}},
  \bibinfo{journal}{Nucl. Phys. A} \textbf{\bibinfo{volume}{772}},
  \bibinfo{pages}{167} (\bibinfo{year}{2006}), \eprint{nucl-th/0511071}.

\bibitem[{\citenamefont{Schnedermann et~al.}(1993)\citenamefont{Schnedermann,
  Sollfrank, and Heinz}}]{Schnedermann:1993ws}
\bibinfo{author}{\bibfnamefont{E.}~\bibnamefont{Schnedermann}},
  \bibinfo{author}{\bibfnamefont{J.}~\bibnamefont{Sollfrank}},
  \bibnamefont{and} \bibinfo{author}{\bibfnamefont{U.~W.} \bibnamefont{Heinz}},
  \bibinfo{journal}{Phys. Rev. C} \textbf{\bibinfo{volume}{48}},
  \bibinfo{pages}{2462} (\bibinfo{year}{1993}), \eprint{nucl-th/9307020}.

\bibitem[{\citenamefont{Andronic et~al.}(2019)\citenamefont{Andronic,
  Braun-Munzinger, K\"ohler, and Stachel}}]{Andronic:2018vqh}
\bibinfo{author}{\bibfnamefont{A.}~\bibnamefont{Andronic}},
  \bibinfo{author}{\bibfnamefont{P.}~\bibnamefont{Braun-Munzinger}},
  \bibinfo{author}{\bibfnamefont{M.~K.} \bibnamefont{K\"ohler}},
  \bibnamefont{and} \bibinfo{author}{\bibfnamefont{J.}~\bibnamefont{Stachel}},
  \bibinfo{journal}{Nucl. Phys. A} \textbf{\bibinfo{volume}{982}},
  \bibinfo{pages}{759} (\bibinfo{year}{2019}), \eprint{1807.01236}.

\bibitem[{\citenamefont{Bondorf et~al.}(1995)\citenamefont{Bondorf, Botvina,
  Ilinov, Mishustin, and Sneppen}}]{Bondorf:1995ua}
\bibinfo{author}{\bibfnamefont{J.~P.} \bibnamefont{Bondorf}},
  \bibinfo{author}{\bibfnamefont{A.~S.} \bibnamefont{Botvina}},
  \bibinfo{author}{\bibfnamefont{A.~S.} \bibnamefont{Ilinov}},
  \bibinfo{author}{\bibfnamefont{I.~N.} \bibnamefont{Mishustin}},
  \bibnamefont{and} \bibinfo{author}{\bibfnamefont{K.}~\bibnamefont{Sneppen}},
  \bibinfo{journal}{Phys. Rept.} \textbf{\bibinfo{volume}{257}},
  \bibinfo{pages}{133} (\bibinfo{year}{1995}).

\bibitem[{\citenamefont{Bratkovskaya et~al.}(2023)\citenamefont{Bratkovskaya,
  Gl\"assel, Kireyeu, Aichelin, Bleicher, Blume, Coci, Kolesnikov, Steinheimer,
  and Voronyuk}}]{Bratkovskaya:2022vqi}
\bibinfo{author}{\bibfnamefont{E.}~\bibnamefont{Bratkovskaya}},
  \bibinfo{author}{\bibfnamefont{S.}~\bibnamefont{Gl\"assel}},
  \bibinfo{author}{\bibfnamefont{V.}~\bibnamefont{Kireyeu}},
  \bibinfo{author}{\bibfnamefont{J.}~\bibnamefont{Aichelin}},
  \bibinfo{author}{\bibfnamefont{M.}~\bibnamefont{Bleicher}},
  \bibinfo{author}{\bibfnamefont{C.}~\bibnamefont{Blume}},
  \bibinfo{author}{\bibfnamefont{G.}~\bibnamefont{Coci}},
  \bibinfo{author}{\bibfnamefont{V.}~\bibnamefont{Kolesnikov}},
  \bibinfo{author}{\bibfnamefont{J.}~\bibnamefont{Steinheimer}},
  \bibnamefont{and} \bibinfo{author}{\bibfnamefont{V.}~\bibnamefont{Voronyuk}},
  \bibinfo{journal}{EPJ Web Conf.} \textbf{\bibinfo{volume}{276}},
  \bibinfo{pages}{03005} (\bibinfo{year}{2023}), \eprint{2208.11802}.

\bibitem[{\citenamefont{Coci et~al.}(2023)\citenamefont{Coci, Gl\"a\ss{}el,
  Kireyeu, Aichelin, Blume, Bratkovskaya, Kolesnikov, and
  Voronyuk}}]{Coci:2023daq}
\bibinfo{author}{\bibfnamefont{G.}~\bibnamefont{Coci}},
  \bibinfo{author}{\bibfnamefont{S.}~\bibnamefont{Gl\"a\ss{}el}},
  \bibinfo{author}{\bibfnamefont{V.}~\bibnamefont{Kireyeu}},
  \bibinfo{author}{\bibfnamefont{J.}~\bibnamefont{Aichelin}},
  \bibinfo{author}{\bibfnamefont{C.}~\bibnamefont{Blume}},
  \bibinfo{author}{\bibfnamefont{E.}~\bibnamefont{Bratkovskaya}},
  \bibinfo{author}{\bibfnamefont{V.}~\bibnamefont{Kolesnikov}},
  \bibnamefont{and} \bibinfo{author}{\bibfnamefont{V.}~\bibnamefont{Voronyuk}},
  \bibinfo{journal}{Phys. Rev. C} \textbf{\bibinfo{volume}{108}},
  \bibinfo{pages}{014902} (\bibinfo{year}{2023}), \eprint{2303.02279}.

\bibitem[{\citenamefont{Moulin and Liu}(1999)}]{Moulin:1999mp}
\bibinfo{author}{\bibfnamefont{P.}~\bibnamefont{Moulin}} \bibnamefont{and}
  \bibinfo{author}{\bibfnamefont{J.}~\bibnamefont{Liu}}, \bibinfo{journal}{IEEE
  Transactions on Information Theory} \textbf{\bibinfo{volume}{45}},
  \bibinfo{pages}{909} (\bibinfo{year}{1999}).

\bibitem[{\citenamefont{Verdoolaege and Scheunders}(2011)}]{2011Geodesics}
\bibinfo{author}{\bibfnamefont{G.}~\bibnamefont{Verdoolaege}} \bibnamefont{and}
  \bibinfo{author}{\bibfnamefont{P.}~\bibnamefont{Scheunders}},
  \bibinfo{journal}{International Journal of Computer Vision}
  \textbf{\bibinfo{volume}{95}}, \bibinfo{pages}{265} (\bibinfo{year}{2011}),
  \urlprefix\url{https://doi.org/10.1007/s11263-011-0448-9}.

\bibitem[{\citenamefont{Verdoolaege and Scheunders}(2012)}]{2012On}
\bibinfo{author}{\bibfnamefont{G.}~\bibnamefont{Verdoolaege}} \bibnamefont{and}
  \bibinfo{author}{\bibfnamefont{P.}~\bibnamefont{Scheunders}},
  \bibinfo{journal}{Journal of Mathematical Imaging and Vision}
  \textbf{\bibinfo{volume}{43}}, \bibinfo{pages}{180} (\bibinfo{year}{2012}),
  \urlprefix\url{https://doi.org/10.1007/s10851-011-0297-8}.

\bibitem[{\citenamefont{Armstrong et~al.}(1999)}]{E864:1999zqh}
\bibinfo{author}{\bibfnamefont{T.~A.} \bibnamefont{Armstrong}}
  \bibnamefont{et~al.} (\bibinfo{collaboration}{E864 Collaboration}),
  \bibinfo{journal}{Phys. Rev. Lett.} \textbf{\bibinfo{volume}{83}},
  \bibinfo{pages}{5431} (\bibinfo{year}{1999}), \eprint{nucl-ex/9907002}.

\bibitem[{\citenamefont{Braun-Munzinger and
  Stachel}(1995)}]{Braun-Munzinger:1994zkz}
\bibinfo{author}{\bibfnamefont{P.}~\bibnamefont{Braun-Munzinger}}
  \bibnamefont{and} \bibinfo{author}{\bibfnamefont{J.}~\bibnamefont{Stachel}},
  \bibinfo{journal}{J. Phys. G} \textbf{\bibinfo{volume}{21}},
  \bibinfo{pages}{L17} (\bibinfo{year}{1995}), \eprint{nucl-th/9412035}.

\bibitem[{\citenamefont{Steinheimer et~al.}(2012)\citenamefont{Steinheimer, Xu,
  Gudima, Botvina, Mishustin, Bleicher, and Stocker}}]{Steinheimer:2012gq}
\bibinfo{author}{\bibfnamefont{J.}~\bibnamefont{Steinheimer}},
  \bibinfo{author}{\bibfnamefont{Z.}~\bibnamefont{Xu}},
  \bibinfo{author}{\bibfnamefont{K.}~\bibnamefont{Gudima}},
  \bibinfo{author}{\bibfnamefont{A.}~\bibnamefont{Botvina}},
  \bibinfo{author}{\bibfnamefont{I.}~\bibnamefont{Mishustin}},
  \bibinfo{author}{\bibfnamefont{M.}~\bibnamefont{Bleicher}}, \bibnamefont{and}
  \bibinfo{author}{\bibfnamefont{H.}~\bibnamefont{Stocker}},
  \bibinfo{journal}{J. Phys. Conf. Ser.} \textbf{\bibinfo{volume}{389}},
  \bibinfo{pages}{012022} (\bibinfo{year}{2012}).

\bibitem[{\citenamefont{Vovchenko et~al.}(2016)\citenamefont{Vovchenko, Begun,
  and Gorenstein}}]{Vovchenko:2015idt}
\bibinfo{author}{\bibfnamefont{V.}~\bibnamefont{Vovchenko}},
  \bibinfo{author}{\bibfnamefont{V.~V.} \bibnamefont{Begun}}, \bibnamefont{and}
  \bibinfo{author}{\bibfnamefont{M.~I.} \bibnamefont{Gorenstein}},
  \bibinfo{journal}{Phys. Rev. C} \textbf{\bibinfo{volume}{93}},
  \bibinfo{pages}{064906} (\bibinfo{year}{2016}), \eprint{1512.08025}.

\bibitem[{\citenamefont{Butler and Pearson}(1963)}]{Butler:1963pp}
\bibinfo{author}{\bibfnamefont{S.~T.} \bibnamefont{Butler}} \bibnamefont{and}
  \bibinfo{author}{\bibfnamefont{C.~A.} \bibnamefont{Pearson}},
  \bibinfo{journal}{Phys. Rev.} \textbf{\bibinfo{volume}{129}},
  \bibinfo{pages}{836} (\bibinfo{year}{1963}).

\bibitem[{\citenamefont{Yu et~al.}(2020)\citenamefont{Yu, Zhang, and
  Luo}}]{Yu:2018kvh}
\bibinfo{author}{\bibfnamefont{N.}~\bibnamefont{Yu}},
  \bibinfo{author}{\bibfnamefont{D.}~\bibnamefont{Zhang}}, \bibnamefont{and}
  \bibinfo{author}{\bibfnamefont{X.}~\bibnamefont{Luo}},
  \bibinfo{journal}{Chin. Phys. C} \textbf{\bibinfo{volume}{44}},
  \bibinfo{pages}{014002} (\bibinfo{year}{2020}), \eprint{1812.04291}.

\bibitem[{\citenamefont{Kolesnikov}(2008)}]{Kolesnikov:2007ps}
\bibinfo{author}{\bibfnamefont{V.~I.} \bibnamefont{Kolesnikov}}
  (\bibinfo{collaboration}{NA49}), \bibinfo{journal}{J. Phys. Conf. Ser.}
  \textbf{\bibinfo{volume}{110}}, \bibinfo{pages}{032010}
  (\bibinfo{year}{2008}), \eprint{0710.5118}.

\bibitem[{\citenamefont{Bearden et~al.}(2002)}]{Bearden:2002ta}
\bibinfo{author}{\bibfnamefont{I.~G.} \bibnamefont{Bearden}}
  \bibnamefont{et~al.}, \bibinfo{journal}{Eur. Phys. J. C}
  \textbf{\bibinfo{volume}{23}}, \bibinfo{pages}{237} (\bibinfo{year}{2002}).

\bibitem[{\citenamefont{Bennett et~al.}(1998)}]{E878:1998vna}
\bibinfo{author}{\bibfnamefont{M.~J.} \bibnamefont{Bennett}}
  \bibnamefont{et~al.} (\bibinfo{collaboration}{E878 Collaboration}),
  \bibinfo{journal}{Phys. Rev. C} \textbf{\bibinfo{volume}{58}},
  \bibinfo{pages}{1155} (\bibinfo{year}{1998}).

\bibitem[{\citenamefont{Adler et~al.}(2005)}]{PHENIX:2004vqi}
\bibinfo{author}{\bibfnamefont{S.~S.} \bibnamefont{Adler}} \bibnamefont{et~al.}
  (\bibinfo{collaboration}{PHENIX Collaboration}), \bibinfo{journal}{Phys. Rev.
  Lett.} \textbf{\bibinfo{volume}{94}}, \bibinfo{pages}{122302}
  (\bibinfo{year}{2005}), \eprint{nucl-ex/0406004}.

\bibitem[{\citenamefont{Adler et~al.}(2001{\natexlab{b}})}]{STAR:2001pbk}
\bibinfo{author}{\bibfnamefont{C.}~\bibnamefont{Adler}} \bibnamefont{et~al.}
  (\bibinfo{collaboration}{STAR Collaboration}), \bibinfo{journal}{Phys. Rev.
  Lett.} \textbf{\bibinfo{volume}{87}}, \bibinfo{pages}{262301}
  (\bibinfo{year}{2001}{\natexlab{b}}), \bibinfo{note}{[Erratum: Phys.Rev.Lett.
  87, 279902 (2001)]}, \eprint{nucl-ex/0108022}.

\bibitem[{\citenamefont{Csernai and Kapusta}(1986)}]{Csernai:1986qf}
\bibinfo{author}{\bibfnamefont{L.~P.} \bibnamefont{Csernai}} \bibnamefont{and}
  \bibinfo{author}{\bibfnamefont{J.~I.} \bibnamefont{Kapusta}},
  \bibinfo{journal}{Phys. Rept.} \textbf{\bibinfo{volume}{131}},
  \bibinfo{pages}{223} (\bibinfo{year}{1986}).

\bibitem[{\citenamefont{Sun et~al.}(2022)\citenamefont{Sun, Zhou, Chen, Ko, Li,
  Wang, and Xu}}]{Sun:2022cxp}
\bibinfo{author}{\bibfnamefont{K.-J.} \bibnamefont{Sun}},
  \bibinfo{author}{\bibfnamefont{W.-H.} \bibnamefont{Zhou}},
  \bibinfo{author}{\bibfnamefont{L.-W.} \bibnamefont{Chen}},
  \bibinfo{author}{\bibfnamefont{C.~M.} \bibnamefont{Ko}},
  \bibinfo{author}{\bibfnamefont{F.}~\bibnamefont{Li}},
  \bibinfo{author}{\bibfnamefont{R.}~\bibnamefont{Wang}}, \bibnamefont{and}
  \bibinfo{author}{\bibfnamefont{J.}~\bibnamefont{Xu}} (\bibinfo{year}{2022}),
  \eprint{2205.11010}.

\bibitem[{\citenamefont{Sun et~al.}(2019)\citenamefont{Sun, Ko, and
  D\"onigus}}]{Sun:2018mqq}
\bibinfo{author}{\bibfnamefont{K.-J.} \bibnamefont{Sun}},
  \bibinfo{author}{\bibfnamefont{C.~M.} \bibnamefont{Ko}}, \bibnamefont{and}
  \bibinfo{author}{\bibfnamefont{B.}~\bibnamefont{D\"onigus}},
  \bibinfo{journal}{Phys. Lett. B} \textbf{\bibinfo{volume}{792}},
  \bibinfo{pages}{132} (\bibinfo{year}{2019}), \eprint{1812.05175}.

\bibitem[{\citenamefont{Nagle et~al.}(1996)\citenamefont{Nagle, Kumar,
  Kusnezov, Sorge, and Mattiello}}]{Nagle:1996vp}
\bibinfo{author}{\bibfnamefont{J.~L.} \bibnamefont{Nagle}},
  \bibinfo{author}{\bibfnamefont{B.~S.} \bibnamefont{Kumar}},
  \bibinfo{author}{\bibfnamefont{D.}~\bibnamefont{Kusnezov}},
  \bibinfo{author}{\bibfnamefont{H.}~\bibnamefont{Sorge}}, \bibnamefont{and}
  \bibinfo{author}{\bibfnamefont{R.}~\bibnamefont{Mattiello}},
  \bibinfo{journal}{Phys. Rev. C} \textbf{\bibinfo{volume}{53}},
  \bibinfo{pages}{367} (\bibinfo{year}{1996}).

\bibitem[{\citenamefont{Das~Gupta and Mekjian}(1981)}]{DasGupta:1981xx}
\bibinfo{author}{\bibfnamefont{S.}~\bibnamefont{Das~Gupta}} \bibnamefont{and}
  \bibinfo{author}{\bibfnamefont{A.~Z.} \bibnamefont{Mekjian}},
  \bibinfo{journal}{Phys. Rept.} \textbf{\bibinfo{volume}{72}},
  \bibinfo{pages}{131} (\bibinfo{year}{1981}).

\end{thebibliography}

\end{document}